\pdfoutput=1

\documentclass[letterpaper,10pt,twoside]{article}
\usepackage[nottoc]{tocbibind}

\usepackage[utf8]{inputenc}
\usepackage[english]{babel} 

\usepackage[T1]{fontenc}
\usepackage{lmodern}

\usepackage{amssymb,amsmath,amsthm,mathtools} 
\usepackage[only,llbracket,rrbracket,llparenthesis,rrparenthesis,boxslash]{stmaryrd} 

\usepackage[hyperref]{xcolor}
\usepackage{transparent}
\usepackage{graphicx}
\graphicspath{{images/}}
\usepackage[font=small,skip=5pt]{caption} 
\usepackage{hyperref} 
\usepackage{geometry} 
\usepackage[frame,arc,knot,matrix]{xy}

\newcommand{\rxy}[1]{{\begin{xy} 0;<1mm,0mm>:<0mm,1mm>::0;0,#1
\end{xy}}}

\newcommand{\im}{\mathrm{i}}
\newcommand{\K}{\mathbb{K}}
\newcommand{\N}{\mathbb{N}}
\newcommand{\R}{\mathbb{R}}
\newcommand{\bC}{\mathbb{C}}
\newcommand{\defeq}{\coloneqq}
\newcommand{\tens}{\otimes}
\newcommand{\bc}{\mathcal{B}^+}
\newcommand{\bcg}{\mathcal{B}}
\newcommand{\pr}{\mathcal{P}}
\newcommand{\prp}{\mathcal{P}^+}
\newcommand{\np}{\boxslash}
\newcommand{\zp}{\mathbf{0}}
\newcommand{\lv}{\llbracket}
\newcommand{\rv}{\rrbracket}
\newcommand{\lb}{\llparenthesis}
\newcommand{\rb}{\rrparenthesis}
\DeclareMathOperator{\pcomp}{\diamond}
\newcommand{\cC}{\mathcal{C}}
\newcommand{\cM}{\mathcal{M}}
\DeclareMathOperator{\id}{id}
\DeclareMathOperator{\ev}{\mathrm{ev}}
\DeclareMathOperator{\coev}{\mathrm{coev}}
\newcommand{\one}{\mathbf{1}}
\DeclareMathOperator{\ip}{\mathrm{ip}}
\DeclareMathOperator{\coip}{\mathrm{coip}}
\newcommand{\toi}{\hookrightarrow}
\newcommand{\cobs}{\mathcal{C}}
\newcommand{\qobs}{\mathcal{O}}
\DeclareMathOperator{\cocomp}{\bullet}
\DeclareMathOperator{\qcomp}{\circ}
\newcommand{\xd}{\mathrm{d}}

\newcommand{\cH}{\mathcal{H}}
\newcommand{\cS}{\mathcal{S}}
\newcommand{\cA}{\mathcal{A}}
\newcommand{\po}{\mathsf{P}}
\newcommand{\op}{\mathcal{B}}
\newcommand{\rop}{\mathcal{B}^{\R}}
\newcommand{\pop}{\mathcal{B}^{+}}
\DeclareMathOperator{\tr}{\mathrm{tr}}
\newcommand{\npe}{T}
\newcommand{\ou}{\mathbf{e}}
\newcommand{\lbl}{\texttt}
\newcommand{\cN}{\mathcal{N}}
\newcommand{\cI}{\mathcal{I}}

\newcommand{\keyword}{\emph}

\theoremstyle{definition}
\newtheorem{dfn}{Definition}[section]

\newtheorem{rem}[dfn]{Remark}
\theoremstyle{plain}

\newtheorem{prop}[dfn]{Proposition}

\allowdisplaybreaks

\begin{document}

\begin{titlepage}
\title{\textbf{A local and operational framework\\ for the foundations of physics}}
\author{Robert Oeckl\footnote{email: robert@matmor.unam.mx}\\ \\
Centro de Ciencias Matemáticas,\\
Universidad Nacional Autónoma de México,\\
C.P.~58190, Morelia, Michoacán, Mexico}
\date{UNAM-CCM-2016-2\\ 27 October 2016\\ 27 October 2017 (v2)\\ 7 November 2019 (v3)}

\maketitle

\vspace{\stretch{1}}

\begin{abstract}

We discuss a novel framework for physical theories that is based on the principles of locality and operationalism. It generalizes and unifies previous frameworks, including the standard formulation of quantum theory, the convex operational framework and Segal's approach to quantum field theory. It is capable of encoding both classical and quantum (field) theories, implements spacetime locality in a manifest way and contains the complete modern notion of measurement in the quantum case. Its mathematical content can be condensed into a set of axioms that are similar to those of Atiyah and Segal. This is supplemented by two basic rules for extracting probabilities or expectation values for measurement processes.
The framework, called the positive formalism, is derived in three completely different ways. One derivation is from first principles, one starts with classical field theory and one with quantum field theory. The latter derivation arose previously in the programme of the general boundary formulation of quantum theory. As in the convex operational framework, the difference between classical and quantum theories essentially arises from certain partially ordered vector spaces being either lattices or anti-lattices.
If we add the ad hoc ingredient of imposing anti-lattice structures, the derivation from first principles may be seen as a reconstruction of quantum theory.
Among other things, the positive formalism suggests a statistical approach to classical field theories with dynamical metric, provides a common ground for quantum information theory and quantum field theory, introduces a notion of local measurement into quantum field theory, and suggests a new perspective on quantum gravity by removing the incompatibility with general relativistic principles.
The positive formalism as a framework for quantum theory is in conflict with various interpretations or modifications of quantum theory, including physical collapse theories, many-worlds interpretations, and non-local hidden variable theories.

\end{abstract}

\vspace{\stretch{1}}
\end{titlepage}

\tableofcontents

\section{Introduction}

\subsection{Frameworks for physical theories}

Physical \emph{theories} or \emph{models} are usually not formulated ab initio, but rely on a \emph{framework} or \emph{meta-theory} into which they are embedded. Such a framework determines basic mathematical structures, relations between these structures and most importantly, rules that prescribe how these structures may be related to the real world.

One such framework is \emph{Hamiltonian mechanics}, dating in its essentials to the 19th century. A principal ingredient of this framework is a notion of \emph{time}. The objective of the framework is the description of the evolution of mechanical systems in time. Another of its principal notions is thus that of a \emph{state space} or \emph{phase space}. A physical system at a given time is represented by a point in this phase space. Dynamics is the description of motion in phase space. There are further mathematical structures to accomplish this. Thus, a phase space is a \emph{symplectic manifold} (or something slightly more general), there is a \emph{Hamiltonian function} on phase space etc.
The relation between mathematical structures and the real world is usually not spelled out explicitly, because it is considered obvious. Namely, the universe (spacetime) is filled with particles and fields. A point in phase space specifies all particle positions and momenta as well as all field values and their derivatives in all of \emph{space}. Thus, a trajectory in phase space corresponds to a possible physical reality. This direct correspondence between mathematical structures and physical reality in spacetime is the hallmark of classical physics.

For the emerging quantum mechanics in the 1920s a new framework had to be developed. This incorporated new mathematical objects, such as a complex Hilbert space and operators on this space. However, the overall structure was kept in close analogy to Hamiltonian mechanics. Notably, there is a notion of time, a notion of state space, a Hamiltonian (now an operator instead of a function) describing infinitesimal evolution. The most crucial difference, unfortunately somewhat overshadowed by this suggestive analogy, lies in the relation between the mathematical structures and the real world. Instead of providing an ``image'' of reality in space(time), quantum mechanics comes with an explicit and non-trivial notion of \emph{measurement}. It is only through measurement processes that an \emph{observer} may relate the mathematical objects of the theory to properties of the real world. This is an instance of the principle of \emph{operationalism}.

\emph{Statistical mechanics} is set in a \emph{statistical} version of the Hamiltonian framework. There states are statistical distributions on phase space. These may describe ensembles or degrees of belief about a ``true state''. \emph{Statistical quantum mechanics} is set in a counterpart of the Hilbert space formalism of quantum mechanics where states are described by \emph{density matrices} or \emph{density operators} acting on Hilbert space. Despite the word ``statistical'', \emph{density operators} are not probability distributions and do not admit a general ensemble interpretation. We also emphasize that an adequate codification of the theory of measurement in quantum mechanics is only achieved in the statistical setting which permits a full formalization of the principle of operationalism. Describing measurement in the setting restricted to states as vectors in Hilbert space is awkward and indirect by comparison.

\begin{figure}
\begin{center}
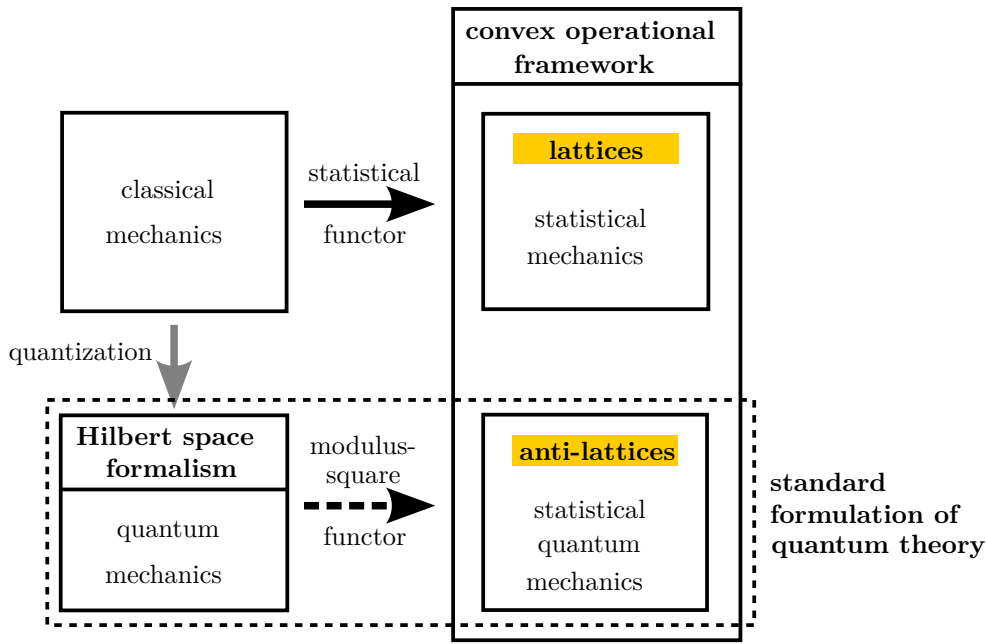
\end{center}
\caption{Time-evolution frameworks for physical theories.}
\label{fig:found_scheme_evol}
\end{figure}

Both classical statistical and quantum statistical frameworks can be brought together into a unified framework known as the \emph{convex operational framework}, see e.g.\ \cite{BaWi:olincatframe}. The central mathematical object here is that of a \emph{partially ordered vector space} taking the role of state space. In the classical case this is a space of functions or measures on phase space while in the quantum case it is the space of self-adjoint operators on Hilbert space. The former can be characterized by being a \emph{lattice} and the latter (roughly) by being an \emph{anti-lattice} (see Subsections~\ref{sec:cspecial} and \ref{sec:qchar}). Crucially, not only the basic mathematical objects, but also the rules for extracting measurable quantities from these can be formulated uniformly, not depending on the theory being classical or quantum.

Figure~\ref{fig:found_scheme_evol} illustrates the different frameworks and their relations. Given a classical mechanical model in the Hamiltonian framework (upper left) we can immediately set up the statistical counterpart (upper right) within the convex operational framework. The mathematical objects of one framework can be transformed directly (that is \emph{functorially}) into those of the other framework. This is denoted in the figure by the upper arrow labeled \emph{statistical functor}. Similarly, given a quantum mechanical model in terms of the Hilbert space formalism (lower left), this can be directly transformed into the statistical version with density operators (lower right), also within the convex operational framework. However, in this case the transformation of the notion of \emph{observable} is not as direct and the statistical version is strictly richer with respect to the measurement theory of quantum mechanics. To illustrate this, the corresponding lower arrow is drawn with a dashed line rather than a straight line. It is labeled \emph{modulus-square functor}, a terminology explained in Subsection~\ref{sec:qstat}. This map was first explicitly described as a functor by Selinger \cite{Sel:cccandcpm} and called the \emph{CPM construction}. There is also a downward arrow on the left hand side, illustrating \emph{quantization} procedures which transform a classical mechanical theory into a quantum mechanical one. The Hilbert space and the density operator frameworks for quantum theory constitute what we shall denote here the \emph{standard formulation of quantum theory}.

All the frameworks represented in Figure~\ref{fig:found_scheme_evol} have a certain basic structure in common. They suppose a fixed notion of time and center on a notion of state space that is meant to represent physics at an instant of time. The main objective is then to describe the evolution of states in time. We call these \emph{time-evolution frameworks}. Not all of physics is formulated in this way of course. With the advent of field theory and the unification of the notions of space and time in special and then general relativity a different perspective on classical physics emerged. In this perspective spacetime is viewed as an entity and the dynamics of a theory is expressed in terms of \emph{partial differential equations} for fields. This realizes and exploits the powerful principle of spacetime \emph{locality}, which means that the physics in a small piece of spacetime only directly depends on the physics immediately surrounding it and not anything further away. This change in perspective and mathematical description does not present any problems in classical physics where the solution of an equation of motion in time or spacetime is meant as a direct image of reality. On the other hand, such a change in perspective is less straightforward in classical statistical physics which continues to rely predominantly on a time-evolution framework.

In the text book approach to \emph{quantum field theory}, e.g., \cite{ItZu:qft} a spacetime perspective is also taken, although this is heavily built on the standard formulation of quantum theory. However, only a fragment of the measurement theory of quantum mechanics carries over to this spacetime framework. In particular, while there is a notion of observable, its connection to a notion of measurement is much more indirect than in the standard formulation. Indeed, there is only a very special type of measurement ever considered, namely that of a final state at asymptotically late time with an initial state prepared at asymptotically early time. There has not been any proper spacetime framework for statistical quantum field theory. The only partial exception is the appearance of a statistical notion of state in the framework of \emph{algebraic quantum field theory} \cite{Haa:lqp}. However, the latter is similar to text book quantum field theory in that any relation to the real world through measurable quantities is handled by regression to the time-evolution framework of the standard formulation.

\subsection{The positive formalism}

\begin{figure}
\begin{center}
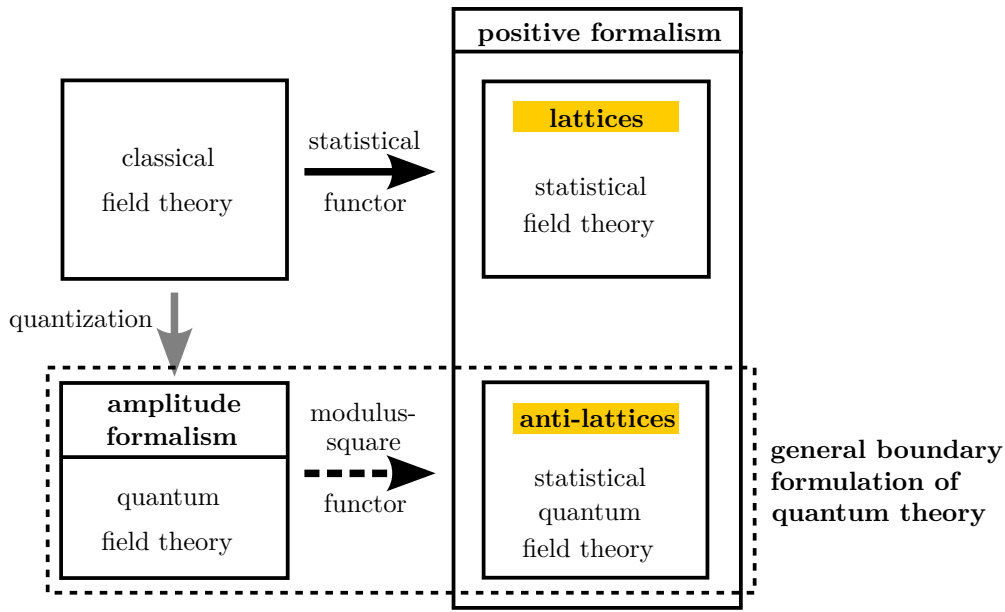
\end{center}
\caption{Spacetime frameworks for physical theories.}
\label{fig:found_scheme_st}
\end{figure}

The present work is concerned with detailing a novel framework for physical theories, called the \emph{positive formalism}. This framework brings together the two key principles of \emph{locality} and \emph{operationalism}. While the framework is novel, it builds on and incorporates previous ideas and results from a wide range of fields, including quantum field theory, operational approaches in quantum foundations, quantum information theory, monoidal category theory, symplectic field theory, diagrammatic calculus, spin foam models of quantum gravity, to name a few.
Figure~\ref{fig:found_scheme_st} illustrates the scope of this framework and its relation to other frameworks. We explain this in the following.

Spacetime is treated as a uniform entity, without any split into space and time at a fundamental level. Locality is implemented without a notion of metric by separating spacetime into \emph{regions} that are in communication through their \emph{boundaries}. This notion of locality emerged with the mathematical framework of \emph{topological quantum field theory (TQFT)} \cite{Ati:tqft} that revolutionized various areas of mathematics starting in the late 1980s. This was originally motivated by insights about quantum field theory, notably properties of the path integral, due mainly to Witten \cite{Wit:physgeom}. The implementation of this notion of locality in quantum field theory is also called \emph{Segal locality} after Graeme Segal who was one of its principal inventors and has since pursued a program to formalize quantum field theory in this way \cite{Seg:cftdef}. This was taken up as the starting point of a program to reformulate quantum theory, including measurement, in a spacetime local way, called the \emph{general boundary formulation} of quantum theory \cite{Oe:boundary,Oe:gbqft}. The resulting framework, called here the \emph{amplitude formalism}, is depicted in Figure~\ref{fig:found_scheme_st} on the lower left hand side. This already has a rich history and applications, see Section~\ref{sec:quantum}. In contrast to the text book approach to quantum field theory or to algebraic quantum field theory this framework already includes certain notions of measurement that go beyond those of the standard formulation of quantum theory.

A predecessor in classical physics of this notion of locality can be found in the symplectic approach to field theory of Kijowski and Tulczyjew \cite{KiTu:symplectic}. A formalization leads to an axiomatic framework for classical (field) theory (e.g. the linear case \cite{Oe:holomorphic}), depicted in on the upper left hand side in Figure~\ref{fig:found_scheme_st}. Quantization schemes that convert a classical theory in this form into a quantum theory in the amplitude formalism have also been developed \cite{Oe:feynobs}, but will not be the subject of the present work. They are indicated in Figure~\ref{fig:found_scheme_st} by the downward arrow on the left hand side.

The positive formalism was first conceived of as a statistical analogue of the amplitude formalism for quantum (field) theory and presented as such in the article \cite{Oe:dmf}. It was meant in particular as a further development of the general boundary formulation of quantum theory. That article also introduced the \emph{modulus-square functor} that maps the amplitude formalism to the positive formalism, indicated by the lower dashed arrow in Figure~\ref{fig:found_scheme_st}. However, it is only with the present work that the full scope and content of the positive formalism becomes clear. This concerns operationalism and (quantum) measurement theory on the one hand. On the other hand, classical statistical (field) theory also fits into the positive formalism with a \emph{statistical functor} leading to the positive formalism, as indicated in the upper half of Figure~\ref{fig:found_scheme_st}.

The structural similarity of Figures~\ref{fig:found_scheme_evol} and \ref{fig:found_scheme_st} is intentional. The spacetime frameworks depicted in the latter are generalizations of the corresponding time-evolution frameworks in the former. More specifically, splitting spacetime into space and time and then restricting to spacetime regions induced by time intervals recovers a time-evolution framework from the corresponding spacetime framework.\footnote{There is usually an additional structure on the statistical time-evolution framework that is not recovered in this way. This comes in the form of normalization conditions implementing causality, see Subsection~\ref{sec:causnorm}.}  Within the positive formalism classical and quantum theories can again be roughly characterized by the structure of certain partially ordered vector spaces being lattices (in the former case) or anti-lattices (in the latter case).

\subsection{A motivation}

Why is a new framework for physical theories necessary or useful? While the time-evolution frameworks depicted in Figure~\ref{fig:found_scheme_evol} have a long and successful history in physics, they present important shortcomings. Most obviously, these frameworks conceptually (but not necessarily historically, see the standard formulation of quantum theory) predate in their treatment of spacetime the special and general relativistic revolutions. It is artificial to describe a special relativistic theory in terms of time-evolution in a particular reference frame. This obscures symmetries and unnecessarily complicates the dynamical laws of the theory. For general relativity the situation is worse, as without already (partially) fixing the dynamics, i.e.\ the metric, it is not even meaningful to talk about time-evolution. We could also say that these frameworks lack an implementation of the principle of spacetime \emph{locality}. These are of course precisely the reasons why classical and quantum field theories are usually not described in this way.

On the other hand, the frameworks that are frequently used to describe classical or quantum field theories lack crucial properties that the time-evolution frameworks depicted on the right hand side of Figure~\ref{fig:found_scheme_evol}, subsumed under the name \emph{convex operational framework}, do possess. This is chiefly an \emph{operationalist} implementation of measurement theory, essential for fully capturing the relation between mathematical objects and reality in quantum physics.

Thus, for quantum theory in particular, a peculiar situation has arisen. Its standard formulation is in an uneasy tension with special relativity and in open conflict with general relativity. Other frameworks for quantum theory such as text book quantum field theory or algebraic quantum field theory are better adapted to special relativity, but are severely incomplete in incorporating only a tiny bit of measurement theory (via the S-matrix). This state of affairs has made it particularly difficult to bring together quantum theory and general relativity into a single theory, commonly called \emph{quantum gravity}.

Most approaches to quantum gravity have navigated this situation in one of two ways: The first is to embrace quantum field theory and accept the necessity for an asymptotic region of spacetime with a fixed classical metric where any measurement is to take place via the conventional S-matrix. This implies a perturbative treatment of the metric around this fixed asymptotic background metric. It must be noted that such a perturbation theory is fundamentally more problematic than the usual perturbation theory of fields that ``live on top of the metric'' as the metric is normally a fixed core structure that quantum field theory is built on. Indeed, naive quantization of general relativity along these lines does not lead to a consistent theory due to lack of renormalizability.
A response has been to propose theories that are distinct from general relativity, but are meant to recover it in some approximation. The most popular of these approaches has been \emph{string theory} \cite{GSW:sstring}.
On the other hand, using the powerful framework of \emph{effective field theory} it is possible to make limited sense of the perturbation expansion of quantum general relativity in spite of the non-renormalizability \cite{DoHo:qgeffective}. This provides probably the most successful tool so far for predicting certain quantum gravitational effects.

The second way is more ambitious in that it aims for a ``full'' theory of quantum gravity without the a priori need for perturbation expansions. It relies on the mathematical content of the standard formulation of quantum theory. Thus, quantization is performed of an initial-data formulation of general relativity. The aim is to construct a Hilbert space and an algebra of observables on it. However, the necessary absence of a predetermined notion of time means that most of the operational content of the standard formulation is lost. It is hoped that once the mathematical objects have been constructed, a new measurement theory relating them to the real world can be worked out. The most popular approaches along these lines have been \emph{quantum geometrodynamics} \cite{Dew:qgrav1,Whe:sspace} and its modern incarnation, \emph{loop quantum gravity} \cite{Rov:qg,Thi:canqgr}.

In spite of many decades of work all attempts have so far failed to produce a theory of quantum gravity. In light of the previous remarks, it is suggestive to attribute this failure at least in part to the lack of a suitable framework. This provides a strong motivation for the present work.

\subsection{Structure of this work}

The positive formalism is developed from three different starting points, all converging on the same framework. The first route, exhibited in Section~\ref{sec:first} is a \emph{derivation} of the the positive formalism from first principles, based on \emph{locality} and \emph{operationalism}. Parts of this were previously announced in \cite{Oe:firstproc}. The second route, exhibited in Section~\ref{sec:classical} starts from classical field theory. Implementing spacetime locality leads to an axiomatic description of field theory. Moving this to a statistical setting turns out to lead to the positive formalism. (Subsections~\ref{sec:ceom}-\ref{sec:cobs} mostly review previous work.) Section~\ref{sec:quantum} deals with the third route. Starting from quantum field theory, properties of the path integral motivate the mathematical framework of \emph{topological quantum field theory}. Applied to quantum field theory this implements Segal locality. Augmenting this with local notions of boundary measurement and observables yields the \emph{amplitude formalism} of the \emph{general boundary formulation} of quantum theory. Moving this to a statistical setting, complemented with a local generalization of \emph{quantum operation} also leads to the positive formalism. This is the way it was originally encountered \cite{Oe:dmf}. (Subsections~\ref{sec:qlocpath}-\ref{sec:qstat} mostly review previous work.)

In Section~\ref{sec:catdiag} some elements of monoidal category theory and related diagrammatic calculus are reviewed. These have provided powerful tools for formalization, calculation and visualizations in topological quantum field theory, state sum models, quantum information theory, the convex operational framework, foundations of quantum theory and other fields. Many of these uses carry over to and are unified in the positive formalism. Use of this is made in subsequent sections.
In Section~\ref{sec:evolution} the convex operational framework (right hand half of Figure~\ref{fig:found_scheme_evol}) is recovered from the positive formalism (right hand half of Figure~\ref{fig:found_scheme_st}). To this end, spacetime is split into space and time, and the usual notion of states and their evolution is recovered. Also, causality is introduced. The classical and quantum sections also contain brief discussions of specialized versions of this time-evolution framework, see Subsection~\ref{sec:cevol} and \ref{sec:qevol} respectively.
While Section~\ref{sec:first} and most of Section~\ref{sec:evolution} are developed in a self-contained fashion, relations to other frameworks are presented in Subsection~\ref{sec:pcomp} in so far as they are not specific to classical or quantum theory. Related works specific to classical or quantum theory are mostly cited within the classical Section~\ref{sec:classical} or the quantum Section~\ref{sec:quantum}. In the latter case Subsection~\ref{sec:qcomp} contains discussion of additional related work.

Section~\ref{sec:examples} serves to illustrate the applicability of the positive formalism far from its origins in quantum field theory using two examples from the literature. The first, (Subsection~\ref{sec:qteleport}) is the \emph{quantum teleportation} protocol, demonstrating applicability in quantum information theory and exhibiting a mixture of classical and quantum degrees of freedom. The second, (Subsection~\ref{sec:indefcausal}) is an example where the implementation of causal structure of the standard formulation of quantum theory is partially relaxed outside of two isolated laboratories, allowing for \emph{indefinite causal structure} in their interaction.

The presentation of the positive formalism in this work should only be seen as provisional, with many adjustments and refinements certainly necessary or desirable. Some such refinements are discussed in Section~\ref{sec:refine}. Subsection~\ref{sec:instruments} refines the notion of \emph{probe} (from Section~\ref{sec:first}) in the spirit of that of \emph{quantum instrument}. Certain refinements necessary in quantum field theory and in fermionic theories are already worked out or implied in previous literature, see Subsections~\ref{sec:refqft} and \ref{sec:fermions}. Abstractions of the notion of locality are considered in Subsection~\ref{sec:absbdy}.
A concluding discussion of this work is presented in Section~\ref{sec:discussion}.

In an effort to make this work accessible to a wider audience certain mathematical details are glossed over in most parts. For the reader interested in those mathematical details this is partially remedied with Subsection~\ref{sec:infinidim}. Also, a few relevant mathematical facts about partially ordered vector spaces are contained in Appendix~\ref{sec:povs}. These are referenced from the main text when required.

The present paper need not be read linearly. It is possible to start reading with either Section~\ref{sec:first} (first principles), Section~\ref{sec:classical} (classical theory) or Section~\ref{sec:quantum} (quantum theory). In case of starting with one of the latter two it is advisable to read Subsection~\ref{sec:spacetime} first. In case of starting with Section~\ref{sec:quantum} also Subsection~\ref{sec:csts} should be read first.

 A mathematical convention that we use throughout this paper is that the word \emph{positive} means non-negative, if not indicated otherwise. In particular, $0$ is positive.

\section{The positive formalism from first principles}
\label{sec:first}

We present in this section a constructive account of the positive formalism as emerging from a first-principles approach to a framework for theories of physics. Parts of the content of this section were announced in the paper \cite{Oe:firstproc}.
The starting point is this question: What form should a fundamental physical theory take? We have in mind here not physical theories for specific phenomena, but are looking for a universal framework, based on generic notions of making experiments or observations and reasoning about them. In constructing such a framework we would like to ignore our actual knowledge of physical laws or phenomena as much as possible, for two reasons. Firstly, we would like the framework to be as generally applicable as possible. Secondly, we would like to avoid the biases and limitations inherent in the often arbitrary or accidental historical choices of structures of known physical theories. Of course, we cannot hope to formulate empirical theories or even frameworks for such theories without taking into account any empirical knowledge. The empirical knowledge we do take into account is reflected in two core principles that we take as the basis for our construction:
\begin{description}
\item[Locality] refers to our experience that forces do not act at a distance, but through fields permeating spacetime. In particular, we can parametrize any known interaction as mediated by ``signals'' connecting the participating objects through paths in spacetime.
\item[Operationalism] refers here to the principle that we should describe the world through our interaction with it, rather than in terms of an abstract and objective reality ``out there''. This is motivated by the key lesson of quantum theory that the classical view of reality in terms of objective trajectories of particles and field configurations in spacetime is unsustainable. Instead, the act of observation or measurement becomes central to our description of physics.
\end{description}
As to the nature of the predictions admitted by the framework, for generality, we should allow these to be of a probabilistic rather than deterministic nature. Moreover, we shall not make any assumptions as to the origin of probabilistic uncertainty; for example whether or to what extent such uncertainty might be attributable to a ``lack of knowledge''.

\subsection{Spacetime}
\label{sec:spacetime}

It is an essential aspect of any modeling of the physical world to be able to isolate certain features of it from all others. Often in the physics literature this is subsumed under a notion of ``system'' as to be distinguished from other systems or the environment. It would be possible to proceed with much of the following construction by postulating some generic notion of ``system'' and its potential interaction with other ``systems''. However, the principle of locality allows us to avoid such a questionable postulate in favor of a more empirically grounded one that turns out to yield more predictive power at the same time. The price to pay is the introduction of a notion of \emph{spacetime}, even though a very weak one.

\begin{figure}
\centering
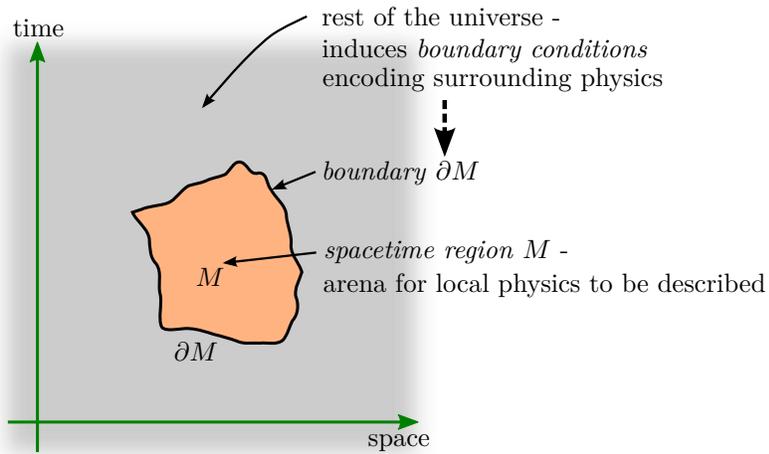
\caption{\emph{Locality} allows to separate the physics of interest from that of the rest of the universe. This is implemented through a notion of spacetime \emph{region}, which communicates with the rest of the universe through its \emph{boundary}.}
\label{fig:st_region}
\end{figure}

\begin{figure}
\centering
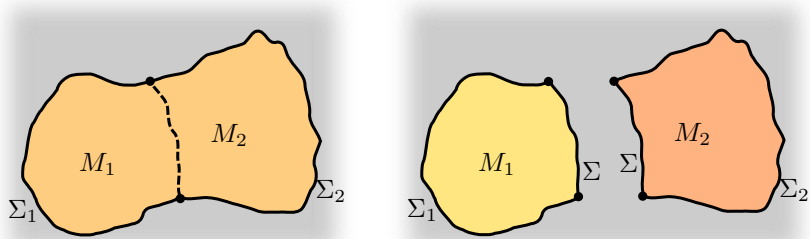
\caption{Composition of spacetime regions $M_1$ and $M_2$ along a common hypersurface $\Sigma$. Left: Composite region $M_1\cup M_2$. Right: component regions $M_1$ and $M_2$.}
\label{fig:st_compose}
\end{figure}

\begin{figure}
\centering
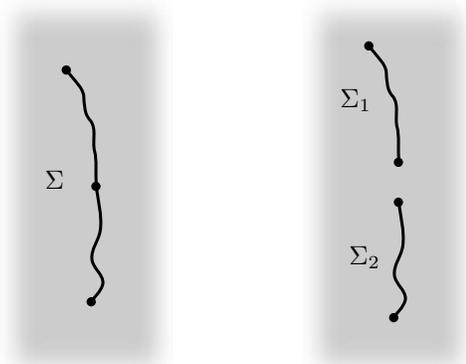
\caption{Decomposition of a hypersurface $\Sigma=\Sigma_1\cup\Sigma_2$ (left) into component hypersurfaces $\Sigma_1$ and $\Sigma_2$ (right).}
\label{fig:st_decomp}
\end{figure}

The first and most basic notion we need is that of a spacetime \emph{region}. This allows for the distinction of the physics inside the region, from that outside, i.e., the rest of the universe. This is illustrated in Figure~\ref{fig:st_region}. Secondly, it is a crucial element of physical theorizing to make statements about the interaction or correlation between ``systems'', i.e., here the physics in spacetime regions. By locality, interaction requires adjacency, in this case adjacency of the two regions in question. To formalize this we need a notion of \emph{composition}, i.e., a way to view two adjacent spacetime regions as a single region. To describe this composition we need in turn another spacetime concept, namely that of a \emph{hypersurface}. In particular, the \emph{boundary} of a region is a hypersurface. Moreover, if we compose two regions together, this happens along a hypersurface which is a piece of the boundary of each of the participating regions, see Figure~\ref{fig:st_compose}. We clarify this taking ``pieces'' of hypersurfaces in terms of a notion of \emph{decomposition} of a hypersurface into hypersurfaces. We take this to mean the presentation of a hypersurface $\Sigma$ in terms of a union of hypersurfaces $\Sigma_1\cup\cdots\cup\Sigma_n$ such that the component hypersurfaces intersect only at their boundaries, see Figure~\ref{fig:st_decomp}.

To summarize, we have notions of spacetime \emph{regions}, and \emph{hypersurfaces} as well as operations of \emph{decomposing} hypersurfaces into pieces and of \emph{composing} two regions along a common hypersurface component of their boundaries. We call this a \emph{spacetime system}. A convenient minimal mathematical formalization is as follows. We take \emph{regions} to be \emph{topological manifolds} of dimension $d$ (the \emph{spacetime dimension}) and \emph{hypersurfaces} to be topological manifolds of dimension $d-1$. The boundary $\partial M$ of a region $M$ is clearly a hypersurface. The composition of regions along a hypersurface is then a \emph{gluing} of manifolds. We refrain from specifying further mathematical details as these are mostly irrelevant for our treatment here. Moreover, any particular physical model one wishes to consider within the framework may add additional structure onto our spacetime elements, such as a metric or an ambient spacetime manifold. However, this does not modify the basic structure of the framework as laid out in this section.

\subsection{Probes}
\label{sec:probes}

\begin{figure}
\centering
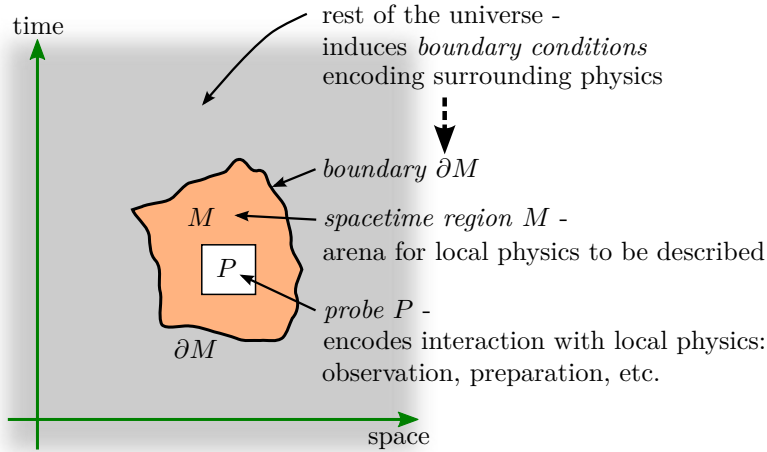
\caption{We associate \emph{probes} with spacetime regions. These represent experiments, observations, preparations etc.}
\label{fig:st_probe}
\end{figure}

Guided by operationalism, we describe the physics inside a spacetime region $M$ through a means of interacting with it. This might be by performing an experiment, making an observation, placing an apparatus or interfering in some other way with the physics in $M$. Crucially, this might involve a question that is asked about this physics. We formalize this through a notion of \emph{probe} in $M$, which is to be thought of provisionally as the specification of an experiment, observation, apparatus etc., see Figure~\ref{fig:st_probe}. Correspondingly, we associate a \emph{space of probes} $\pr_M$ to the spacetime region $M$. If no experiment is performed, no observation made, etc., we represent this as a special probe that we call the \emph{null probe}, denoted by $\np$.

The notion of \emph{composition} extends from spacetime regions to probes contained in them. Given adjacent spacetime regions $M$ and $N$ and their composite $M\cup N$ there is a map $\pcomp:\pr_M\times \pr_N\to \pr_{M\cup N}$ that describes this composition. Physically this just encodes the simple fact that performing two experiments together is itself an experiment. When the experiments involve questions, however, then this map also involves a specific way how these questions are combined to a new question of the composite experiment. We shall come back to this point once we introduce quantitative features into our framework.

\subsection{Boundary conditions}

By locality, any influence of the outside of a region $M$ on the inside and vice versa should be mediated by signals traversing the boundary $\partial M$ of the region. In particular, knowing the signal at the boundary should be sufficient to determine this influence or interaction or communication completely. We capture this in terms of a notion of \emph{boundary condition}, see Figure~\ref{fig:st_region}. That is, a boundary condition is a complete parametrization of the influence of the rest of the universe on the physics in a spacetime region or vice versa. More generally, we associate a \emph{space of boundary conditions} $\bc_{\Sigma}$ to any hypersurface $\Sigma$ (not only to boundaries). The latter is to be thought of as describing all possible communications between potential regions adjacent to the hypersurface in so far as they are mediated by signals passing through this hypersurface (as opposed to through other parts of the boundaries of the regions).
For the moment, we take the space of boundary conditions to merely have the structure of a set. Pushing the principle of locality further, we shall assume for now that if we cut or \emph{decompose} a hypersurface $\Sigma$ into pieces $\Sigma_1,\dots, \Sigma_{n}$, the spaces of boundary conditions $\bc_{\Sigma}$ decomposes into a corresponding direct product $\bc_{\Sigma}=\bc_{\Sigma_1}\times\cdots\times\bc_{\Sigma_n}$. This assumption turns out to be rather strong and we shall drop it in favor of a weaker assumption in due course.

\subsection{Values and pairing}

A further ingredient is necessary in our framework in order to allow for quantitative predictions. To each spacetime region $M$, probe $P\in\pr_M$ in $M$ and boundary condition $b\in\bc_{\partial M}$ we associate a \emph{real number} $\lv P,b\rv_M$, called \emph{value}. Roughly speaking, this value encodes the answer to the question posed by the probe $P$ in $M$ under the specific ambient conditions given by the boundary condition $b$. Following tradition in theoretical physics, we suppose that any measurable quantity can be represented in terms of a real number or a multiple of real numbers. (In the latter case one would use a multiple of probes.)

We note that the values for a given spacetime region yield a real valued pairing $\lv\cdot,\cdot\rv_M:\pr_M\times\bc_{\partial M}\to\R$. In this way a probe $P\in\pr_M$ can be seen as a real valued \emph{function} on the set $\bc_{\partial M}$ of boundary conditions. Similarly, a boundary condition $b\in\bc_{\partial M}$ can be seen as a real valued function on the space of probes $\pr_M$. Now, the space of real valued functions on a set inherits the structure of a \emph{real vector space} from the real numbers. Moreover, it inherits a \emph{partial order}, also from the real numbers. (That is, a function is larger or equal to another one if its value at every point is larger or equal to the value of the other function.)

\subsection{From values to predictions}
\label{sec:valpred}

The relation between values and predictions of measurement outcomes is more complicated than we have suggested so far. The reason is that apart from an explicit question we might pose in an experiment, there is an implicit question arising as soon as we consider a probe together with a boundary condition. This is the question of the \emph{compatibility} between the experimental setup encoded by the probe and the boundary condition. Crucially, this question even arises in the absence of any experimental setup or observation, i.e., for the null probe. How can a question of compatibility of a boundary condition arise when apparently there is nothing it has to be compatible with? The answer is that there is such a thing and that it is the \emph{presence} of the spacetime region itself. By locality, a space of boundary conditions only ``knows'' about the hypersurface to which it is associated. Declaring that this hypersurface is actually the boundary of a region $M$ is an additional non-trivial piece of information. For $b\in\bc_{\partial M}$ the value $\lv\np,b\rv_M$ thus is a measure for the degree of compatibility between the boundary condition $b$ and the presence of the spacetime region $M$. While it is clear that precisely in the case of incompatibility this value should be zero, the (positive) value it takes in general can usually only be ascribed a relative meaning, as we shall see.

We proceed to consider the more interesting situation of the presence of an actual experimental setup in the region $M$. The first consequence of this presence is that the physics in $M$ is generically \emph{altered}. This is modeled by a probe $Q\in\pr_M$ so that the value $\lv Q,b\rv_M$ is a measure for the compatibility of the boundary condition $b\in\bc_{\partial M}$ with the altered physics in $M$. Usually, what we actually want to do with the experiment is ask an explicit question. For simplicity we consider an experiment with a \emph{binary outcome}, which we symbolize for definiteness by \lbl{YES}/\lbl{NO}. Imagine for example an apparatus with a single light, that will show \lbl{RED} (for \lbl{YES}) or \lbl{GREEN} (for \lbl{NO}) after the experiment. The precise question we will ask is this: How probable is the outcome \lbl{YES} (light \lbl{RED}) of the experiment given a boundary condition $b\in\bc_{\partial M}$? To model this we need another probe $P\in\pr_M$, which represents the experimental setup with the answer \lbl{YES}. In the example this is the apparatus present and with its light showing \lbl{RED}. Compared to the probe $Q$, the probe $P$  ``filters out'' the case of the response \lbl{NO} (light \lbl{GREEN}). Because of this we also say that the probe $P$ is \emph{selective}. We say that the probe $Q$, in contrast, is \emph{non-selective}. The probability $\Pi$ that constitutes the answer to the question is given by the following \emph{quotient},
\begin{equation}
 \Pi=\frac{\lv P,b\rv_M}{\lv Q,b\rv_M} .
\label{eq:elemprob}
\end{equation}
We can think of both values involved as representing degrees of compatibility. That is, $\lv Q,b\rv_M$ is the compatibility of the boundary condition with the experimental setup, while $\lv P,b\rv_M$ is the compatibility of the boundary condition with the experimental setup where the answer is \lbl{YES} (the light shows \lbl{RED}). This also clarifies the meaning of the situation where the denominator is zero and $\Pi$ is thus undefined. The value $\lv Q,b\rv_M$ being zero signifies precisely that the experimental setup and the boundary condition are mutually incompatible, i.e., their joint occurrence is unphysical. It should be stressed, however, that in general the numerical value of numerator and denominator in the expression (\ref{eq:elemprob}) do not have any direct physical meaning individually, only the quotient $\Pi$ does.

\subsection{Primitive probes and hierarchies}
\label{sec:hierarchies}

\begin{figure}
\centering
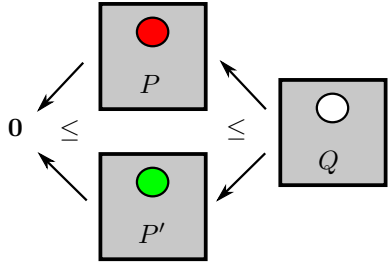
\caption{An apparatus with an indicator light with two possible outcome states \lbl{RED} and \lbl{GREEN}. The mere presence of the apparatus with undetermined light state is depicted on the right hand side. The apparatus with defined outcome states is depicted in the middle. The corresponding probes form a partially ordered set representing a hierarchy.}
\label{fig:hier1}
\end{figure}

The probes $P$ and $Q$ are part of a special class of probes, that we shall call \emph{primitive probes}, denoted $\prp_M\subseteq\pr_M$. A primitive probe corresponds to an experiment with an (explicit) \emph{binary outcome}, which we symbolize as before by \lbl{YES}/\lbl{NO}. Obviously, the selective probe $P$ we considered before is such a primitive probe. However, the non-selective probe $Q$ we considered is also a primitive probe. This is because in such a given experiment we can always simply ignore the actual outcome and replace it with the outcome \lbl{YES}. In this way, any non-selective probe that merely represents the presence of an experimental setup is a primitive probe. So is even the null probe, by considering the absence of any apparatus or intervention also as a valid experiment. In particular, the null probe may be considered a non-selective probe.

\begin{figure}
\centering
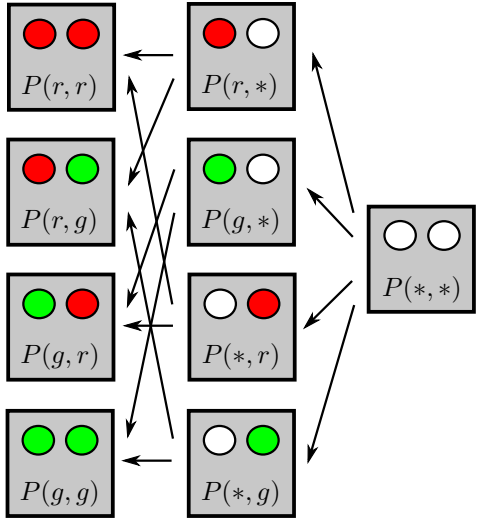
\caption{An apparatus with two indicator light yields a more complicated hierarchy of probes.}
\label{fig:hier2}
\end{figure}

It is clear that we have the inequality $0\le \lv P,b\rv_M \le \lv Q,b\rv_M$ for any boundary condition $b\in\bc_{\partial M}$. This guarantees in particular that $\Pi$ (if defined) takes values between $0$ and $1$ as necessary for a probability. In terms of the partial order structure on the space of probes this induces the inequality 
\begin{equation}
 \zp\le P\le Q ,
\end{equation}
where $\zp\in\pr_M$ is the probe that returns the value zero for any boundary condition. The inequality between $P$ and $Q$ has another interesting interpretation here. It represents a \emph{hierarchy of generality} in the space of primitive probes. In the case at hand, the probe $P$ represents a situation that is \emph{more special} than that represented by $Q$. Namely, in addition to representing the mere presence of the apparatus it represents the apparatus indicating the answer \lbl{YES} (i.e., with the light showing \lbl{RED}). Conversely, $Q$ is \emph{more general} than $P$. Similarly, we can define a (selective) probe $P'\in\prp_M$ that corresponds to the same experimental setup, but with the answers \lbl{YES} and \lbl{NO} interchanged. In the example with the light we are asking for the probability that the light shows \lbl{GREEN} after the experiment. We have then an analogous hierarchy and inequality $\zp\le P'\le Q$ for this probe, see Figure~\ref{fig:hier1}. What is more, in this case the selective probes are complementary with respect to the non-selective probe $Q$ and satisfy the additive relation $Q=P+P'$. The additive structure here is simply the additive law for the combination of probabilities of exclusive outcomes.
With increasing complexity of the experimental setup the hierarchies that occur become richer. The example of an apparatus  with two lights, each of which shows either \lbl{RED} or \lbl{GREEN} after the experiment is depicted in Figure~\ref{fig:hier2}.

Boundary conditions form similar hierarchies of generality. It is intuitively clear for example that the boundary condition ``the ambient temperature is between 10 and 20 degrees Celsius'' is more general than the boundary condition ``the ambient temperature is between 10 and 15 degrees Celsius''. Associated to these are inequalities that hold for pairings with primitive probes (not general probes!). What is more, the inequalities can be used to obtain quantities that can be interpreted as probabilistic predictions for boundary conditions. Say, in a spacetime region $M$ we have an apparatus represented by a primitive probe $Q\in\prp_M$ and we know that a certain boundary condition $c\in \bc_{\partial M}$ holds. Now we may ask what is the probability $\Pi$ that a more special boundary condition $b\in \bc_{\partial M}$ with $\zp\le b\le c$ holds? The answer is the quotient,
\begin{equation}
 \Pi=\frac{\lv Q,b\rv_M}{\lv Q,c\rv_M} .
\label{eq:elembdy}
\end{equation}
One has to remember here that a boundary condition encodes as much the influence of the exterior on the interior as the influence of the interior on the exterior. Also, this makes even sense for $Q$ being the null probe. However, recalling our operationalist principles we might be cautious about these types of predictions. They seem to rely on a preexisting understanding of the physical meaning of the boundary conditions which would have to be acquired in a previous step through the use of probes for which we expressly assume such an understanding.

\subsection{Expectation values}
\label{sec:peval}

The prediction of \emph{expectation values} of measurements with real valued outcomes is achieved in a manner very similar to the case of binary outcomes. Consider an apparatus with a pointer device and a continuous scale providing a reading in a range $[r,s]\subseteq\R$ after the experiment. On the one hand there is a (non-selective) primitive probe $Q\in\prp_M$ that represents the mere presence of the apparatus. On the other hand there is a probe $P\in\pr_M$ that represents the apparatus together with the numerical outcome. Then, the predicted expectation value $E$, given a boundary condition $b\in\bc_{\partial M}$ is the quotient,
\begin{equation}
 E=\frac{\lv P,b\rv_M}{\lv Q,b\rv_M} .
\end{equation}
To better understand the nature of the probe $P$ we note that it can be approximated in terms of a linear combination of (selective) primitive probes. To this end consider the (selective) primitive probe that represents the apparatus together with the question whether the measurement outcome is in the subrange $[x,y]\subseteq [r,s]$. We denote this probe by $P[x,y]\in\prp_M$. Note that $Q=P[r,s]$. We proceed to approximate $P$ by linear combinations of probes of this type as follows. Let $n$ be an integer to determine the precision of the approximation in terms of an equal spaced subdivision of the scale. Define the probe $P_n\in\pr_M$ as,
\begin{equation}
 P_n= \sum_{k=0}^{n-1} \left(r+\left(k+\tfrac12\right)l\right) P\left[r+kl,r+(k+1)l\right] ,
\end{equation}
where $l=(s-r)/n$. The approximate expectation value for the pointer position on the scale given a boundary condition $b\in\bc_{\partial M}$ is then $\lv P_n,b \rv_M/\lv Q,b \rv_M$. In the limit $n\to\infty$ the probes $P_n$ approximate the probe $P$ in the sense of uniform convergence of expectation values. More precisely, we have the estimate,
\begin{equation}
 \left|\frac{\lv P-P_n,b\rv_M}{\lv Q,b\rv_M}\right|\le \frac{1}{2} l .
\end{equation}

We have seen that values themselves have no direct physical meaning, but only their quotients, both in the case of probabilistic predictions (primitive probes) and for expectation values (general probes). The scaling of the values and thus the scaling of boundary conditions and of probes is thus arbitrary. This implies that the spaces of primitive probes $\prp_M$, of probes $\pr_M$ and of boundary conditions $\bc_{\Sigma}$ are closed under multiplication by positive real numbers, corresponding to rescalings. We also take approximability to mean that any probe can be arbitrarily approximated by linear combinations of primitive probes. Moreover, for \emph{positive} probes, i.e., probes that do not take negative values for any boundary condition this can be achieved with positive coefficients. Assuming completeness, this implies that a positive probe is in fact a primitive probe and vice versa.

\subsection{Probabilities and convex combinations}

Through the pairing with values, primitive probes and boundary conditions can be seen as functions with values in (relative and conditional) probabilities. But probabilities for exclusive events can be combined via convex combinations. An analogous statement holds for general probes and expectation values. We shall take this as sufficient evidence that the space $\prp_M$ of primitive probes, the space $\pr_M$ of all probes and the space of boundary conditions $\bc_{\Sigma}$ are closed under convex combinations. That is, given (primitive) probes $P_1,\dots,P_n\in\prp_M$ and probabilities $p_1,\dots,p_n$ summing to $1$, we consider the convex linear combination
\begin{equation}
P\defeq \sum_k p_k P_k 
\end{equation}
also to be a (primitive) probe. Similarly for boundary conditions.

To justify this further we consider the following example of an apparatus equipped with an input dial with positions $1,\dots,n$ and one indicator light that is either turned on or off after the experiment. We suppose that the dial position does not affect the physics in $M$, except for the final state of the light. That is, the mere presence of the apparatus is represented by a (non-selective) probe $Q\in\prp_M$ which is the same, independent of the dial position. On the other hand the (selective) probe $P_k\in\prp_M$, where in addition we ask for the probability for the light state to be on, does depend on the dial position $k$. Suppose now that an experimenter is putting the dial randomly in position $k$ with probability $p_k$ at the beginning of the experiment. Given boundary condition $b\in\bc_{\partial M}$, the probability for the light to be switched on after the experiment is thus,
\begin{equation}
 \sum_k p_k \frac{\lv P_k,b\rv_M}{\lv Q,b\rv_M}=\frac{\lv\sum_k p_k P_k, b\rv_M}{\lv Q,b\rv_M} .
\end{equation}

Since we have taken the space $\prp_M$ of primitive probes to be closed under convex combinations as well as under scalar multiplication by positive numbers it is in fact closed under \emph{conical combinations}. That is, it is closed under taking arbitrary linear combinations with positive coefficients. The same is true for the space $\pr_M$ of general probes. However, in the latter we can do more arbitrary operations corresponding to the freedom to redefine numerical measurement outcomes. This freedom includes for example the freedom to multiply the output value with a non-zero real number. But combined with convex combinations this is sufficient to make $\pr_M$ into a \emph{real vector space}. Thus, the subspace $\prp_M$ of primitive probes is precisely the \emph{cone of positive elements} in the \emph{partially ordered vector space} $\pr_M$. (Compare Definitions~\ref{dfn:povs} and \ref{dfn:cone} in the Appendix.) Moreover, the approximability of probes by primitive probes (see Subsection~\ref{sec:peval}) means that $\prp_M$ is a \emph{generating cone}.

Similarly, the space of boundary conditions $\bc_{\Sigma}$ is closed under conical combinations. We define the space $\bcg_{\Sigma}$ of \emph{generalized boundary conditions} to be the real vector space \emph{generated} by $\bc_{\Sigma}$. For any region $M$ we extend the pairing $\lv\cdot,\cdot\rv_M:\pr_M\times\bc_{\partial M}\to\R$ to a map $\pr_M\times\bcg_{\partial M}\to\R$ by linearity. For a given probe $P\in\pr_M$ the pairing $\lv P,b\rv_M$ with any boundary condition $b\in\bc_{\partial M}$ is positive precisely if this probe is positive, i.e., a primitive probe. Extending this to arbitrary hypersurfaces by definition, it means that $\bc_{\Sigma}$ is the \emph{cone of positive elements} in the \emph{partially ordered vector space} $\bcg_{\Sigma}$. By construction $\bc_{\Sigma}$ is a \emph{generating cone}. (Compare Definition~\ref{dfn:cone} in the Appendix.)

We turn to the interplay between the notion of generalized boundary condition and that of hypersurface decomposition. Say we have a hypersurface $\Sigma$ decomposing into the union $\Sigma_1\cup\Sigma_2$. Based on a strong interpretation of locality, we have so far made the working assumption that there is an associated isomorphism $\bc_{\Sigma_1}\times\bc_{\Sigma_2}\to\bc_{\Sigma}$ between the corresponding sets of boundary conditions. This implies that the extension to generalized boundary conditions, $\bcg_{\Sigma_1}\times\bcg_{\Sigma_2}\to\bcg_{\Sigma}$ is a bilinear map. We can rewrite this as a linear map via the tensor product, $\tau:\bcg_{\Sigma_1}\tens\bcg_{\Sigma_2}\to\bcg_{\Sigma}$. We now replace our original assumption by the following one: The map $\tau$ is an isomorphism of real vector spaces and it is positive, i.e., tensor products of positive elements (boundary conditions) are mapped to positive elements (boundary conditions). This appears to be a natural implementation of the idea of locality in the probabilistic setting. However, this assumption is indeed weaker than the original one. It is implied by the original assumption, but not the other way round.\footnote{It turns out that the weakened assumption is still too strong to capture some physical theories of interest, compare Section~\ref{sec:refqft}. However, it is adequate to capture the key conceptual issues and a further weakening at this point would needlessly complicate the following account.}

\subsection{Slice regions and inner product}
\label{sec:sregip}

\begin{figure}
\centering
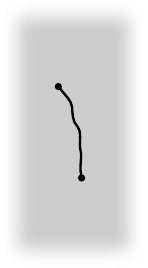
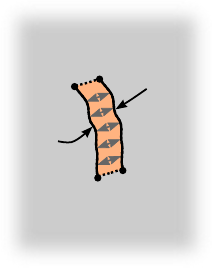
\caption{For each hypersurface (left) we define a \emph{slice region} (right) by infinitesimal thickening.}
\label{fig:st_sreg}
\end{figure}

The principle of locality can be taken further advantage of by introducing a special type of degenerate region that we shall term \emph{slice region}. Given a hypersurface $\Sigma$, there corresponds to it a slice region $\hat{\Sigma}$ that we can think of as arising from an infinitesimal thickening of $\Sigma$, see Figure~\ref{fig:st_sreg}. In particular, the boundary $\partial \hat{\Sigma}$ can be decomposed into a union of $\Sigma$ with another copy of $\Sigma$ ``on the other side''.
This decomposition induces the isomorphism $\tau:\bcg_{\Sigma}\tens\bcg_{\Sigma}\to \bcg_{\partial \hat{\Sigma}}$. Now, $\hat{\Sigma}$ is treated as a region and we therefore have probes associated with it and a value pairing $\lv\cdot,\cdot\rv_{\hat{\Sigma}}:\pr_{\hat{\Sigma}}\times\bcg_{\partial \hat{\Sigma}}\to\R$. Composing the two maps and evaluating with the null probe yields a \emph{bilinear map} $\lb \cdot,\cdot \rb_{\Sigma}: \bcg_{\Sigma}\times\bcg_{\Sigma}\to\R$,\footnote{Here as in the following we often omit writing $\tau$ for the decomposition map. That is, we write $b\tens c$ instead of $\tau(b\tens c)$.}
\begin{equation}
  \lb b,c\rb_{\Sigma} \defeq \lv \np,b\tens c\rv_{\hat{\Sigma}} .
  \label{eq:ipfrompairing}
\end{equation}
This bilinear map should be \emph{symmetric} because the two ``sides'' $\Sigma$ of the slice region $\hat{\Sigma}$, where the (generalized) boundary conditions reside, are interchangeable as they reside at the same place in spacetime. The bilinear map should also be \emph{non-degenerate}. Otherwise there would be (generalized) boundary conditions on $\Sigma$ that do not correspond to any communication between potential regions on the two sides that one copy of $\Sigma$ might interface.

What is more, the bilinear map should induce an isomorphism of the space of generalized boundary conditions $\bcg_{\Sigma}$ with its dual space $\bcg_{\Sigma}^*$ via $b\mapsto \lb b,\cdot\rb_{\Sigma}$. We justify this physically by the dual role of boundary conditions as parametrizing the influence of one side on the other as well as the response of the former on the latter. A conjecture of DeMarr \cite{dem:orderisomdual} then implies that the bilinear map is positive-definite. We assume positive-definiteness henceforth and refer to the bilinear map as an \emph{inner product}. Independent of that it is immediately clear that the inner product between boundary conditions (that is, positive elements) in $\bc_{\Sigma}$ must be positive. We say that the inner product is \emph{positive}. The isomorphism between $\bcg_{\Sigma}$ and $\bcg_{\Sigma}^*$ implies something stronger: If an element $c\in\bcg_{\Sigma}$ yields a positive inner product $\lb c, b\rb_{\Sigma}\ge 0$ with any positive element $b\in\bc_{\Sigma}$ then it must be positive itself. We then say that the inner product is \emph{sharply positive}, compare Definition~\ref{dfn:spos}. From here onward we use the term \emph{partially ordered inner product space} to refer to a partially ordered vector space that carries a sharply positive and positive-definite inner product, compare Definition~\ref{dfn:poips}. In particular, the spaces $\bcg_{\Sigma}$ of generalized boundary conditions are thus partially ordered inner product spaces.

\begin{figure}
  \centering
  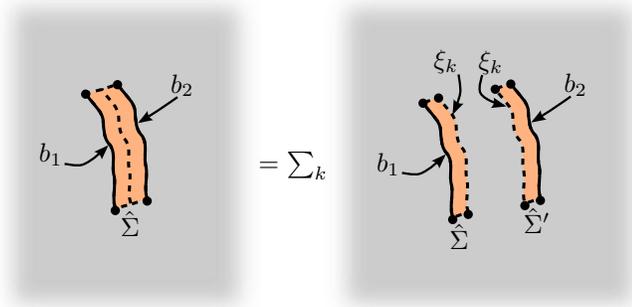
  \caption{Composition of slice regions via completeness of inner product.}
  \label{fig:st_completeness}
\end{figure}

Let $\{\xi_k\}_{k\in I}$ be an orthonormal basis of $\bcg_{\Sigma}$. Given generalized boundary conditions $b_1,b_2\in\bcg_{\Sigma}$ we have the completeness relation,
\begin{equation}
 \lb b_1,b_2 \rb_{\Sigma} = \sum_{k\in I} \lb b_1,\xi_k\rb_{\Sigma} \lb \xi_k,b_2\rb_{\Sigma} .
\label{eq:complip}
\end{equation}
Geometrically this corresponds precisely to the gluing of two identical slice regions (right hand side) to one slice regions (left hand side), see Figure~\ref{fig:st_completeness}. A single generalized boundary condition $b\in\bcg_{\Sigma}$ can be expanded in terms of the basis as,
\begin{equation}
  b = \sum_{k\in I} \lb \xi_k,b\rb_{\Sigma}\, \xi_k .
\label{eq:baseexp}
\end{equation}

\subsection{Composition of probes}
\label{sec:probecomp}

\begin{figure}
  \centering
  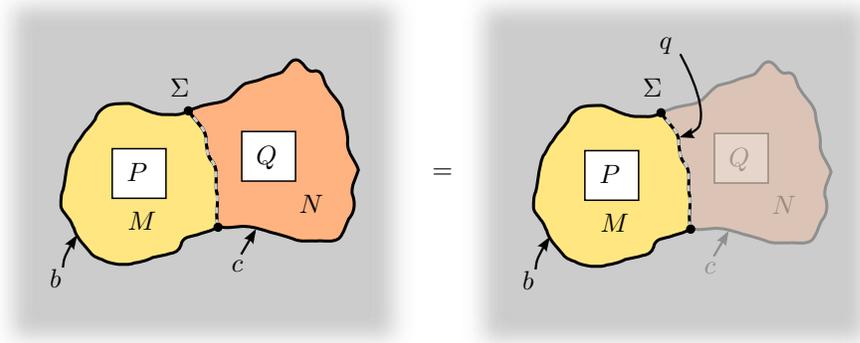
  \caption{Composite probe $P\pcomp Q$ in region $M\cup N$ with generalized boundary condition $b\tens c$ (left). We may replace the region $N$ with probe $Q$ and generalized boundary condition $c$ with a single generalized boundary condition $q$ (right).}
  \label{fig:comp_to_bdy}
\end{figure}

The correspondence between the completeness relation and the gluing of slice regions has a remarkable generalization. This is a formula that yields an explicit description of a composite probe in terms of the component probes. Consider regions $M$ and $N$ with decomposable boundaries $\partial M=\Sigma_1\cup\Sigma$ and $\partial N=\Sigma_2\cup\Sigma$, glued along the common hypersurface $\Sigma$. Moreover, let $P\in\pr_M$ and $Q\in\pr_N$ be probes in the respective regions. We are looking for the value of the composite probe $P\pcomp Q\in \pr_{M\cup N}$ given the boundary condition $\tau(b\tens c)\in\bcg_{\partial (M\cup N)}$, see Figure~\ref{fig:comp_to_bdy}, left hand side. First we use the fact that there must be a generalized boundary condition $q\in\bcg_{\Sigma}$ that encodes the effect that region $N$ with probe $Q$ and partial boundary condition $c$ has on region $M$ through hypersurface $\Sigma$. That is, for any $P$ and $b$,
\begin{equation}
 \lv P\pcomp Q, b\tens c\rv_{M\cup N}=\lv P, b\tens q\rv_M .
\label{eq:PQq}
\end{equation}
This is depicted in Figure~\ref{fig:comp_to_bdy}, right hand side.

\begin{figure}
  \centering
  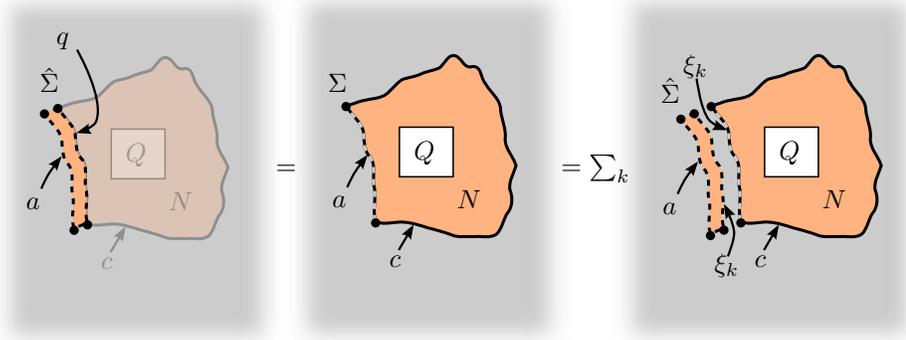
  \caption{Calculation of the generalized boundary condition $q$.}
  \label{fig:probe_to_bc}
\end{figure}

\begin{figure}
  \centering
  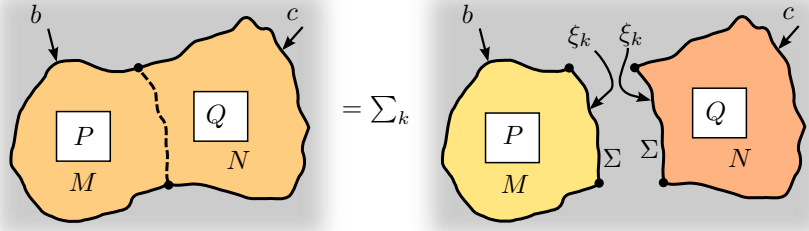
  \caption{Composition of probes: binary composition identity.}
  \label{fig:comp_probes}
\end{figure}

In order to work out how $q$ can be calculated in terms of $Q$ and $c$ we remove the region $M$ and put an arbitrary generalized boundary condition $a\in\bcg_{\Sigma}$ in its place, see Figure~\ref{fig:probe_to_bc}, left hand side. Then we replace $q$ in the inner product between $a$ and $q$ on the interfacing hypersurface $\Sigma$ in terms of the region $N$ with probe $Q$ and partial boundary condition $c$, see Figure~\ref{fig:probe_to_bc}, middle. Mathematically, this is the identity
\begin{equation}
 \lb a, q\rb_{\Sigma}=\lv Q, a\tens c\rv_N .
\label{eq:ipaq}
\end{equation}
Now, we insert for $a$ a basis expansion according to identity (\ref{eq:baseexp}) and use linearity of the value to obtain the equality,
\begin{equation}
 \lv Q, a\tens c\rv_N= \sum_{k\in I} \lb a,\xi_k\rb_{\Sigma} \lv Q, \xi_k\tens c\rv_N .
\label{eq:aqexp}
\end{equation}
The corresponding gluing operation is depicted in Figure~\ref{fig:probe_to_bc} on the right hand side. Combining equations (\ref{eq:ipaq}) and (\ref{eq:aqexp}) and using the non-degeneracy of the inner product yields,
\begin{equation}
 q= \sum_{k\in I} \lv Q, \xi_k\tens c\rv_N\,  \xi_k .
\end{equation}
Inserting this in turn on the right hand side in equation (\ref{eq:PQq}) yields the remarkable \emph{composition identity} for probes,
\begin{equation}
 \lv P\pcomp Q, b\tens c\rv_{M\cup N}=\sum_{k\in I} \lv P, b\tens \xi_k\rv_M \lv Q, \xi_k\tens c\rv_N .
\label{eq:bincompid}
\end{equation}
This is illustrated in Figure~\ref{fig:comp_probes}.

\subsection{Axiomatization}
\label{sec:paxioms}

It is useful to summarize the structures found by condensing them into an \emph{axiomatic system}. We do this in the following, using freely some of the notation already introduced. We assume that a \emph{spacetime system} is given.

\begin{itemize}
\item[\textbf{(P1)}] Associated to each hypersurface $\Sigma$ there is a real vector space $\bcg_{\Sigma}$. A generating cone $\bc_{\Sigma}\subset\bcg_{\Sigma}$ of positive elements makes $\bcg_{\Sigma}$ into a partially ordered vector space. For $\Sigma$ the empty set, $\bcg_{\Sigma}=\R$ with the canonical order.
\item[\textbf{(P2)}] Given a hypersurface $\Sigma$ decomposing into a union $\Sigma_1\cup\cdots\cup\Sigma_n$, there is a positive isomorphism of vector spaces $\tau:\bcg_{\Sigma_1}\tens\cdots\tens\bcg_{\Sigma_n}\to\bcg_{\Sigma}$. This is required to be associative in the obvious way.
\item[\textbf{(P4)}] Associated to each region $M$ there is a real vector space $\pr_M$. A generating cone $\prp_M$ of positive elements makes $\pr_M$ into a partially ordered vector space. There is a special non-zero element $\np\in\prp_M$. There is a positive bilinear map $\lv\cdot,\cdot\rv_M:\pr_M\times\bcg_{\partial M}\to\R$.
\item[\textbf{(P3x)}] Let $\Sigma$ be a hypersurface and $\hat{\Sigma}$ the associated slice region. Then, the bilinear map $\lb\cdot,\cdot\rb_{\Sigma}:\bcg_{\Sigma}\times\bcg_{\Sigma}\to\R$ given by $\lb b_1,b_2\rb_{\Sigma}\defeq \lv\np,b_1\tens b_2 \rv_{\hat{\Sigma}}$ is positive-definite, symmetric and sharply positive, making $\bcg_{\Sigma}$ into a partially ordered inner product space.
\item[\textbf{(P5a)}] Let $M_1$ and $M_2$ be regions and $M=M_1\sqcup M_2$ their disjoint union. Then, there is an isomorphism of partially ordered vector spaces $\pcomp:\pr_{M_1}\times \pr_{M_2}\to \pr_M$ such that, for all $P_1\in\pr_{M_1}$, $P_2\in\pr_{M_2}$, $b_1\in\bcg_{\partial M_1}$, $b_2\in\bcg_{\partial M_2}$,
\begin{equation}
  \lv P_1\pcomp P_2, b_1\tens b_2 \rv_{M}=\lv P_1, b_1\rv_{M_1} \lv P_2, b_2\rv_{M_2} .
\end{equation}
Moreover, $\np\pcomp\np=\np$.
\item[\textbf{(P5b)}] Let $M$ be a region with its boundary decomposing as $\partial M=\Sigma_1\cup\Sigma\cup\Sigma'$, where $\Sigma'$ is a copy of $\Sigma$. Let $M_1$ denote the gluing of $M$ to itself along $\Sigma,\Sigma'$ and suppose it is a region. Then, there is a positive linear map $\pcomp_{\Sigma}:\pr_M\to\pr_{M_1}$ such that the following is true. Given any orthonormal basis $\{\xi_k\}_{k\in I}$ of $\bcg_{\Sigma}$ we have, for all $P\in\pr_M$ and $b\in\bcg_{\partial M_1}$,
\begin{equation}
\lv \pcomp_{\Sigma} P,b\rv_{M_1}=\sum_{k\in I} \lv P, b\tens \xi_k\tens \xi_k\rv_{M} .
\end{equation}
Also, $\pcomp_{\Sigma}\np=\np$.
\end{itemize}

\begin{figure}
  \centering
  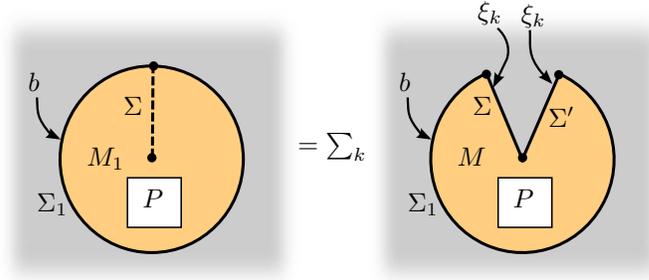
  \caption{Composition of probes: self-composition identity.}
  \label{fig:comp_probes_single}
\end{figure}

We remark on several aspects of the axioms that go beyond or differ from our discussion of structures up to this point. Firstly, in order to allow for the occurrence of hypersurfaces that are empty sets in the notion of decomposition of hypersurfaces we have to associate a space of (generalized) boundary conditions also to the latter. For consistency this needs to be a copy of $\R$, see Axiom~(P1). Secondly, the composition of probes is encoded into two axioms instead of just one. Instead of an axiom corresponding to the composition of two probes through a gluing of the two underlying regions as in the identity (\ref{eq:bincompid}) we have introduced two composition axioms, (P5a) and (P5b). The first of these, (P5a) encodes a composition of two probes, but without an actual gluing occurring on the underlying regions. That is, the regions are combined into a single region simply through a \emph{disjoint} union. The second of these, (P5b), encodes a composition of a probe \emph{with itself}, through a gluing of the underlying region to itself along a duplicated boundary component. The gluing as in the identity (\ref{eq:bincompid}), see Figure~\ref{fig:comp_probes}, can still be performed, although in two steps. First, the probes $P$ and $Q$ are glued to a single probe $P\pcomp Q$ in the region $M\sqcup N$ that is the disjoint union of $M$ and $N$, according to Axiom~(P5a). Then, the probe $P\pcomp Q$ is composed with itself to the probe $\pcomp_{\Sigma}(P\pcomp Q)$ while the region $M\sqcup N$ is glued to itself along $\Sigma,\Sigma'$ resulting in $M\cup N$, according to Axiom~(P5b). When no confusion can arise we write $P\pcomp Q$ as before to mean $\pcomp_{\Sigma}(P\pcomp Q)$. Using these two axioms instead of one has various advantages. One is that each of the two is simpler than the binary composition identity (\ref{eq:bincompid}). Another is that certain compositions allowed through the two axioms cannot be described in terms of binary compositions. The prime example is the self-composition of a probe according to Axiom~(P5b) in a region that is connected, see Figure~\ref{fig:comp_probes_single}.

\subsection{An example}
\label{sec:pf_example}

\begin{figure}
  \centering
  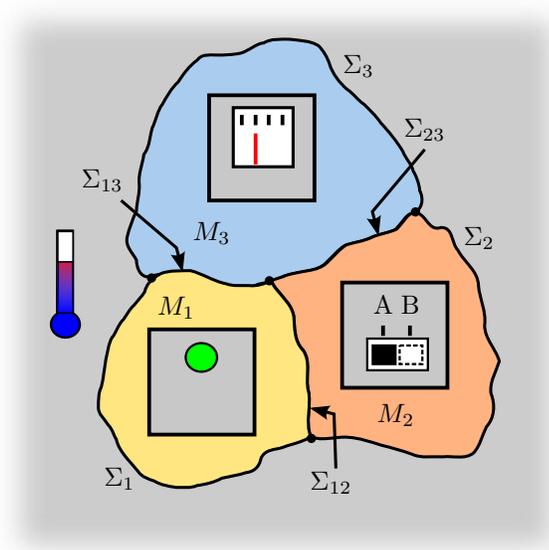
  \caption{Experimental setup with three apparatuses and a boundary condition.}
  \label{fig:pf_example}
\end{figure}

To illustrate the framework developed so far we consider an example setup. Consider three adjacent spacetime regions $M_1,M_2,M_3$ with interfacing hypersurfaces $\Sigma_{1 2},\Sigma_{1 3},\Sigma_{2 3}$ and external boundary hypersurfaces $\Sigma_1,\Sigma_2,\Sigma_3$, see Figure~\ref{fig:pf_example}. In each of the regions we place a measurement apparatus: In $M_1$ we place an apparatus with a single light that turns either \lbl{RED} or \lbl{GREEN} in the experiment. In $M_2$ we place an apparatus that has a switch which can be put either in position A or in position B. In $M_3$ we place an apparatus with a pointer device and a scale. We denote the composite region by $M=M_1\cup M_2\cup M_3$. On the outer boundary $\partial M=\Sigma_1\cup\Sigma_2\cup\Sigma_3$ we apply some boundary condition $b=\sum_l b_1^l \tens b_2^l\tens b_3^l\in\bcg_{\partial M}$, indicated in Figure~\ref{fig:pf_example} by a thermometer.\footnote{In this subsection we use upper indices as summation indices.}

We associate the following probes with the regions:
\begin{itemize}
\item[$M_1$:] Probe $P(r)$ represents the apparatus with outcome \lbl{RED}, $P(g)$ with outcome \lbl{GREEN}, and $P(*)$ without considering the outcome.
\item[$M_2$:] Probe $Q(A)$ represents the apparatus with switch in position A and $Q(B)$ with switch in position B.
\item[$M_3$:] Probe $R[a,b]$ represents the apparatus with pointer in the range $[a,b]$, and probe $R[*]$ without considering the outcome. $R$ denotes the probe that provides the pointer reading.
\end{itemize}
Probes $P(*)$, $Q(A)$, $Q(B)$, and $R[*]$ are non-selective while probes $P(r)$, $P(g)$ and $R[a,b]$ are selective. Probe $R$ is the only probe that is not necessarily primitive.
In order to predict the outcomes of different experiments that can be described with those probes we need to form expressions involving compositions of the probes in $M$ and their values for the boundary condition $b$. Say we take in $M_1$ the probe $P(*)$ corresponding to the presence of the apparatus with the indicator light, in $M_2$ the probe $Q(A)$ corresponding to the apparatus with switch in position A, and in $M_3$ the probe $R[*]$ corresponding to the presence of the apparatus with the pointer device. The composite probe $P(*)\pcomp Q(A)\pcomp R[*]$ and its value for $b$ can be calculated in terms of the values of the individual probes using iterations of the composition identities of Axioms~(P5a) and (P5b):
 \begin{multline}
 \lv P(*)\pcomp Q(A)\pcomp R[*],b\rv_M
 =\sum_{i,j,k,l} \lv P(*),b_1^l\tens \xi_{1 2}^i\tens\xi_{1 3}^j\rv_{M_1} \\
 \lv Q(A),b_2^l\tens \xi_{1 2}^i\tens\xi_{2 3}^k\rv_{M_2} \lv R[*],b_3^l\tens \xi_{1 3}^j\tens\xi_{2 3}^k\rv_{M_3} .
 \label{eq:example_pcomp}
 \end{multline}
Here we use upper indices to enumerate the elements of orthonormal basis of the various spaces of boundary conditions. The expression is analogous for different choices of probes in the regions $M_1, M_2, M_3$.

We remark that the boundary condition $b$ can be replaced with a probe $S$ in a region $X$ that comprises all of the ``exterior'' and whose boundary thus coincides with that of $M$, i.e., $\partial X=\partial M$. For the probe $S$ to induce the boundary condition $b$ we would need for any $c\in\bcg_{\partial M}$,
\begin{equation}
 \lv S,c \rv_X = \lb b,c\rb_{\partial M} .
\end{equation}
Then, the above value can be rewritten as,
\begin{equation}
 \lv P(*)\pcomp Q(A)\pcomp R[*],b\rv_M = P(*)\pcomp Q(A)\pcomp R[*]\pcomp S .
\end{equation}
Other probe compositions can be rewritten analogously. It might seem strange that we do not have any pairing with a boundary condition on the right hand side. This is simply due to the fact that the composite region $M\cup X$ does not have any boundary. Thus, the composite probe directly evaluates to a real number. Alternatively, and in more strict adherence to the axioms, this can be viewed as the pairing with the canonical boundary condition $1\in\R$ for the boundary that is an empty set. To make this explicit we could write,
\begin{equation}
 P(*)\pcomp Q(A)\pcomp R[*]\pcomp S\defeq \lv P(*)\pcomp Q(A)\pcomp R[*]\pcomp S,1 \rv_{M\cup X} .
\label{eq:bctoextprobe}
\end{equation}
In the following we keep with viewing $b$ as a boundary condition, but the notational change involved in considering the probe $S$ instead would be trivial.

Out of the different possible experiments that can be performed with this setup we consider a few illustrative cases.
\begin{enumerate}
\item
  The switch is set to position A. We disregard the pointer device. The probability $p$ to find the light in state \lbl{GREEN} is,
  \begin{equation}
    p=\frac{\lv P(g)\pcomp Q(A)\pcomp R[*],b\rv_M}{\lv P(*)\pcomp Q(A)\pcomp R[*],b\rv_M} .
  \end{equation}
\item
  The switch is set to position $A$. We ignore the light. The expected value $v$ shown on the pointer device is,
   \begin{equation}
    v=\frac{\lv P(*)\pcomp Q(A)\pcomp R,b\rv_M}{\lv P(*)\pcomp Q(A)\pcomp R[*],b\rv_M} .
  \end{equation}
\item
  The switch is set to position $A$. We want to know the expected value $v$ shown by the pointer device given that the light is in state \lbl{GREEN}. (That is we select only those instances of the experiment where the light shows \lbl{GREEN}.) This is,
   \begin{equation}
    v=\frac{\lv P(g)\pcomp Q(A)\pcomp R,b\rv_M}{\lv P(g)\pcomp Q(A)\pcomp R[*],b\rv_M} .
  \end{equation}
\item
  We introduce an agent who randomly sets the switch to position A or to position B in each instance of the experiment with probability $\mu$ for position A and $(1-\mu)$ for position B. Alternatively, we can think of the agent as realized in terms of a machine connected to the apparatus that performs the switching. We ignore the pointer device. The probability for the light state to turn out \lbl{GREEN} is,
  \begin{equation}
    p=\mu\frac{\lv P(g)\pcomp Q(A)\pcomp R[*],b\rv_M}{\lv P(*)\pcomp Q(A)\pcomp R[*],b\rv_M}
      +(1-\mu)\frac{\lv P(g)\pcomp Q(B)\pcomp R[*],b\rv_M}{\lv P(*)\pcomp Q(B)\pcomp R[*],b\rv_M} .
  \end{equation}
\item
  The switch is operated as before. We ignore the pointer device. We are interested in the fraction $p$ of the experiments where the light being \lbl{GREEN} coincides with the switch in position A and the light being \lbl{RED} coincides with the switch in position B. This is,
  \begin{equation}
    p=\mu\frac{\lv P(g)\pcomp Q(A)\pcomp R[*],b\rv_M}{\lv P(*)\pcomp Q(A)\pcomp R[*],b\rv_M}
      +(1-\mu)\frac{\lv P(r)\pcomp Q(B)\pcomp R[*],b\rv_M}{\lv P(*)\pcomp Q(B)\pcomp R[*],b\rv_M} .
  \end{equation}
\item
  The experiment is as before. However, we now want to know the expectation value $v$ of the pointer device for those outcomes where the coincidence (\lbl{GREEN} - A, \lbl{RED} - B) occurs. This is,
   \begin{equation}
    v = \frac{\mu\frac{\lv P(g)\pcomp Q(A)\pcomp R,b\rv_M}{\lv P(*)\pcomp Q(A)\pcomp R[*],b\rv_M}
      +(1-\mu)\frac{\lv P(r)\pcomp Q(B)\pcomp R,b\rv_M}{\lv P(*)\pcomp Q(B)\pcomp R[*],b\rv_M}}
     {\mu\frac{\lv P(g)\pcomp Q(A)\pcomp R[*],b\rv_M}{\lv P(*)\pcomp Q(A)\pcomp R[*],b\rv_M}
      +(1-\mu)\frac{\lv P(r)\pcomp Q(B)\pcomp R[*],b\rv_M}{\lv P(*)\pcomp Q(B)\pcomp R[*],b\rv_M}} .
  \label{eq:garbeval}
  \end{equation}
\item
  We change the behavior of the agent controlling the switch. Instead of a random selection the switch is now put into position A if the light state is \lbl{GREEN} and put into position B if the light state is \lbl{RED}. This prescription carries a hidden assumption for it to make sense. The assumption is that the light state does not itself depend on the position of the switch. One can interpret this as the absence of a \emph{causal influence} of the switch on the light. For the light state \lbl{GREEN}, this assumption is encoded in the equality
  \begin{equation}
    \frac{\lv P(g)\pcomp Q(A)\pcomp R[*],b\rv_M}{\lv P(*)\pcomp Q(A)\pcomp R[*],b\rv_M}
    =\frac{\lv P(g)\pcomp Q(B)\pcomp R[*],b\rv_M}{\lv P(*)\pcomp Q(B)\pcomp R[*],b\rv_M} .
  \label{eq:lindeps}
  \end{equation}
  The corresponding equality for the state \lbl{RED} follows since $P(g)+P(r)=P(*)$. With this setup we can ask for example for the expectation value $v$ of the pointer device. This is,
  \begin{equation}
    v=\frac{\lv P(g)\pcomp Q(A)\pcomp R,b\rv_M}{\lv P(*)\pcomp Q(A)\pcomp R[*],b\rv_M}
      +\frac{\lv P(r)\pcomp Q(B)\pcomp R,b\rv_M}{\lv P(*)\pcomp Q(B)\pcomp R[*],b\rv_M} .
  \end{equation}
  Remarkably, (given identity (\ref{eq:lindeps})) this expectation value exactly equals the value (\ref{eq:garbeval}) for the case $\mu=1/2$. That is, we obtain exactly the same predictions if instead of an agent setting the switch depending on the light state we have an agent setting the switch randomly with equal probability and then selecting the instances where light and switch are correlated correctly.
\end{enumerate}
We add the general remark that for all considered cases the denominators in expressions for probabilities or expectation values may turn out to be zero, making the respective expression undefined. This invariably signals the physical impossibility of the setup, recall the discussion in Subsection~\ref{sec:valpred}.

\subsection{A diagrammatic representation}
\label{sec:pfcd}

\begin{figure}
  \centering
  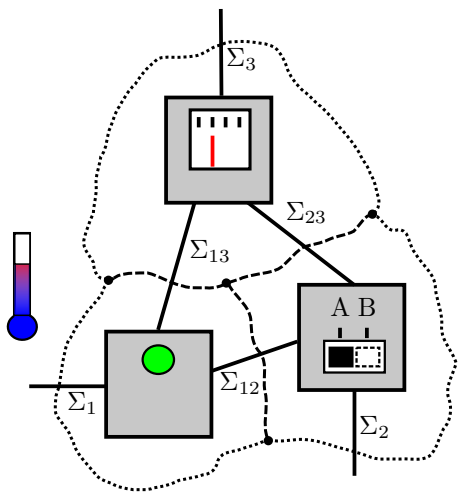
  \caption{Simplified graphical representation of experimental setup via the dual 1-complex. Links are represented as thick straight lines. They are dual to internal hypersurfaces (dashed lines) and external hypersurfaces (dotted lines).}
  \label{fig:pf_dual_example}
\end{figure}

For the graphical representation of an experimental setup such as that depicted in Figure~\ref{fig:pf_example} it is convenient to introduce some simplifications. Since we represent probes or apparatuses by boxes we can dispose with the explicit representation of spacetime regions, as long as there is exactly one box per region. That means that we also should draw a box representing the null probe if the region is empty. The boundaries of regions are important though, since these are the interfaces where probes interact. There are two types of boundary hypersurfaces, \emph{internal} and \emph{external} ones. The internal ones are boundary components where a gluing takes place and which consequently end up in the interior of the composite region. These are depicted by dashed lines in Figure~\ref{fig:pf_dual_example}. The external ones are the ones to which nothing is glued and that consequently end up as boundary components of the composite region. These are depicted by dotted lines in Figure~\ref{fig:pf_dual_example}. We now draw one line, called \emph{link}, for each hypersurface as follows. If the hypersurface is internal the line connects the boxes that represent the probes or apparatuses in the adjacent regions. If the hypersurface is external then the line connects the box in the interior region with the exterior, with an open end. This is depicted in Figure~\ref{fig:pf_dual_example}. We use the same label that indicates a hypersurface for the corresponding link. We may now omit the hypersurfaces from the drawing as the relevant information is contained in the links. In mathematical language, we are drawing the \emph{1-skeleton} of the \emph{cell complex} that is \emph{dual} to the cell complex formed by the spacetime regions and hypersurfaces. The boxes are the 0-cells dual to regions ($d$-cells) and the links are 1-cells dual to hypersurfaces ($d-1$-cells).

\begin{figure}
  \centering
  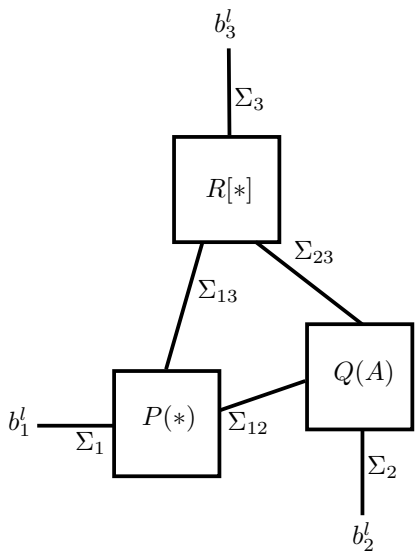
  \caption{From spacetime arrangement to circuit diagram: The diagram is obtained as the dual 1-complex.}
  \label{fig:pf_diag_example}
\end{figure}

For purposes of carrying out calculations of relevant probabilities or expectation values the resulting graphical representation (Figure~\ref{fig:pf_dual_example} with dashed and dotted lines removed) of the experimental setup contains as much information as the original spacetime representation (Figure~\ref{fig:pf_example}). We can make the relation between graphical representation and calculation into a formal correspondence by labeling the boxes with corresponding probes and putting boundary conditions on the open ends of links. For the same example setup, taking probes $P(*)$, $Q(A)$ and $R[*]$ as well as the external boundary condition $b=\sum_l b_1^l\tens b_2^l\tens b_3^l$ (compare Subsection~\ref{sec:pf_example}) results in the diagram depicted in Figure~\ref{fig:pf_diag_example}. (The summation over the index $l$ is not depicted.) This represents the value $\lv P(*)\pcomp Q(A)\pcomp R[*],b\rv_M$.

These diagrams representing values of probe compositions are a special version of \emph{circuit diagrams} to be discussed in Section~\ref{sec:catdiag}. The corresponding calculation in terms of the composition identities for probes can be read off from the diagram as follows. For each box we have a probe, in the example case $P(*)$, $Q(A)$ and $R[*]$. For each link we have a space of generalized boundary conditions, here $\bcg_{\Sigma_{1 2}},\bcg_{\Sigma_{1 3}},\bcg_{\Sigma_{2 3}}$ for internal links, and $\bcg_{\Sigma_1},\bcg_{\Sigma_2},\bcg_{\Sigma_3}$ for external (open) links. For each internal link we take an orthonormal basis of the corresponding space of generalized boundary conditions, here $\{\xi_{1 2}^i\}, \{\xi_{1 3}^j\}, \{\xi_{2 3}^k\}$. Then, for each internal link we choose one basis element. Also, we choose one summand for the external boundary condition. For each probe we combine the generalized boundary conditions for all of its links: basis elements for internal links and external boundary conditions for external links. We pair the probe with the combined generalized boundary condition. In the example, for the probe $P(*)$ we get the value $\lv P(*),b_1^l\tens\xi_{1 2}^i\tens\xi_{1 3}^j\rv_M$. Finally, we take the product of the values for all the probes and sum over the basis for all links and sum over the index for the external boundary condition. In the example this yields exactly the right hand side of equation (\ref{eq:example_pcomp}).

It is tempting to view the diagrams as more than calculational tools. Indeed, they suggest to do away with the spacetime concepts of region and hypersurface. In their place we could put the simpler concept of link, both between probes and from probes to the ``exterior''. Physically, a link represents an interaction or communication interface between spacetime regions, or to avoid the latter notion completely, between apparatuses or \emph{processes}. We may even reformulate the axioms of Subsection~\ref{sec:paxioms} straightforwardly in this language. There are, however, two important reasons for why we have not followed this route from the beginning. The first reason is that the potent physical principle of locality requires some notion of spacetime. Without this we would have had a harder time to get sufficient guidance or justification in building the present framework. The second reason is that the present framework, as developed up to this point, turns out to need refinements to successfully describe certain physical theories. Some of these refinements require additional structure that depends on boundaries of hypersurfaces and on how exactly hypersurfaces are joined in a decomposition. This information is lost when we take only the dual 1-skeleton formed by the links. We shall give more details on some of these refinements in Subsections~\ref{sec:refqft} and \ref{sec:fermions}. However, much of the following exposition relies only on the framework as developed up to this point and we shall mostly use the spacetime and diagrammatic points of view interchangeably.

\begin{figure}
  \centering
  \begin{tabular}{cc}
  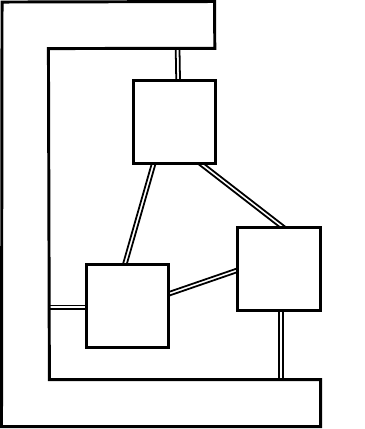 &
  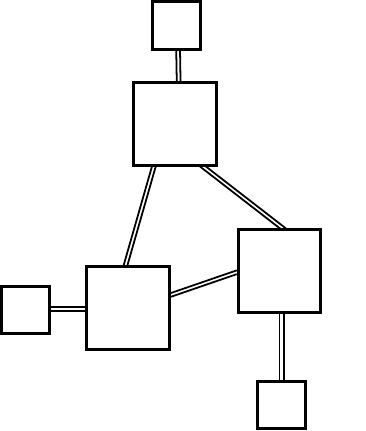 \\ \\
  (a) & (b)
  \end{tabular}
  \caption{(a) Boundary condition converted to a probe $S$. (b) Factorizing boundary condition with each component converted to a probe.}
  \label{fig:pf_diag_example_uc}
\end{figure}

For further simplicity it might be desirable to restrict to closed diagrams only, i.e., diagrams that have no links with open ends. At the same time this permits to do away with a notion of external boundary condition. We have seen already in Subsection~\ref{sec:pf_example} how this can be accomplished. Namely we can convert the external boundary condition into a probe associated to a region that comprises all of the exterior. In Figure~\ref{fig:pf_dual_example} this would translate into drawing a box around the thermometer (representing this external probe) and connecting the three open links (labeled $\Sigma_1, \Sigma_2, \Sigma_3$) to this box. As compared to Figure~\ref{fig:pf_diag_example} this yields an additional box with the label $S$ (compare Subsection~\ref{sec:pf_example}) with the open links connected to it, see Figure~\ref{fig:pf_diag_example_uc}.a. The corresponding value then writes as on the right hand side of equation (\ref{eq:bctoextprobe}). If the boundary condition $b=\sum_l b_1^l\tens b_2^l\tens b_3^l$ factorizes as $b=b_1\tens b_2\tens b_3$ it might instead be convenient to model each component as a separate probe. The corresponding diagram is depicted in Figure~\ref{fig:pf_diag_example_uc}. For later use, in Figures~\ref{fig:pf_diag_example_uc}.a and b double lines have been used instead of single lines.

Physically it is often unnecessary to model the ``exterior'' of an experimental setup at all. That is, the experiment can be considered reasonably insensitive to changes in external boundary conditions. Alternatively, we might view this as choosing a ``standard'' boundary condition which is incorporated into the (internal) probes. In the spacetime setting this can be implemented by removing any external boundary hypersurface ``to infinity''. In the diagrammatic setting we simply remove any external (i.e.\ open) links. The treatment of the example setup of Figure~\ref{fig:pf_example} in Subsection~\ref{sec:pf_example} was such that always the same boundary condition was chosen for all considered experiments. Thus, removing this boundary condition does not modify the respective statements about probabilities or expectation values except for replacing values of the type
\begin{equation}
 \lv P(*)\pcomp Q(A)\pcomp R[*],b\rv_M\quad \text{by values of the type}\quad P(*)\pcomp Q(A)\pcomp R[*] .
\end{equation}

\subsection{Infinite dimensions}
\label{sec:infinidim}

We have so far pretended that all mathematical expressions, including predicted expectation values (or their ingredients), are perfectly well defined and finite. Experience tells us that this is not to be expected in physical theories once we deal with infinitely many degrees of freedom such as in field theory. Indeed, we shall see this expectation confirmed in the sections on classical theory (Section~\ref{sec:classical}) and on quantum theory (Section~\ref{sec:quantum}). To deal with this we need to make some functional analytic refinements to the framework so far developed.

A key issue is that the pairing between probes and generalized boundary conditions may not be well defined everywhere, due to ``infinities''. As one consequence, the inner product on the space of generalized boundary conditions may also not be well defined everywhere. Traditionally in functional analysis, one deals with such ``unbounded'' maps by defining them on subspaces which ideally are dense with respect to a convenient topology. In the present setting positivity allows for a simpler approach that does not explicitly require additional topological structure. The relevant concepts are developed in the appendix starting with the notion of \emph{unbounded positive linear} map, see Definition~\ref{dfn:wposmap}. The latter generalizes the notion of (real valued) \emph{positive linear map}, see Definition~\ref{dfn:posmap}. The core idea is very simple: To know a real valued positive linear map it is enough to know it on positive elements of the partially ordered vector space if its cone is generating. But restricting the map to the positive cone it only takes positive values and we may allow it to take values not only in $[0,\infty)$ but in $[0,\infty]$.

In this way we take the pairing to be an \emph{unbounded positive bilinear map} (Definition~\ref{dfn:wposbmap}). Consequently, the inner product on the space of generalized boundary conditions becomes a \emph{positive-definite unbounded sharply positive symmetric bilinear map}. The space of generalized boundary conditions itself is then a \emph{partially ordered unbounded inner product space} (Definition~\ref{dfn:wpoip}). Up to these modifications the axioms of Subsection~\ref{sec:paxioms} remain the same. There are a few associated adjustments, however, that are worth pointing out. One important issue is the use of orthonormal basis of the space of generalized boundary conditions. With the inner product unbounded it becomes unclear what an orthonormal basis is supposed to be. What is meant is in fact an orthonormal basis of a maximal partially ordered subspace on which the inner product is well-defined (in the sense of Proposition~\ref{prop:wpbmapdef}). Such a subspace has to be chosen for every space of generalized boundary conditions. In turns out that in both classical and quantum theory there are natural choices for these, see Sections~\ref{sec:classical} and \ref{sec:quantum}. Another issue is that tensor products of vector spaces such as that appearing in Axiom (P2) are to be understood in a suitably completed sense.

There are also minor implications for physical predictability. The denominator (and possibly also the numerator) in probability expressions (\ref{eq:elemprob}) and (\ref{eq:elembdy}) may turn out to be infinite. The physical interpretation of this occurring is that the condition encoded in the denominator is insufficient, i.e., not sufficiently stringent to determine a meaningful measurement.

\section{Diagrams and categories}
\label{sec:catdiag}

\subsection{Categories and circuit diagrams}
\label{sec:catcd}

The correspondence between the composition of a diagram and the composition of the calculation of the associated value fits into a more general mathematical framework, best formulated in terms of the theory of \emph{monoidal categories}. We proceed to give an elementary account of this, tailored to the specific needs of the present paper. We warn the reader that this account is condensed and not fully accurate, leaving out important mathematical details in order not to overload this paper.

While diagrams have been used for a long time to represent mathematical calculations and their compositional structure especially in physics, Penrose \cite{Pen:appndtens} gave a systematic account of this around 1970. Almost 20 years later this diagrammatic language was put on a categorical foundation, making it much more powerful with variants for many different purposes developed. This happened as part of the revolution that brought together the fields of category theory, knot theory, low dimensional topology, quantum groups, algebraic topology and (topological) quantum field theory, see Subsection~\ref{sec:tqft}. A small selection of relevant references is \cite{FrYe:braidcat,Yet:reps,ReTu:ribboninv,JoSt:geomtcalc}. For a survey we refer the reader to Selinger's paper \cite{Sel:surveygraphcat}. Our presentation is mostly adapted from the book \cite{Oe:tqft}.

We start with a collection $\cC=\{V,W,\dots\}$ of vector spaces over $\K$, which is either the real numbers $\K=\R$ or the complex numbers $\K=\bC$. To avoid irrelevant complications, we take these vector spaces to be finite dimensional. In addition we have, for each pair $(V,W)$ of vector spaces a set $\cM(V,W)$ of linear maps $V\to W$, called \emph{morphisms}. These sets are closed under composition. That is given $f\in\cM(V,W)$ and $g\in\cM(W,X)$ we have $g\circ f\in\cM(V,X)$. Also, $\cM(V,V)$ contains the identity map $\id_V:V\to V$. What we have described so far is a \emph{category} of vector spaces, which we also denote by $\cC$.

\begin{figure}[h]
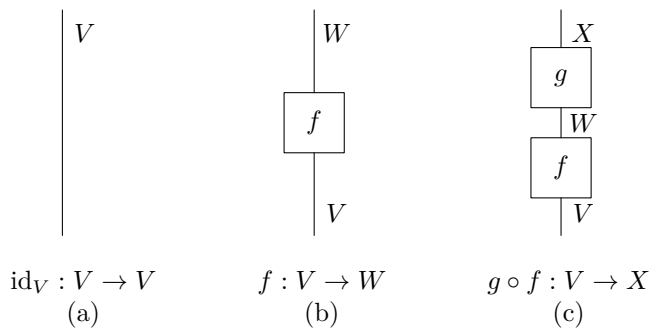

\centering
\begin{tabular}{cp{5mm}cp{5mm}cp{5mm}c}
\input{figures/elem_id} && \input{figures/elem_mor}
&& \input{figures/elem_comp} \\
\\
$\id_V:V\to V$ &&  $f:V\to W$ && $g\circ f:V\to X$ \\
(a) && (b) && (c)
\end{tabular}
\caption{Elementary diagrams associated to a category: (a) identity morphism, (b)
  general morphism, (c) composition of morphisms.}
\label{fig:diagelem}
\end{figure}

We proceed to introduce a diagrammatic language to represent morphisms of our category, see Figure~\ref{fig:diagelem}. The simplest diagram is a vertical line, representing the identity morphism $\id_V:V\to V$, Figure~\ref{fig:diagelem}.a. The \emph{label} on the line identifies the vector space $V$. An arbitrary morphism $f:V\to W$ is represented by a \emph{box} labeled by the morphism $f$, with an incoming line at the bottom labeled by the domain $V$ and an outgoing line at the top, labeled by the range $W$, see Figure~\ref{fig:diagelem}.b. This also illustrates that these diagrams are to be read from bottom to top.\footnote{The direction from bottom to top is conventional and chosen here to coincide with an interpretation in terms of an arrow of time for certain applications. A large part of the literature uses the opposite convention, including \cite{Oe:tqft}.} A composition of two morphism is thus represented by two boxes, connected with a vertical line, see Figure~\ref{fig:diagelem}.c. We also refer to the lines in the diagrams as \emph{wires} and to the diagrams themselves as \emph{circuit diagrams}.

\begin{figure}[h]
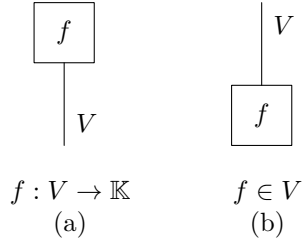

  \centering
  \begin{tabular}{cp{5mm}cp{5mm}c}
  \input{figures/elem_mor_in} && \input{figures/elem_mor_out} \\
  \\
  $f:V\to\K$ &&  $f\in V$ \\
  (a) && (b)
  \end{tabular}
  \caption{Diagrams involving the unit object: (a) a linear function, (b) an element of a vector space.}
  \label{fig:diagunit}
\end{figure}

A linear function $f$ on a vector space $V$ with values in $\K$ is considered as a special morphism $f\in\cM(V,\one)$. $\one$ is the \emph{unit object} in the category, identified with the ground field $\K$, viewed as a vector space of dimension $1$. By convention, the wire corresponding to the unit object is omitted in the diagrams. So $f$ is represented as shown in Figure~\ref{fig:diagunit}.a. Similarly, a linear map $f$ from $\K$ to a vector space $V$ is considered a morphism $f\in\cM(\one,V)$. What is more, such a map is in one-to-one correspondence to an element $\tilde{f}$ in $V$ via $f(\lambda)=\lambda \tilde{f}$. In the following, we do not distinguish between such maps and the corresponding elements, using the same notation. Thus, an element $f$ in a vector space $V$ is represented as shown in Figure~\ref{fig:diagunit}.b.

\begin{figure}[h]
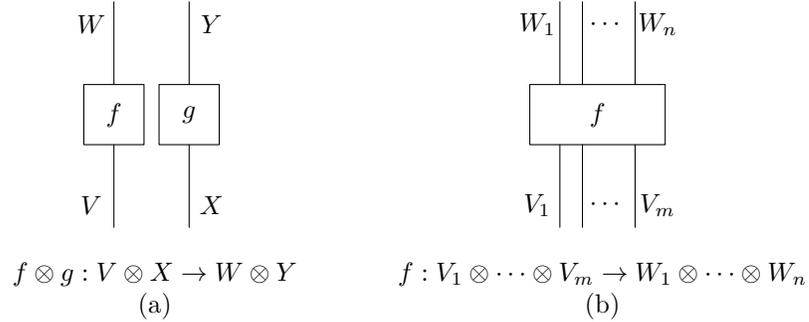

  \centering
  \begin{tabular}{cp{5mm}c}
  \input{figures/elem_tp} && \input{figures/elem_mor_tp} \\
  \\
  $f\tens g:V\tens X\to W\tens Y$ && $f:V_1\tens\cdots\tens V_m\to W_1\tens\cdots\tens W_n$\\
  (a) && (b)
  \end{tabular}
  \caption{Diagrams associated to a tensor category: (a) tensor product of morphisms, (b) morphism between tensor products.}
  \label{fig:diagtp}
\end{figure}

Another important operation that we are going to consider is the tensor product of vector spaces. That is, given $V,W\in\cC$ the tensor product $V\tens W$ is also in $\cC$. Moreover, this operation of taking the tensor product is assumed compatible with the morphisms. In particular, we can take the tensor product of morphisms. This simply means here that given morphisms $f:V\to W$ and $g:X\to Y$ the tensor product $f\tens g:V\tens X\to W\tens Y$ is determined by $v\tens x\mapsto f(v)\tens g(x)$. The category equipped with the tensor product is called a \emph{monoidal category} (or sometimes \emph{tensor category}). The tensor product is diagrammatically represented by horizontal juxtaposition. For the tensor product of two morphisms this is illustrated in Figure~\ref{fig:diagtp}.a. We also consider morphisms that are directly defined between tensor products, see Figure~\ref{fig:diagtp}.b.

\begin{figure}[h]
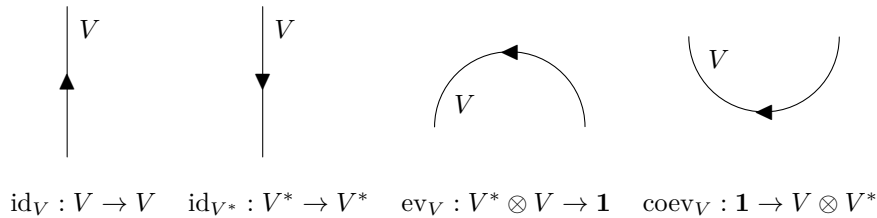

  \centering
  \begin{tabular}{cccc}
  \input{figures/rigid_id} & \input{figures/rigid_idd}
  & \input{figures/rigid_ev} & \input{figures/rigid_coev} \\
  \\
  $\id_V:V\to V$ & $\id_{V^*}:V^*\to V^*$
  & $\ev_V:V^*\tens V\to\one$ & $\coev_V:\one\to V\tens V^*$
  \end{tabular}
  \caption{Diagrams with arrows allow to represent duals.}
  \label{fig:diagdual}
\end{figure}

Let $V$ be a vector space and $V^*$ the dual vector space, i.e., the vector space of linear functions on $V$. We view the evaluation map $\ev_V:V^*\tens V\to\K$ given by $\phi\tens v\mapsto \phi(v)$ as a morphism in $\cM(V^*\tens V,\one)$. Let $\{v_i\}_{i\in I}$ be a basis of $V$ and $\{\phi_i\}_{i\in I}$ a dual basis with the property $\phi_i(v_j)=\delta_{i j}$. Then consider the element $\coev_V\defeq \sum_i v_i\tens \phi_i$ in the tensor product space $V\tens V^*$. As explained previously we can view this as a morphism $\coev_V\in\cM(\one,V\tens V^*)$. It is advantageous to represent these special morphisms diagrammatically not as boxes, but as arches. For this to be consistent, we slightly modify the diagrams introduced so far, by equipping all wires with \emph{arrows}. The arrows to be added to the wires in the diagrams considered so far (as in Figures~\ref{fig:diagelem}, \ref{fig:diagunit}, \ref{fig:diagtp}) are all pointing upwards. A wire with a downward arrow and labeled by $V\in\cC$ represents not $V$, but its dual $V^*$. This allows in particular the representation of $\ev_V$ and $\coev_V$ as arches, see Figure~\ref{fig:diagdual}. We assume in the following that any vector space $V$ is canonically identified with its bi-dual $V^{**}$. (In general such an identification is called a \emph{pivotal} structure.)

\begin{figure}[h]
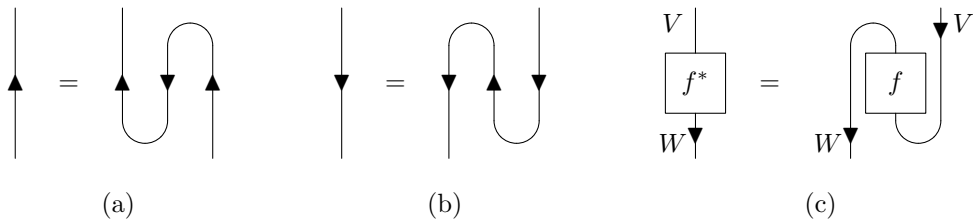

  \centering
  \begin{tabular}{cp{5mm}cp{5mm}c}
  \input{figures/rigid_ident2} && \input{figures/rigid_ident1} && \input{figures/rigid_identd} \\
  \\
  (a) && (b) && (c)
  \end{tabular}
  \caption{Identities involving $\ev_V$ (a), and $\coev_V$ (b). (The object label $V$ on the wires is omitted.) Generating a new morphism $f^*$ from a given one $f$ by dualization (c).}
  \label{fig:diagdident}
\end{figure}

The morphisms $\ev_V$ and $\coev_V$ satisfy elementary identities that can be diagrammatically represented as shown in Figure~\ref{fig:diagdident}.a and b. For example the diagram of Figure~\ref{fig:diagdident}.a represents the identity $\id_V=(\id_V\tens\ev_V)\circ(\coev_V\tens\id_V)$. Expressed in terms of a basis and dual basis of $V$ as above, this is nothing but the identity $v=\sum_i \phi_i(v)\, v_i$. On the other hand, the diagrams suggest an interpretation in terms of a ``straightening''. That is, a wire that involves curved segments and backtracking is equivalent to its ``straightened'' version. We can also use $\ev_V$ and $\coev_V$ to generate a new morphism from a given one. For example, we can use $\ev_V$ to remove a tensor factor of $V$ from the range of the morphism and add a tensor factor of $V^*$ to its domain. For a morphism $f:V\to W$ the composition with $\coev_V$ and $\ev_W$ to obtain a new morphism $f^*:W^*\to V^*$ is illustrated in Figure~\ref{fig:diagdident}.c.

\begin{figure}[h]
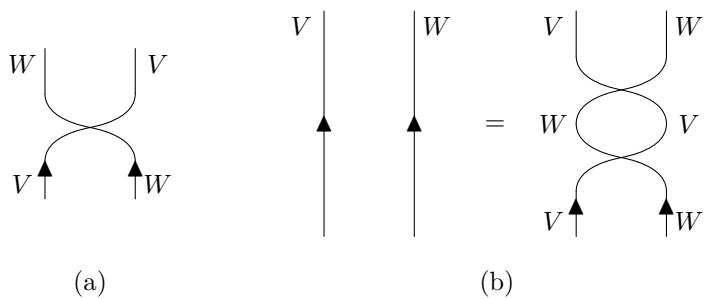

  \centering
  \begin{tabular}{cp{5mm}c}
  \input{figures/braid_sym} && \input{figures/braid_sym_id} \\
  \\
  (a) && (b)
  \end{tabular}
  \caption{Diagram representing the symmetric structure (a) and its self-inverseness identity (b).}
  \label{fig:diagsym}
\end{figure}

Given vector spaces $V,W\in\cC$ we can form their tensor product in two ways, depending on the order, $V\tens W$ and $W\tens V$. Suppose that the category comes equipped for any such pair with a special morphism $\psi_{V,W}:V\tens W\to W\tens V$ that identifies these two tensor products. If the morphism is self-inverse in the sense $\psi_{W,V}\circ\psi_{V,W}=\id_V\tens\id_W$, it is called a \emph{symmetric structure} and the category is called a \emph{symmetric (monoidal) category}. Naively, one might think that this morphism should just be the flip map $v\tens w\mapsto w\tens v$. Indeed, this is the case for the applications in most of this paper. However, when working with graded vector spaces as in the case of fermions (see Subsection~\ref{sec:fermions}), there are factors of $-1$ coming in.\footnote{More generally, there are interesting monoidal categories with maps $\psi$ that are not symmetric. These play an important role in the construction of invariants of knots and 3-manifolds. They arise for example as representation categories of quantum groups, see e.g.\ \cite{Oe:tqft}.} Figure~\ref{fig:diagsym} shows the diagrammatic representation of the symmetric structure in terms of a crossing as well as the self-inverseness identity.

There are additional important compatibility conditions and identities that we do not mention here, but the upshot concerning the circuit diagrams for a symmetric category is a \emph{combinatorial invariance property}: Suppose we have two circuit diagrams that have (1) the same wire endings with the same labels on the top line and on the bottom line, (2) contain exactly the same boxes with the same labels, and (3) the wire endings on boxes and top and bottom line are connected in exactly the same way (i.e., which end is connected with which other is the same). Then, the two diagrams are equivalent, i.e., they represent the same morphism.

This invariance property suggests to also abandon the diagrammatic distinction between morphisms that are related to each other by composition with the dualization maps $\ev_V$ and $\coev_V$. For example the morphisms $f$ and $f^*$ in Figure~\ref{fig:diagdident} would receive the same label, say $f$. We adopt this convention in the following. Diagrammatically, it then no longer matters whether we attach a wire to the bottom line or top line of the box representing the respective morphism. Since the transformation from one to the other as linear maps is unique, this is well defined. What still matters diagrammatically is whether the arrow on a wire ending on the box representing a morphism points towards the box or away from it.

Suppose that the vector spaces in our category are additionally equipped with positive-definite inner products $\lb\cdot,\cdot\rb$ in a way compatible with the other structures (that we are not going to elaborate on). Given a vector space $V$ the inner product yields an identification with its dual space $V^*$ via $V\to V^*$ given by $v\mapsto \lb v,\cdot\rb$. (By the Riesz Representation Theorem this works also in the infinite-dimensional case.) If we are working over the real numbers, i.e., $\K=\R$ this identification is a linear isomorphism. We can use this to introduce a version of the circuit diagrams without arrows, but still allowing for arches analogous to the ones of Figure~\ref{fig:diagdual}. That is, we interpret the diagrams as if all arrows where pointing up. Moreover, a downward arch represents the morphism $\ip_V:V\tens V\to\one$ given by the inner product, while an upward arch represents the dual morphism $\coip_V:\one\to V\tens V$. Let $\{\xi_k\}_{k\in I}$ be an orthonormal basis of $V$ with $\ip_V(\xi_i\tens\xi_j)=\delta_{i j}$. Then $\coip_V$ is given by the element $\sum_{i\in I} \xi_i\tens\xi_i$ of $V\tens V$. To distinguish this version of the diagrams we refer to them as \emph{undirected} circuit diagrams (and by contrast call the other ones \emph{directed}) and represent them by double lines, see Figure~\ref{fig:diagud}.

\begin{figure}[h]
  \centering
  \begin{tabular}{ccc}
  \input{figures/rigid_ud_id}
  & \input{figures/rigid_ud_ev} & \input{figures/rigid_ud_coev} \\
  \\
  $\id_V:V\to V$ & $\ip_V:V\tens V\to\one$ & $\coip_V:\one\to V\tens V$
  \end{tabular}
  \caption{Undirected diagrams for categories of real inner product spaces.}
  \label{fig:diagud}
\end{figure}

Crucially, the combinatorial invariance property of the directed circuit diagrams also holds for the undirected ones. What is more, it is possible and convenient also in the undirected setting to treat wires connected to the bottom or to the top of a box representing a morphism as equal. This means here that we no longer distinguish diagrammatically between morphisms that can be converted into each other by attaching (undirected!) arch diagrams. For example a morphism $f:V\to W$ is then considered equivalent to the morphism $\tilde{f}:V\tens W\to\R$ given by $v\tens w\mapsto \ip_W(f(v)\tens w)$. We adopt this convention in the following.

Let us remark that over the complex numbers, i.e., $\K=\bC$, the identification of a vector space with its dual induced by an inner product is not complex linear, but complex conjugate linear. Because of this we cannot in this case obtain an undirected version of the diagrammatics with analogous properties. Further modifications would be required. We do not consider this in the present paper, but reserve the use of undirected diagrams to categories of \emph{real} inner product spaces.

\subsection{Application to the positive formalism}
 
It is not difficult to see now that the diagrams used in Subsection~\ref{sec:pfcd} are in fact \emph{undirected circuit diagrams}. Given a theory in the positive formalism, the associated category is a category of partially ordered inner product spaces with positive linear maps as morphisms. The objects of the category are the partially ordered inner product spaces of generalized boundary conditions $\bcg_{\Sigma}$ associated to hypersurfaces $\Sigma$. The morphisms of the category arise from probes and boundary conditions. For boundary conditions this is rather straightforward: As discussed above, given a vector space $\bcg_{\Sigma}$ in the category, an element $b\in\bcg_{\Sigma}$ can be viewed as a linear map $\R=\one\to\bcg_{\Sigma}$. In particular, it is a morphism that may be represented as in Figure~\ref{fig:diagunit}.b (but with a double line).
Probes give rise to morphisms in a similar way. Let $M$ be a region and $\partial M=\Sigma_1\cup\cdots\cup\Sigma_n$ a decomposition of its boundary. Given a probe $P\in\pr_M$, its value yields a linear map $\bcg_{\Sigma_1}\tens\cdots\tens\bcg_{\Sigma_n}\to \one=\R$ via, $b_1\tens\cdots\tens b_n\mapsto \lv P, b_1\tens\cdots\tens b_n\rv_M$.

The way we have identified boundary conditions as morphisms these would be represented as boxes with wires attached at the top. Similarly, probes would be boxes with wires attached at the bottom. However, as explained, we follow the convention that wires may be attached arbitrarily as long as the underlying morphisms are transformed correspondingly. It then does not even matter if we depict wires as ending on sides of boxes instead of on the top or the bottom. We may thus read the diagrams depicted in Figures~\ref{fig:pf_diag_example_uc}.a and b of Subsection~\ref{sec:pfcd} as undirected circuit diagrams. Moreover, it is easy to see that the rules for their evaluation coincide precisely with those already presented in that subsection.

\section{Time evolution and causality}
\label{sec:evolution}

The traditional way to describe the dynamics of physical systems is in terms of the evolution of \emph{states} in \emph{time}. This is frequently the most practical and useful description for non-relativistic systems and also for some relativistic systems. The present section is dedicated to describing how this point of view is recovered from the positive formalism as developed so far. For simplicity, we disregard the refinement introduced in Subsection~\ref{sec:infinidim}. We use the notion of \emph{system} freely in the following in the way it is commonly used in the literature.

\subsection{States and evolution}
\label{sec:statesevol}

We are now necessarily working in a setting where we have a fixed global spacetime background with a foliation into equal-time hypersurfaces. For simplicity we take spacetime to be Minkowski spacetime $\R\times\R^3$, although we do not consider its relativistic aspects and work in a fixed inertial reference frame. We denote the equal-time hypersurface of all events at time $t$ by $\Sigma_t$. Moreover, we denote the associated space of generalized boundary conditions by $\bcg_t$ as a shorthand for $\bcg_{\Sigma_t}$. Similarly, we take $[t_1,t_2]$ as a shorthand to designate the spacetime region $M=[t_1,t_2]\times\R^3$ which is obtained by extending the time interval $[t_1,t_2]$ over all of space $\R^3$.

\begin{figure}
\centering
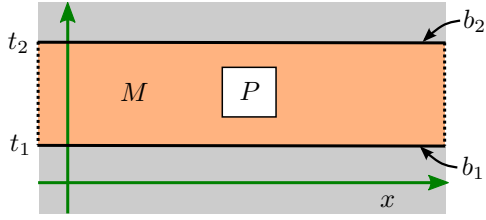
\caption{Time interval region $M=[t_1,t_2]\times\R^3$ with probe $P$ and boundary condition $b_1\tens b_2$.}
\label{fig:tintpf}
\end{figure}

Consider a measurement taking place between the initial time $t_1$ and the final time $t_2$. In the positive formalism this is represented by one or several probes $P\in\pr_{[t_1,t_2]}$ associated to the spacetime region $[t_1,t_2]$.
A prediction for measurement outcomes in terms of probabilities or expectation values is obtained by additionally specifying a boundary condition $b\in\bc_{\partial [t_1,t_2]}\subseteq\bcg_1\tens\bcg_2$. For simplicity we consider a factorizing boundary condition $b=b_1\tens b_2$ (otherwise we would have a linear combination of factorizing boundary conditions), see Figure~\ref{fig:tintpf}. By locality, $b_1$ encodes any knowledge about the past of the system before $t_1$ in so far as it might be relevant for the prediction of any measurement outcome in $[t_1,t_2]$. Similarly, $b_2$ encodes corresponding information about the future after $t_2$.
A concise definition of \emph{state} of a system at a fixed time $t$ can be derived from this prototypical example. A \emph{state} is a parametrization of all the knowledge we have about the past of a system in so far as this may be relevant for the prediction of the outcome of any measurement that can be effected on the system in the future. This may include information about preparations, measurement outcomes etc.\ in the past. With this justification we say that $\bc_t$ is the \emph{state space} of the system at time $t$ and $\bcg_t$ is the generated vector space of \emph{generalized states}. In the following we adopt the customs of the physics literature and freely use expressions such as ``the system is at time $t_1$ in state $b_1$''. However, we emphasize that we remain agnostic as to any ontological interpretation such language might suggest.

\begin{figure}
\centering
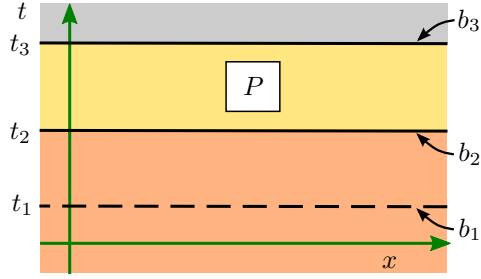
\caption{Evolving state $b_1$ from $t_1$ to $t_2$ and then doing measurement $P$ before $t_3$ should give the same result as doing the measurement $P$ on $b_2$.}
\label{fig:tintpf2}
\end{figure}

Consider the following situation. At an initial time $t_1$ a system is in a state $b_1\in\bc_{t_1}$. There is no intervention until time $t_2$. Then, between time $t_2$ and $t_3$ a measurement takes place, see Figure~\ref{fig:tintpf2}. The quantities used for predicting the measurement outcome are then values of the form
\begin{equation}
 \lv \np\pcomp P, b_1\tens b_3\rv_{[t_1,t_3]},
\label{eq:val1}
\end{equation}
where $P\in\pr_{[t_2,t_3]}$ is a probe associated to the measurement and $b_3$ is a final boundary condition, i.e., a final state. Alternatively, there should be a state $b_2\in\bc_{t_2}$ containing the information on the system encoded in $b_1$ as well as the information about the free evolution between time $t_1$ and $t_2$ such that all posterior measurements yield the same predictions with this state. That is, the value
\begin{equation}
 \lv P, b_2\tens b_3\rv_{[t_2,t_3]}
\end{equation}
must be identical to the value (\ref{eq:val1}) for any $P\in\pr_{[t_2,t_3]}$ and any $b_3\in\bc_{t_3}$. From the composition rule (\ref{eq:bincompid}) this implies,
\begin{equation}
 b_2=\sum_{k\in I} \lv \np,b_1\tens\xi_k\rv_{[t_1,t_2]} \xi_k ,
\end{equation}
where $\{\xi_k\}_{k\in I}$ is an orthonormal basis of $\bcg_{t_2}$. Thus, $b_2$ can be considered the state of the system resulting at time $t_2$ from the free evolution from the initial state $b_1$ at $t_1$.

This prompts us to define the \emph{time evolution map} $\npe_{[t_1,t_2]}:\bcg_{t_1}\to\bcg_{t_2}$ in terms of the null probe through the relation,
\begin{equation}
 \lv \np, b_1\tens b_2\rv_{[t_1,t_2]}=\lb \npe_{[t_1,t_2]}(b_1), b_2 \rb_{t_2} .
\label{eq:npechar}
\end{equation}
This relation also makes it clear that the map $\npe_{[t_1,t_2]}$ is \emph{positive}, i.e., maps states to states (compare Definition~\ref{dfn:posmap}). This follows from positivity of the null probe together with sharp positivity of the inner product (compare Definition~\ref{dfn:spos}). Using an orthonormal basis we can make the definition more explicit,
\begin{equation}
 \npe_{[t_1,t_2]}(b)\defeq\sum_{k\in I} \lv \np,b\tens\xi_k\rv_{[t_1,t_2]} \xi_k .
\label{eq:npedef}
\end{equation}
The composition rule (\ref{eq:bincompid}) for the null probe translates for time evolution maps to a simple composition of maps. Given $t_1\le t_2\le t_3$ we have,
\begin{equation}
\npe_{[t_1,t_3]}=\npe_{[t_2,t_3]}\circ\npe_{[t_1,t_2]} .
\label{eq:npecomp}
\end{equation}
Note also $T_{[t,t]}=\id_t$.
In the same way as the null probe we may express any probe $P\in\pr_{[t_1,t_2]}$ in terms of a corresponding \emph{probe map} $\tilde{P}:\bcg_{t_1}\to\bcg_{t_2}$ via the relation,
\begin{equation}
 \lv P, b_1\tens b_2\rv_{[t_1,t_2]}=\lb \tilde{P}(b_1), b_2 \rb_{t_2} .
\label{eq:pmap}
\end{equation}
Equivalently, we can make a direct definition using an orthonormal basis,
\begin{equation}
 \tilde{P}(b)\defeq\sum_{k\in I} \lv P,b\tens\xi_k\rv_{[t_1,t_2]} \xi_k .
\label{eq:pmapdef}
\end{equation}
As for the null probe, any primitive probe $P$ gives rise to a positive probe map $\tilde{P}$. Note that the converse is not necessarily true. That is, positivity of the probe map does not necessarily imply positivity of the probe. We say that a probe map is \emph{boundary positive} if and only if the corresponding probe is positive. The relation between positivity and boundary positivity for probe maps is intimately related with the hypersurface decomposition rule, Axiom (P2), see Subsection~\ref{sec:paxioms}. Consider for a time interval region $[t_1,t_2]$ the map for the decomposition of the boundary state space $\bcg_{t_1}\tens\bcg_{t_2}\to\bcg_{t_1 \sqcup t_2}$. If any positive element in $\bcg_{t_1 \sqcup t_2}$ arises as a conical combination of tensor products of positive elements in $\bcg_{t_1}$ with positive elements in $\bcg_{t_2}$ then positivity and boundary positivity for probe maps are the same. Otherwise, boundary positivity is strictly stronger.\footnote{It turns out that in classical theory boundary positivity and positivity are the same, compare Subsection~\ref{sec:cevol}. In quantum theory, boundary positivity is precisely \emph{complete positivity} and thus stronger than mere positivity, see Subsections~\ref{sec:qmesop} and \ref{sec:qevol}.}
The composition in time of probes according to the rule (\ref{eq:bincompid}) translates to the composition of the corresponding probe maps, generalizing the case of the null probe of expression (\ref{eq:npecomp}). In the present setting it is convenient to work completely in terms of time evolution maps and probe maps rather than using the original pairing map. The latter can always be recovered via relations (\ref{eq:npechar}) and (\ref{eq:pmap}).

An important class of physical systems are time-translation symmetric. That implies that all the state spaces $\bcg_t$ are isomorphic as partially ordered inner product spaces. It is then often convenient and natural to explicitly identify these spaces with a single copy $\bcg$, which is then considered \emph{the} state space of the system. Time evolution maps are then maps from this state space to itself. Moreover, time-translation symmetry implies then $T_{[t_1,t_1+\Delta]}=T_{[t_2,t_2+\Delta]}$ for any times $t_1,t_2$ and duration $\Delta$. That is, the time evolution map only depends on the difference $\Delta$ between initial and final time. This leads us to define the \emph{time evolution operator} $T_{\Delta}:\bcg\to\bcg$ for any $\Delta\ge 0$ such that $T_{[t,t+\Delta]}=T_{\Delta}$ for any $t$. These time evolution operators satisfy $T_0=\id$ and
\begin{equation}
 T_{\Delta_1+\Delta_2}=T_{\Delta_2}\circ T_{\Delta_1}
\end{equation}
for any $\Delta_1,\Delta_2\ge 0$. That is, the time evolution operators form a \emph{one-parameter semigroup} of boundary positive operators.

In many physical systems time-evolution preserves the state space. That is, the time evolution maps $T_{[t_1,t_2]}:\bc_{t_1}\to\bc_{t_2}$ are not only positive linear maps, but are isomorphisms of partially ordered inner product spaces. If we have this property combined with time-translation symmetry, the time evolution operators form a one-parameter group of isomorphisms. In that case we have the option to use these operators themselves rather than any other means for identifying the state spaces $\bcg_t$ at different times. Doing so would render the time evolution operators trivial by construction, $T_{\Delta}=\id$. In quantum theory this is called working in the \emph{Heisenberg picture}.

\subsection{State evolution and Bayesian updating}
\label{sec:sevolbayes}

We consider now the evolution of a state that is subject to a measurement. Suppose we perform a binary measurement between time $t_1$ and time $t_2$ on a system that we take to be in an initial state $b\in\bc_{t_1}$. The measurement is described by a non-selective probe $P_*\in\prp_{[t_1,t_2]}$ and selective probes $P_r,P_g\in\prp_{[t_1,t_2]}$ with $P_r+P_g=P_*$ that select the two possible outcomes. To test the final state of this measurement we perform a second, similar one between time $t_2$ and time $t_3$ described by a non-selective probe $Q_*\in\prp_{[t_2,t_3]}$ and selective probes $Q_r,Q_g\in\prp_{[t_2,t_3]}$ with $Q_r+Q_g=Q_*$. Say, we pair with a final state $c\in\bc_{t_3}$ at time $t_3$. We wish to predict the probability for the second experiment to yield the outcome corresponding to probe $Q_r$. If we do not know (or do not wish to take into account) anything about the outcome of the first experiment, this probability is given according to our rules by
\begin{equation}
 \Pi(Q_r)=\frac{\lb c,\tilde{Q}_r\tilde{P}_* b\rb_{t_3}}{\lb c,\tilde{Q}_*\tilde{P}_* b\rb_{t_3}} =\frac{\lb c,\tilde{Q}_r b_*\rb_{t_3}}{\lb c,\tilde{Q}_* b_*\rb_{t_3}} .
\end{equation}
Here, $b_*\defeq \tilde{P}_* b\in\bc_{t_2}$. That is, the system behaves exactly as if it is in the state $b_*$ at time $t_2$.

Suppose now, however, that we know the outcome of the first experiment and that this corresponds to probe $P_r$. Then, we can improve our prediction for the second experiment and the probability for outcome $Q_r$ given that knowledge is,
\begin{equation}
 \Pi(Q_r|P_r)=\frac{\lb c,\tilde{Q}_r\tilde{P}_r b\rb_{t_3}}{\lb c,\tilde{Q}_*\tilde{P}_r b\rb_{t_3}} =\frac{\lb c,\tilde{Q}_r b_r\rb_{t_3}}{\lb c,\tilde{Q}_* b_r\rb_{t_3}} .
\end{equation}
Here, $b_r\defeq \tilde{P}_r b\in\bc_{t_2}$. That is, the system behaves exactly as if it is in the state $b_r$ at time $t_2$.

This clearly illustrates that a state in the present framework is not an objective property of the system, but rather represents knowledge about the system. In particular, we may assign different states depending on how much or which knowledge we take into account. Different assignments are not necessarily more or less correct, but they may lead to better or worse predictions of future measurement outcomes. The replacement of state $b$ (or $b_*$) by either $b_r$ or $b_g$, depending on the outcome of the first measurement is in the context of quantum theory sometimes called a \emph{collapse} and in certain interpretations thought of as a spontaneous physical process. However, in the present framework it simply reflects a \emph{Bayesian updating} of our knowledge about the system. Indeed the replacement of the \emph{prior} probability $\Pi(Q_r)$ by the \emph{posterior} probability $\Pi(Q_r|P_r)$ satisfies Bayes' rule,
\begin{gather}
\Pi(Q_r|P_r)=\Pi(Q_r)\frac{\Pi(P_r|Q_r)}{\Pi(P_r)},\quad\text{where}\\
 \Pi(P_r|Q_r)=\frac{\lb c,\tilde{Q}_r\tilde{P}_r b\rb_{t_3}}{\lb c,\tilde{Q}_r\tilde{P}_* b\rb_{t_3}}\quad\text{and}\quad
 \Pi(P_r)=\frac{\lb c,\tilde{Q}_*\tilde{P}_r b\rb_{t_3}}{\lb c,\tilde{Q}_*\tilde{P}_* b\rb_{t_3}} .
\end{gather}

\subsection{The state of maximal uncertainty}
\label{sec:pou}

We introduce at this point a new object into the formalism that traditionally forms an important part of essentially any description of measurement in dynamical systems, either explicitly or implicitly. Recall from Subsection~\ref{sec:hierarchies} that boundary conditions form hierarchies of generality and that this is the principal physical significance of the partial order structure on the spaces of boundary conditions. It is thus conceivable that there exists a ``maximally general'' boundary condition. Let us denote it by $\ou\in\bc_{\Sigma}$. Physically, this boundary condition would encode the fact that we have no knowledge at all about the physics on one side of the hypersurface and its impact on the other side. Mathematically, this would imply that any other boundary condition $b\in\bc_{\Sigma}$ satisfies the inequality $b\le\lambda\ou$ for some $\lambda>0$. Recall that we have no overall notion of normalization in $\bcg_{\Sigma}$, prompting the need for the relative normalization factor $\lambda$.
An element in a partially ordered vector space with this property is called an \emph{order unit}, compare Definition~\ref{dfn:ou} of the appendix. Mathematically, an order unit is not unique as for example any other positive element can be added to an order unit to yield another order unit. Of course we are contemplating here the situation where the physical meaning singles out one specific order unit, up to normalization.
In the following we shall specifically consider such a special boundary condition in the context of states, i.e., for equal-time hypersurfaces. We call it then the \emph{state of maximal uncertainty}.

Consider a single measurement that takes place in a time interval $[t_1,t_2]$ as illustrated in Figure~\ref{fig:tintpf}. Usually we are interested in a question of the following type: Given the details of the experimental setup or measurement apparatus and given an initial state $b_1$, what is the probability of a certain measurement outcome? Even though we do not specify any final state, this question makes perfect sense. Indeed, it implies that we do not know or do not wish to assume anything about the system after the measurement has concluded at time $t_2$. But this is precisely the meaning of the state of maximal uncertainty as a final state. That is, we have implicitly chosen $b_2=\ou$ as the final state.
More concretely, suppose the measurement apparatus is described by the primitive probe $Q\in\prp_{[t_1,t_2]}$ while the apparatus with selected outcome is described by the primitive probe $P\in\prp_{[t_1,t_2]}$. Then, specifying only the initial state $b\in\bc_{t_1}$, the probability $\Pi$ for a positive outcome is,
\begin{equation}
 \Pi=\frac{\lv P, b\tens\ou\rv_{[t_1,t_2]}}{\lv Q, b\tens\ou\rv_{[t_1,t_2]}}
 = \frac{\lb \tilde{P}(b),\ou\rb_{t_2}}{\lb \tilde{Q}(b),\ou\rb_{t_2}} .
\end{equation}

A measurement where explicitly a final state $b_2$ is chosen that is distinct from the state of maximal uncertainty $\ou$ is called a measurement with \emph{post-selection}.
It is also legitimate to consider the state of maximal uncertainty as an initial state. This corresponds to a context in which we do not know anything about the system before the measurement. However, due to the notion of causality that we use when reasoning about the world the precise meaning of ``not knowing anything about the system before the measurement'' is much less clear than the corresponding statement concerning the system after the measurement. We come back to the question of the meaning of the state of maximal uncertainty in the more concrete contexts of classical or quantum theory, see Subsections~\ref{sec:cevol} and \ref{sec:qevol} respectively.

If time evolution preserves the state space, then time evolution maps should not only be isomorphisms of partially ordered inner product spaces, but also preserve the state of maximal uncertainty. We assume this from here onward and also assume the identification of (generalized) state spaces at all times, denoted $\bcg$, without any label.

\subsection{Causality and normalization}
\label{sec:causnorm}

An order unit $\ou$ in a partially ordered inner product space has an interesting property with respect to the inner product. Consider a strictly positive element $b$. Then, $b\le\lambda\ou$ for some $\lambda>0$. So $\lambda \ou - b\ge 0$. By positivity of the inner product we thus have $\lb \lambda \ou - b, b\rb\ge 0$ and $\lb b,b\rb> 0$ by positive-definiteness. This implies $\lb \lambda\ou,b\rb>0$ and also $\lb \ou,b\rb>0$. That is, the inner product of the order unit with any strictly positive element is strictly positive.

This suggests a manner to normalize states. Namely, we say that a state $b\in\bc$ is \emph{normalized} if its inner product with the state of maximal uncertainty is $1$, i.e., $\lb \ou,b \rb=1$. By multiplying with some strictly positive number any non-zero state can be normalized. It then appears natural to say that a probe map $\tilde{Q}$ corresponding to a probe $Q\in\pr_{[t_1,t_2]}$ is \emph{normalization preserving} if it maps normalized states to normalized states, i.e., if it preserves the inner product with the state of maximal uncertainty. That is, for any $b\in\bcg$, (by linearity we need not restrict to $b\in\bc$)
\begin{equation}
 \lb \ou, \tilde{Q}(b)\rb = \lb \ou,b\rb .
\label{eq:pmnorm}
\end{equation}

\begin{figure}
\centering
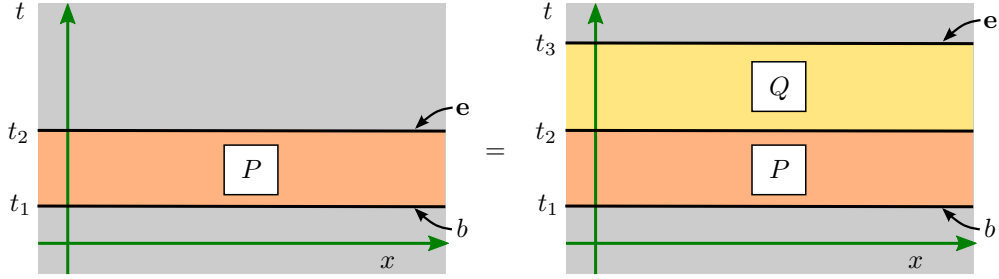
\caption{Causal probe: The outcome of measurement $P$ does not depend on whether or not a subsequent measurement $Q$ is performed as long as we do not select for the outcome of $Q$. This is ensured by the non-selective probe $Q$ satisfying \emph{forward causality}.}
\label{fig:tintcausal}
\end{figure}

The condition to be normalization preserving has a remarkable physical interpretation in terms of causality. (This is of course well known in the context of quantum theory, see Subsection~\ref{sec:qevol}.) By \emph{causality} we mean here the statement that the outcome of an experiment is independent of the choice (but not outcome!) of any future experiment we may perform on the system later.\footnote{We do not consider here any notion of \emph{relativistic} causality.} We illustrate this with the following setting, depicted in Figure~\ref{fig:tintcausal}. We take an initial state $b\in\bc$ at time $t_1$. We perform an experiment between time $t_1$ and time $t_2$. The corresponding apparatus is represented by a non-selective probe $P_*\in\prp_{[t_1,t_2]}$ and a selective probe $P_r\in\prp_{[t_1,t_2]}$ where we impose a desired outcome. We proceed in two alternative ways: Either the experiment is finished, i.e., we evaluate without post-selection by pairing with the state of maximal uncertainty. Or, we perform a second experiment on the system, between time $t_2$ and $t_3$, represented by a non-selective probe $Q\in\prp_{[t_2,t_3]}$. In this case we are not interested in the outcome and thus do not need to specify any selective probe. At time $t_3$ the experiment is finished and we pair with the state of maximal uncertainty. Now, causality would dictate that the outcome of the first experiment should not depend on the mere performance or not of the second experiment. That is, the probability for the positive outcome of the first experiment should be the same in both cases,
\begin{equation}
\frac{\lb \ou,\tilde{P}_r(b)\rb}{\lb \ou,\tilde{P}_*(b)\rb}=\frac{\lb \ou,\tilde{Q}(\tilde{P}_r(b))\rb}{\lb \ou,\tilde{Q}(\tilde{P}_*(b))\rb} .
\label{eq:probcausaleq}
\end{equation}
Comparing with expression (\ref{eq:pmnorm}) we see that this is satisfied in particular if the probe map $\tilde{Q}$ is normalization preserving. This motivates us to say that a non-selective probe is \emph{forward causal} if the corresponding probe map is normalization preserving. It is easy to see that if the non-selective probes for all involved experiments are forward causal, causality holds for any chain of experiments. Note that selective probes cannot then be required to be normalization preserving. However, any selective probe $P_r$ is less general than the corresponding non-selective probe $P_*$, i.e.\ $P_r\le P_*$. This implies for any $b\in\bc$,
\begin{equation}
 \lb \ou,\tilde{P}_r(b) \rb \le \lb\ou,\tilde{P}_*(b)\rb = \lb\ou, b\rb .
\end{equation}
That is, non-selective probe maps are then \emph{normalization decreasing}.\footnote{Note that the word ``decreasing'' is understood here to include the possibility of preserving the normalization.} We call a model \emph{(forward) causal} if all the selective and non-selective probe maps for experiments have these normalization properties. It is then convenient to choose initial states to be normalized. As is easily seen in expression (\ref{eq:probcausaleq}) this makes the denominators in probability expressions equal to $1$, simplifying them considerably. Of course, this simplification applies only as long as we do not condition on intermediate measurement outcomes and refrain from performing post-selection.

The property of being normalization preserving is time-asymmetric. This becomes immediately visible by rewriting expression (\ref{eq:pmnorm}) in the standard notation of the positive formalism,
\begin{equation}
 \lv Q, b\tens\ou\rv_{[t_1,t_2]} = \lv \np, b\tens\ou\rv_{[t_1,t_2]} .
\label{eq:pmnormst}
\end{equation}
The time-reversed version of this property is,
\begin{equation}
 \lv Q, \ou\tens b\rv_{[t_1,t_2]} = \lv \np, \ou\tens b\rv_{[t_1,t_2]} .
\label{eq:pmnormstrev}
\end{equation}
We call a non-selective probe satisfying this condition \emph{backward causal}. Note that the two conditions are distinct, but do not exclude each other. Physically, ``forward (or backward) causal'' merely implies compatibility with a forward (or backward) causal interpretation. It does not mean that any other interpretation is excluded. There could indeed be a large class of non-selective probes that have both properties. As is evident from expressions (\ref{eq:pmnormst}) and (\ref{eq:pmnormstrev}) the null probe and thus all time-evolution maps have both properties. As we shall discuss later, also all measurements in quantum theory induced by observables have both properties, see Subsection~\ref{sec:qevol}.

Note that in the general spacetime version of the positive formalism as presented in Section~\ref{sec:first} the notions of \emph{selectiveness} and \emph{non-selectiveness} for probes were only introduced to help describe their role with respect to a specific measurement process. There is no implication that these notions correspond to any intrinsic property of a probe, in contrast to the notion of primitiveness for example. This changes in the present temporal setting with the implementation of forward causality. Here, the notions of selectiveness and non-selectiveness are elevated to intrinsic properties of probes.

\subsection{Composite systems}
\label{sec:evolcompsys}

The simplification of the positive formalism we have considered in this section so far is rather drastic. We have reduced all notions of spacetime to merely a linear notion of time. The spacetime system was thus effectively reduced to the real line with closed intervals as the only connected regions and points as the only connected hypersurfaces. Instead, we may consider an intermediate situation where we take into account a non-trivial structure for space, but keep a fixed directed notion of time for any probe. A convenient way to do this is to impose on any hypersurface a \emph{temporal orientation}, i.e., declare one side to be past facing and the other side future facing. Physically, this would mean that any hypersurface must be a spacelike hypersurface. As this seems rather restrictive we may alternatively declare that the space of generalized boundary conditions for any timelike hypersurface is trivial, i.e., isomorphic to $\R$. This has the effect that no transmission of signals through these hypersurfaces is considered and a temporal orientation for them does not matter. In general, such a restriction would be unphysical, in particular if considering fundamental theories. However, we are limiting ourselves here to models compatible with this restriction. That is, we consider models where any signal flow between probes has a strict temporal direction.
We take the spacelike nature of relevant hypersurfaces as a justification to call the associated (generalized) boundary conditions \emph{(generalized) states}, extending the nomenclature introduced in Subsection~\ref{sec:statesevol}.

\begin{figure}
  \centering
  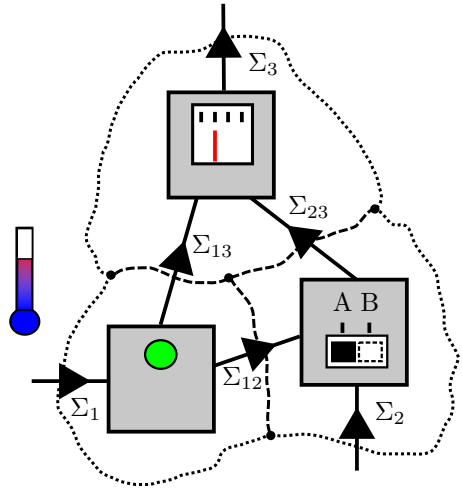
  \caption{Simplified graphical representation of experimental setup via dual 1-complex, with choices of orientations on links.}
  \label{fig:pf_dual_example_oriented}
\end{figure}

\begin{figure}
  \centering
  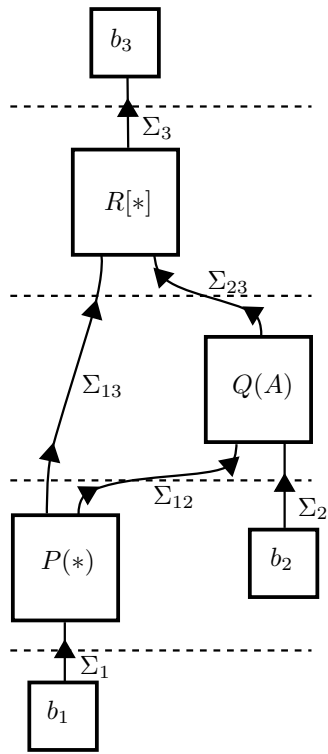
  \caption{Directed circuit diagram for the example setup with factorizing boundary conditions. The orientations correspond to those depicted in Figure~\ref{fig:pf_dual_example_oriented}. The dashed lines indicate a suitable slicing.}
  \label{fig:pf_diag_example_or}
\end{figure}

To illustrate the setting we return to the example of Subsection~\ref{sec:pf_example}. Recall the setup depicted in Figure~\ref{fig:pf_example} and its simplified graphical representation via the dual 1-complex, depicted in Figure~\ref{fig:pf_dual_example}. As a first step we choose temporal orientations for hypersurfaces. In the dual 1-complex these can be represented as directional arrows on links. We fix such a choice as depicted in Figure~\ref{fig:pf_dual_example_oriented}. We assume for simplicity that the boundary condition factorizes as $b=b_1\tens b_2\tens b_3$ and view the component boundary conditions $b_1,b_2,b_3$ as probes as explained in Subsection~\ref{sec:pfcd}. Now, instead of extracting the corresponding undirected circuit diagram (Figure~\ref{fig:pf_diag_example_uc}.b) we extract a \emph{directed circuit diagram} using the chosen orientations. This is depicted in Figure~\ref{fig:pf_diag_example_or}. There we have also taken care to move the ends of links to the top or bottom of boxes, depending on the temporal orientation of the link.

The evaluation of the directed circuit diagram of Figure~\ref{fig:pf_diag_example_or} follows the rules of the positive formalism and the diagrammatic evaluation rules laid out in Subsection~\ref{sec:catcd}. We proceed to make this explicit. To this end we extend the notion of probe map from Subsection~\ref{sec:statesevol} to the present setting. That is, a probe map codifies a probe as a linear map from the tensor product of state spaces associated to the initial hypersurfaces to the tensor product of state spaces associated to the final hypersurfaces. The difference to Subsection~\ref{sec:statesevol} lies merely in the fact that that we allow for more than one tensor factor on the initial and the final side. For example for the probe $P(*)$ the probe map $\tilde{P}(*):\bcg_1\to\bcg_{1 3}\tens\bcg_{1 2}$ is given by,
\begin{equation}
 \tilde{P}(*)(c)\defeq\sum_{k,l} \lv P,c\tens\xi_{1 3}^k\tens\xi_{1 2}^l\rv_{M_1} \xi_{1 3}^k\tens\xi_{1 2}^l .
\end{equation}
The probe maps $\tilde{Q}(A)$ and $\tilde{R}[*]$ are defined similarly. The initial boundary conditions $b_1\in\bcg_1$ and $b_2\in\bcg_2$ correspond to linear maps from $\R$ to the respective state space. However, these may be represented as elements of the respective state space, i.e., as themselves (recall Figure~\ref{fig:diagunit}.b in Subsection~\ref{sec:catcd}). In contrast, the final boundary condition $b_3\in\bcg_3$ gives rise to the probe map $\tilde{b}_3:\bcg_3\to\R$ given by,
\begin{equation}
 \tilde{b}_3(c)\defeq \lb b_3,c\rb_3 .
\end{equation}
According to Subsection~\ref{sec:catcd} the evaluation of the directed circuit diagram of Figure~\ref{fig:pf_diag_example_or} may now proceed as follows. The diagram is sliced horizontally, so that each slice consists of a horizontal juxtaposition of boxes and lines. Each slice corresponds to a map. This map is obtained as the tensor product of all the maps corresponding to boxes and free lines (i.e., lines that within the slice are not connected to any box). For each box the corresponding map is the probe map, for each free line it is the identity. The maps for all the slices are composed in order. Since the diagram is closed (i.e., there are no lines with free ends) this yields a real number, the value of the diagram. In the present case, a convenient slicing is indicated by dashed lines in Figure~\ref{fig:pf_diag_example_or}. The corresponding evaluation is,
\begin{equation}
\tilde{b}_3\circ \tilde{R}[*]\circ \left(\id_{1 3}\tens \tilde{Q}(A)\right)
\circ\left(\tilde{P}(*)\tens b_2\right)(b_1) .
\label{eq:exadiagoreval}
\end{equation}
As it should, this value coincides with that given by expression (\ref{eq:example_pcomp}) for the case of a factorizing boundary condition.

\begin{figure}
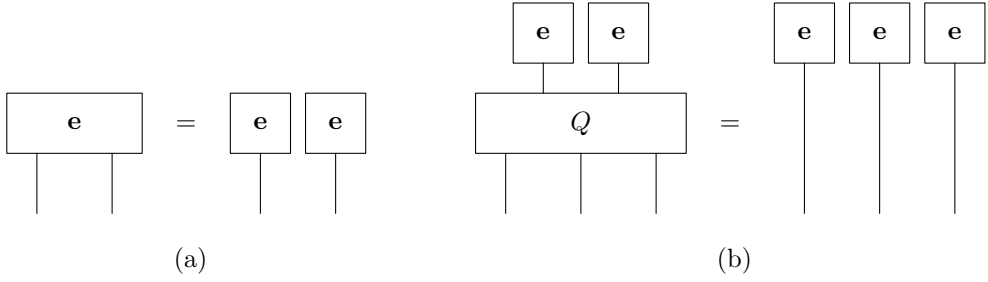

  \centering
  \begin{tabular}{cp{5mm}c}
  \input{figures/ou_tp_ident} && \input{figures/np_tp_ident} \\
  \\
  (a) && (b)
  \end{tabular}
  \caption{(a) Tensor product identity for the state of maximal uncertainty $\ou$. (b) The property to be normalization preserving for the probe $Q$.}
  \label{fig:diagouprop}
\end{figure}

So far we have not used the fact that we want to interpret the orientation of the hypersurfaces (or links) as temporal. In fact, we could have taken the original setup (Figures~\ref{fig:pf_example} and \ref{fig:pf_dual_example}) given an arbitrary orientation to each hypersurface/link, generated the corresponding directed circuit diagram and the corresponding probe maps and evaluated the diagram with the same result. Indeed, it is not a particular manner of evaluating diagrams that interests us here, but additional structure that we may introduce thanks to the notion of order provided by time. In particular, this order allows us to impose forward (or backward) causality, extending the notion introduced in Subsection~\ref{sec:causnorm}. To do this, we first need to further elaborate the notion of state of maximal uncertainty. We assume that the generalized state space $\bcg_{\Sigma}$ for any hypersurface $\Sigma$ contains such a state. We denote it either by $\ou$ when no confusion can arise or by $\ou_{\Sigma}$ to indicate that it belongs to the generalized state space of hypersurface $\Sigma$. Further, we need to assume that the states of maximal uncertainty are compatible with hypersurface decomposition. Say we have a hypersurface $\Sigma$ decomposing as a union $\Sigma=\Sigma_1\cup\Sigma_2$. We then require that the corresponding states of maximal uncertainty are related by the tensor product, i.e.,
\begin{equation}
\ou_{\Sigma}=\ou_{\Sigma_1}\tens\ou_{\Sigma_2} .
\label{eq:tpou}
\end{equation}
The diagrammatic representation of this identity is shown in Figure~\ref{fig:diagouprop}.a.

We may now impose forward causality by extending the corresponding concept of Subsection~\ref{sec:causnorm}. Consider a probe $Q$ with associated probe map $\tilde{Q}:\bcg_{\text{in},1}\tens\cdots\tens\bcg_{\text{in},m}\to\bcg_{\text{out},1}\tens\cdots\tens\bcg_{\text{out},n}$. We say that $\tilde{Q}$ is \emph{normalization preserving} if for any $b_1\in\bcg_{\text{in},1}$, \dots, $b_m\in\bcg_{\text{in},m}$ we have,
\begin{equation}
\lb \ou\tens\cdots\tens\ou, \tilde{Q}(b_1\tens\cdots\tens b_m)\rb=\lb \ou\tens\cdots\tens\ou, b_1\tens\cdots\tens b_m\rb .
\end{equation}
The diagrammatic representation of this identity (for $m=3$, $n=2$) is shown in Figure~\ref{fig:diagouprop}.b.
Similarly, we say that $\tilde{Q}$ is \emph{normalization decreasing} if,
\begin{equation}
\lb \ou\tens\cdots\tens\ou, \tilde{Q}(b_1\tens\cdots\tens b_m)\rb\le\lb \ou\tens\cdots\tens\ou, b_1\tens\cdots\tens b_m\rb .
\end{equation}
Also recall that the inner product is by construction compatible with hypersurface (and therefore tensor product) decomposition. In particular we have,
\begin{equation}
\lb \ou\tens\cdots\tens\ou, b_1\tens\cdots\tens b_m\rb = \lb \ou, b_1\rb \cdots \lb \ou, b_m\rb .
\end{equation}
It remains to mention that the tensor product of normalization preserving (decreasing) maps is normalization preserving (decreasing). Also recall that the time evolution maps (represented by free lines in the diagrams) are normalization preserving. Thus, it makes sense to say as before that a model satisfies \emph{forward causality} if the probe maps of its non-selective probes are normalization preserving and those of the selective probes are normalization decreasing.

At this point we recall considerations presented in Subsection~\ref{sec:pfcd} as to the option of doing away with the spacetime concepts of regions and hypersurfaces, replacing these with more abstract notions of \emph{processes} (instead of regions) and their communication interfaces (instead of hypersurfaces). In the present setting we have an analogous option. Generalizing the usage of the term \emph{system} in this section we may call these communication interfaces systems. The hypersurface compositions then become abstract compositions of systems, encoded in terms of the tensor product. Thus, we may talk about processes, systems and their compositions. We will freely employ this point of view and language in the following. As in Subsection~\ref{sec:pfcd}, however, we must warn the reader that there are important classes of models that can be described in the spacetime context, but not in the abstract one, due to the loss of structure involved in the transition, see also Subsection~\ref{sec:absbdy}.

There is also a crucial difference between the abstraction from spacetime concepts discussed in Subsection~\ref{sec:pfcd} and the present one. While the abstraction as considered there was complete, here it is limited by the fact that we retain a partial order on the set of processes. The links with their arrows establish a relation between processes and consistency with causality demands precisely that this relation is a partial order. In the present context this order arises from a notion of time although one can abstract from this, retaining only the order itself. It must be emphasized that even in this abstracted form the present setting is severely limited in its physical applicability as it only permits uni-directional inter-process communication. If we wanted to get rid of this limitation we would have to return to the more general setting of Section~\ref{sec:first}. However, we would have to give up those structures that depend explicitly on the process order. This is in particular the implementation of forward causality via normalization preserving and decreasing maps.

\subsection{Causality in the example}

To illustrate the implementation of forward causality we return to the example experimental setup considered in Subsection~\ref{sec:pf_example}. We recall that it consists of three apparatuses: one with a light showing either \lbl{RED} or \lbl{GREEN}, one with a switch with positions A and B, and one with a pointer device, see Figure~\ref{fig:pf_example}. We restrict our attention to one particular experimental procedure with this setup, listed in Subsection~\ref{sec:pf_example} as number 7. In this procedure an agent sets the switch on the second apparatus depending on the light state of the first apparatus. More precisely, the switch is put to position A if the light shows \lbl{GREEN} and to position B if the light shows \lbl{RED}. For this to make sense the light state must not itself depend on the position of the switch. Recall that this independence condition is equivalent to the equality (\ref{eq:lindeps}). In a causal world the independence should be a simple consequence if the reading of the light takes place \emph{before} the setting of the switch. This temporal order is realized in the choice of temporal orientations for this experiment depicted in Figure~\ref{fig:pf_dual_example_oriented}. (It is the direction of the arrow on the link labeled $\Sigma_{1 2}$.)

Our implementation of forward causality indeed ensures the validity of equality (\ref{eq:lindeps}) for this temporal ordering with the additional assumption that no post-selection is performed. The latter means that the final state has to be taken to be the state of maximal uncertainty $b_3=\ou$. We can then rewrite the equality (\ref{eq:lindeps}) by using the corresponding probe maps as in expression (\ref{eq:exadiagoreval}),
\begin{equation}
\frac{\tilde{\ou}\circ \tilde{R}[*]\circ \left(\id_{1 3}\tens \tilde{Q}(A)\right)
\circ\left(\tilde{P}(g)\tens b_2\right)(b_1)}{\tilde{\ou}\circ \tilde{R}[*]\circ \left(\id_{1 3}\tens \tilde{Q}(A)\right)
\circ\left(\tilde{P}(*)\tens b_2\right)(b_1)} 
=\frac{\tilde{\ou}\circ \tilde{R}[*]\circ \left(\id_{1 3}\tens \tilde{Q}(B)\right)
\circ\left(\tilde{P}(g)\tens b_2\right)(b_1)}{\tilde{\ou}\circ \tilde{R}[*]\circ \left(\id_{1 3}\tens \tilde{Q}(B)\right)
\circ\left(\tilde{P}(*)\tens b_2\right)(b_1)} .
\label{eq:lindepsor}
\end{equation}
The probe maps $\tilde{P}(*)$, $\tilde{Q}(A)$, $\tilde{Q}(B)$, and $\tilde{R}[*]$ are normalization preserving since the corresponding probes are non-selective. All relevant calculations can easily be performed in terms of diagrams, specifically taking advantage of the identity depicted in Figure~\ref{fig:diagouprop}. The denominators on both sides of equation (\ref{eq:lindepsor}) are thus seen to equal $\tilde{\ou}(b_1)\cdot\tilde{\ou}(b_2)$. If the initial states are normalized this is just equal to $1$. For the numerators we get for both sides,
\begin{equation}
\left(\tilde{\ou}\tens\tilde{\ou}\right)\circ \tilde{P}(g)(b_1) \cdot \tilde{\ou}(b_2).
\end{equation}
In particular, the equality holds as claimed.

\subsection{Categorical and functorial point of view}

Let us focus on the monoidal category (in the sense of Section~\ref{sec:catdiag}) underlying the time-evolution version of the positive formalism. Its objects are partially ordered inner product spaces with order unit. These are the generalized state spaces associated to systems. Morphisms are probe maps that represent processes. The tensor product of objects represents the composition of systems.

As already discussed, the diagrammatics most adapted to this present setting is that of \emph{directed circuit diagrams}, where the temporal order on links between processes is encoded in the arrows on the links. A peculiar consequence of this is that the inner product on the generalized state spaces is actually not needed when evaluating the diagrams (compare Subsection~\ref{sec:catcd}). It comes in when we need to convert a state to a functional on state space. As we have seen, this is particularly important for the state of maximal uncertainty, i.e., the order unit $\ou$.

On the other hand, starting from the outset in a time-evolution setting, it would have seemed quite natural not to require that the state space should carry an inner product which identifies it with its dual. Rather, one might have considered the state space and its dual to be different and separate entities. In particular, rather than a state $\ou$ of maximal uncertainty one would have considered the corresponding functional $\tilde{\ou}$ as an entity of central importance. Indeed, this perspective has prevailed in much previous related work, see Subsection~\ref{sec:pcomp}.

The categorical setting also allows for an appealing conceptual separation of physical and mathematical objects. To this end we consider physical systems themselves as forming a monoidal category. Morphisms are physical processes and the tensor product is the composition of systems. The present formalism is then a \emph{functor} from this ``physical'' category to the category of partially ordered vector spaces, positive maps, etc. While this construction might appear overly artificial or abstract it is in essence in this way how category theory initially came to meet quantum (field) theory, see Subsection~\ref{sec:tqft}.

\subsection{Relation to other frameworks}
\label{sec:pcomp}

Up to this point we have striven to present the positive formalism in a self-contained fashion. This was done in order to emphasize the inner logic of its development and to minimize possible distractions and confusions of the reader by imperfect comparisons to notions in the literature. As emphasized in the introduction, the positive formalism is very much related to and inspired by other approaches and builds on much previous work. In the present subsection we focus on a few of these connections. We exclude here those related approaches exclusive to quantum theory which instead are treated in Subsection~\ref{sec:qcomp}.

The positive formalism forms part of a history of operational frameworks that aim to describe fundamental physical theories without being specific to classical or to quantum theory. A pioneering proposal for an operational axiomatic framework was advanced by Mackey in his book on the foundations of quantum theory \cite{Mac:matfoundquant}. This was intended in particular to pin down the difference between classical and quantum theory in a precise way. It lacked, however, an adequate account of the transformation of states due to measurement as well as an account of composition.
Subsequently, more powerful approaches remedied this and developed into what is known as the \emph{convex operational framework}. For a few selected references, see \cite{Lud:axiomaticqm1,DaLe:opapquantprob,FouRan:emplogictp,Hol:probstatqt,Har:fiveaxioms,Bar:infogpt,BaWi:olincatframe,Har:opstructgpt}. More recently this has also been known as the framework for \emph{general(ized) probabilistic theories}. This centers on a notion of state space as the positive cone in a partially ordered vector space and its transformations. It is closely modeled on the standard formulation of quantum theory and in particular embedded into a time-evolution setting. The comparison with the positive formalism is thus most appropriate in its time-evolution derivative form of the present section.

In the face of variations of the convex operational framework existing in the literature we pick a recent review paper of Barnum and Wilce \cite{BaWi:olincatframe} as reference for definiteness. To ease comparison we adopt their notation here. Thus, associated to a system is a partially ordered vector space $A$ with a generating cone $A^+$. There is a positive linear form $u_A:A\to\R$ which is strictly positive on $A^+\setminus\{0\}$.
A \emph{normalized state} is an element in $\alpha\in A^+$ such that $u_A(\alpha)=1$. The set of normalized states is denoted $\Omega_A$. The form $u_A$ induces a \emph{norm} on $A$ determined by $\|\alpha\|=u(\alpha)$ for $\alpha\in A^+$. The dual vector space $A^*$ is partially ordered with the positive cone being the positive linear forms on $A$. The set of elements $a\in A^*$ such that $0\le a\le u_A$ is called the set of \emph{effects}. These are taken to represent physical \emph{events}. A physical \emph{process} with initial state space $A$ and final state space $B$ is represented by a positive linear map $\tau:A\to B$ which moreover satisfies the normalization condition $u_B(\tau(\alpha))\le u_A(\alpha)$ for all $\alpha\in A^+$. The state space of a composite of systems with state spaces $A$ and $B$ is modeled on the tensor product $A\tens B$. The positive cone in $A\tens B$ is left theory dependent, but two extreme cases are considered in particular. In one case, called the \emph{minimal tensor product} the positive elements, denoted $(A\tens_{\text{min}} B)^+$, are precisely the positive linear combinations of tensor products $\alpha\tens\beta$, where $\alpha$ and $\beta$ are positive. In the other case, called the \emph{maximal tensor product} the positive elements, denoted $(A\tens_{\text{max}} B)^+$, are all elements that when paired with $a\tens b$ for any effects $a$ on $A$ and $b$ on $B$ are positive. As is easy to see we have $(A\tens_{\text{min}} B)^+\subseteq (A\tens_{\text{max}} B)^+$.
The simplest rules for predicting probabilities in this setting are the following. For $a\in \Omega_A$ and $a$ an effect on $A$, $a(\alpha)$ is the probability for the event represented by the effect $a$ to take place. Given $a\in\Omega_A$ and $\tau:A\to B$ a process, $u_B(\tau(\alpha))$ is the probability for the process given the state $\alpha$. Let us call this scheme the \emph{generic convex operational framework}.

We compare this to the version of the positive formalism of the present section with forward causality, keeping the same notation. Thus, $A$ is the partially ordered vector space of generalized states of a system. In addition, we have an inner product on $A$. With the \emph{state of maximal uncertainty} $\ou\in A$ this gives rise to the form $u_A:A\to\R$ with the desired properties via $u_A(\alpha)\defeq \lb\ou,\alpha\rb_A$. A state $\alpha\in A^+$ is normalized when $\lb\ou,\alpha\rb_A=1$ which is thus the same as $u_A(\alpha)=1$. Effects $a:A\to\R$ can be identified with states $a'\in A^+$ via the inner product, $a(\alpha)=\lb a',\alpha\rb_A$. The condition $0\le a\le u_A$ for an effect $a$ is thus the same as $0\le a'\le\ou$ for the corresponding state $a'$. A \emph{selective probe} with initial state space $A$ and final state space $B$ corresponds to a probe map $\tau:A\to B$ which is positive and by forward causality is normalization decreasing $\lb \ou,\tau(\alpha)\rb_B\le \lb\ou,\alpha\rb_A$. The latter condition can be written equivalently as $u_B(\tau(\alpha))\le u_A(\alpha)$. Thus, $\tau$ determines a process in the generic framework. As for the composition of systems, the generalized state space of the composite system is the tensor product $A\tens B$ of those of the subsystems $A$, $B$ as vector spaces. The cone of positive elements in $A\tens B$ is not uniquely determined by the axioms (Subsection~\ref{sec:paxioms}). It turns out, however, that Axiom (P2) precisely implies $(A\tens B)^+\supseteq (A\tens_{\text{min}} B)^+$ and by dualization also $(A\tens B)^+\subseteq (A\tens_{\text{max}} B)^+$. That is, the rules for the composition of systems are precisely those of the generic framework, with the same freedom.

Given a state $a\in\Omega_A$ and en effect $a:A\to\R$ the formula for the associated probability in the generic setting is easily seen to be a special case of formula (\ref{eq:elembdy}) of Subsection~\ref{sec:hierarchies},
\begin{equation}
 a(\alpha)=\lb a',\alpha \rb_A=\frac{\lb a',\alpha \rb_A}{\lb \ou,\alpha \rb_A}=\frac{\lv\np,\alpha\tens a' \rv}{\lv\np,\alpha\tens\ou\rv} .
\end{equation}
Given a state $a\in\Omega_A$ and a selective probe map $\tau:A\to B$ the probability formula in the generic framework for the associated process is a special case of formula (\ref{eq:elemprob}) of Subsection~\ref{sec:valpred},
\begin{equation}
 u_A(\tau(\alpha))=\lb \ou,\tau(\alpha) \rb_B=\frac{\lb \ou,\tau(\alpha) \rb_B}{\lb \ou,\tau_0(\alpha) \rb_B}=\frac{\lv\tau',\alpha\tens\ou \rv}{\lv\tau_0',\alpha\tens\ou\rv} .
\end{equation}
Here $\tau_0:A\to B$ is the non-selective probe map corresponding to the selective probe map $\tau$. However, as forward causality is implemented, $\tau_0$ is normalization preserving so that the denominator in the above expression is equal to $1$. That is, it does not actually matter here what $\tau_0$ is and we do not need to know it. $\tau'$and $\tau_0'$ are the probes corresponding to the probe maps $\tau$ and $\tau_0$.

We see that the time-evolution variant of the positive formalism with forward causality fits perfectly into the generic convex operational framework. In two important respects the positive formalism yields a more restrictive setting than the generic framework. The first lies in the fact that states and effect are really the same objects (up to normalization), living in the same space. Mathematically this is implemented through the inner product on the generalized state space. We are thus in a \emph{self-dual} setting (in the terminology of Barnum and Wilce \cite{BaWi:olincatframe}). This is inherited, of course, from the more general spacetime version of the positive formalism. In the latter, a \emph{time} or a \emph{time direction} does not play any fundamental role. Therefore there is no fundamental distinction between initial and final states.

There is a second, more striking and more unusual (from the time-evolution point of view) restriction arising in the positive formalism. This comes from the fact that there is no fundamental difference between the composition of systems at the same time and of a system with itself at different times. Consequently, a probe map $\tau:A\to B$ need not only be positive as a map $A\to B$, but also positive when understood as a map from the total boundary state space $A\tens B$ to the real numbers. We called this property \emph{boundary positive} in Subsection~\ref{sec:statesevol}. In the present notation we can write this condition as
\begin{equation}
 \sum_i \lb\beta_i,\tau(\alpha_i)\rb_B \ge 0\quad\text{for all}\quad \sum_i \alpha_i\tens\beta_i \in (A\tens B)^+ .
\end{equation}
This is equivalent to ordinary positivity only if $(A\tens B)^+=(A\tens_{\text{min}} B)^+$. Otherwise it is stronger. Remarkably, in quantum theory this is equivalent to \emph{complete positivity}, see Subsections~\ref{sec:qmesop} and \ref{sec:qevol}. Thus, the natural notion of (selective) process coming from the positive formalism with forward causality requires boundary positivity and the trace decreasing property.

The categorical formalization of the generic convex operational framework was pioneered by Selinger \cite{Sel:cccandcpm} and brought into a form similar to the one used in the present Section~\ref{sec:evolution} by Barnum, Duncan and Wilce \cite{BaDuWi:catcom}. Independently, Hardy developed a diagrammatic calculus for the generic convex operational framework, the \emph{duotensor} approach \cite{Har:duotensor}. This can also be fit into the categorical setting of Section~\ref{sec:catdiag}.

A pioneering and inspiring proposal for a framework for fundamental physics that takes a spacetime approach and aims not to rely on a predetermined causal structure is Hardy's \emph{Causaloid formalism} \cite{Har:causaloid}. As in the positive formalism a main focus are measurements performed within spacetime regions. The basic mathematical objects are sets specifying these measurements and their outcomes as well as vectors of probabilities for such outcomes. Similarly to the positive formalism, probabilities are in general conditional and take the form of quotients. In contrast to the positive formalism, however, the account of the interaction between the physics in a region and its outside is implicit and not by construction complete. For example, the causaloid product that can be used to compose measurements in disjoint regions depends in general implicitly on the structure of spacetime outside of the regions and on the physics in it. This precludes a strictly local description of physics as well as a universal and explicit rule of composition.

\section{Classical theory}
\label{sec:classical}

In the present section we develop a framework for classical (field) theory which is based on the same notion of spacetime locality that we employed in the foundation of the positive formalism. In contradistinction to the development of the latter in Section~\ref{sec:first} we take for granted here all the ingredients that are specific of classical physics. These ingredients rather than abstract considerations about measurement and observation will guide the development. Operationalism plays a much less prominent role. Nevertheless, at the end of this path we will recover a specific version of the positive formalism.

\subsection{Spacetime}
\label{sec:csts}

We base the framework on the same notion of spacetime locality as the positive formalism. The first ingredient we need is thus a notion of spacetime in the form of a \emph{spacetime system}. Recall from Subsection~\ref{sec:spacetime} and Figure~\ref{fig:st_region} that what we require specifically are notions of spacetime \emph{regions} and \emph{hypersurfaces}. Additionally we require regions and hypersurfaces here to be \emph{oriented}. Note that given a region, its boundary inherits an orientation from the region. Orientation must be respected in compositions of hypersurfaces and regions.

For many theories of interest spacetime is modeled as \emph{Minkowski spacetime}. Then, regions are 4-dimensional submanifolds (with boundary) of Minkowski spacetime and hypersurfaces are 3-dimensional submanifolds (with boundary).\footnote{It turns out that to capture interesting physical theories it is often necessary to consider hypersurfaces that are ``enriched'' e.g.\ with the germ of a 4-manifold. However, this is largely irrelevant to the conceptual perspective we are taking here.} We need not necessarily consider all possible submanifolds, but might restrict to classes that are sufficiently well-behaved or regular. These submanifolds come with their induced \emph{metrics}. Gluings of regions and decompositions of hypersurfaces can be realized in the submanifold setting. This makes additional matching conditions on gluings and decompositions coming from the metric automatically satisfied.

Another important example for a notion of spacetime arises in general relativity. In that case we can either fix a global topology, the simplest being $\R^4$. Then, regions would be 4-dimensional submanifolds (with boundary) and hypersurfaces 3-dimensional submanifolds (with boundary) of $\R^4$ as a \emph{differentiable} manifold. Alternatively, we can avoid a global choice of topology by considering regions and hypersurfaces as differentiable manifolds (with boundary) in their own right, of dimension 4 and 3 respectively. Compared to the bare topological setting of Subsection~\ref{sec:spacetime} the manifolds carry a differentiable structure in addition to the topological one, but no metric as in the Minkowski example. Note that in order to describe truly local physics we need to be able to glue two regions with the topology of a 4-ball to a region of the same topology, as in Figure~\ref{fig:st_compose}. This implies that me need to allow for \emph{corners} in the differentiable structure of regions. (In Figure~\ref{fig:st_compose} these corners are marked as black dots.)

\subsection{Equations of motions and solution spaces}
\label{sec:ceom}

The content of a classical theory can be roughly described in terms of two ingredients. The first one is a specification of the objects of the theory, typically particles and fields. The conceivable configurations of these objects can be encoded in terms of configuration spaces. The second ingredient are the equations of motion. The \emph{solutions} of these are the configurations that are physically allowed or realizable in the theory. Since we want to describe physics locally, the key object of interest is the space $L_M$ of solutions in a spacetime region $M$. This is the space of particle trajectories and field configurations in $M$ that satisfy the equations of motion restricted to $M$. For these solutions it does not matter how or even if they continue outside of $M$. In particular, even if $M$ is a submanifold of a global spacetime manifold there are typically elements of $L_M$ that do not arise as restrictions of global solutions.

For describing the interaction between the physics in adjacent spacetime regions the key object is the space $L_{\Sigma}$ of \emph{germs of solutions} on the interfacing hypersurface $\Sigma$. Essentially, these are solutions that are defined in a small neighborhood of $\Sigma$. We can think of this as a generalization of the notion of ``initial data'' to arbitrary hypersurfaces. The space $L_\Sigma$ parametrizes the possible ``signals'' that communicate adjacent regions. Again, even if $\Sigma$ is the submanifold of a global spacetime manifold there are typically elements of $L_{\Sigma}$ that do not arise from a restriction of global solutions.

\begin{figure}
\centering
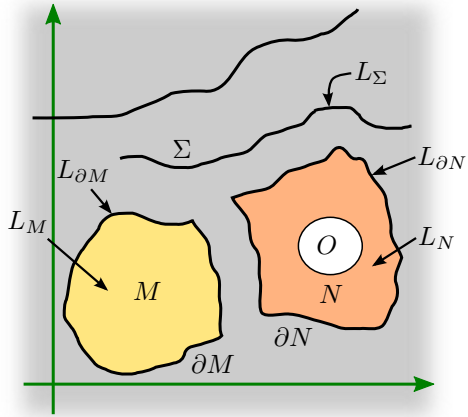
\caption{Assignment of solution spaces to regions and hypersurfaces and an observable in a classical theory: $L_M$ and $L_N$ are the solution spaces assigned to the regions $M$ and $N$. $L_\Sigma$, $L_{\partial M}$, $L_{\partial N}$ are solution spaces assigned to the hypersurfaces $\Sigma$, $\partial M$, $\partial N$. $O$ is an observable on $L_N$.}
\label{fig:st_class_sol_obs}
\end{figure}

For our purposes the classical theory is thus encoded in terms of assignments of solution spaces to hypersurfaces and to regions, see Figure~\ref{fig:st_class_sol_obs}. The solution spaces assigned to different regions and hypersurfaces are of course not independent, but related to each other. This gives rise to additional structure. In particular, if we decompose a hypersurface $\Sigma$ into a union $\Sigma_1\cup\cdots\cup\Sigma_n$ the solution space $L_{\Sigma}$ associated to $\Sigma$ is by locality the product $L_{\Sigma_1}\times\cdots\times L_{\Sigma_n}$ of the solution spaces associated to the components.\footnote{There are important classes of theories where this decomposition of solution spaces takes a more complicated form, notably \emph{gauge theories}. For some remarks on this in the quantum context see Subsection~\ref{sec:refqft}. From the present conceptual perspective, these complications are of minor importance and will be ignored.} Another important structure arises from taking the boundary $\partial M$ of a region $M$. Restricting a solution in $M$ to (a neighborhood of) the boundary of $M$ yields (a germ of) a solution on $\partial M$. This defines a boundary map $r_M:L_M\to L_{\partial M}$.

A further important structure arises in the composition of regions. Consider regions $M_1$ and $M_2$ that can be glued to a joint region $M_1\cup M_2$, see Figure~\ref{fig:st_compose}. Restricting solutions in the joint region $M_1\cup M_2$ to the individual regions $M_1$ and $M_2$ defines a map $L_{M_1\cup M_2} \to L_{M_1}\times L_{M_2}$ between the associated solution spaces. For each of the solution spaces $L_{M_1}$ and $L_{M_2}$ we then consider the corresponding boundary map, $r_{M_1}:L_{M_1}\to L_{\Sigma_1 \cup \Sigma}=L_{\Sigma_1}\times L_{\Sigma}$ and $r_{M_2}:L_{M_2}\to L_{\Sigma_2 \cup \Sigma}=L_{\Sigma_2}\times L_{\Sigma}$. Forgetting the parts of solutions living on $\Sigma_1$ and $\Sigma_2$, we get maps $L_{M_1}\to L_{\Sigma}$ and $L_{M_2}\to L_{\Sigma}$. With these ingredients we can express a key property of the composition in terms of the following \emph{exact sequence},
\begin{equation}
  L_{M_1\cup M_2} \to L_{M_1}\times L_{M_2} \rightrightarrows L_{\Sigma} .
\end{equation}
The two arrows on the right hand side represent the two maps $L_{M_1}\to L_{\Sigma}$ and $L_{M_2}\to L_{\Sigma}$ where in each case the other component in the product $L_{M_1}\times L_{M_2}$ is simply ignored. In plain language the sequence being exact means that if we have a solution in $M_1$ and a solution in $M_2$ then these arise as restrictions of one solution in the joint region $M_1\cup M_2$ if and only if they coincide in (a neighborhood of) the gluing hypersurface $\Sigma$.

Besides the spaces of solutions there are further ingredients that the classical theory may provide. We mention one of them here due to its crucial importance in classical dynamics. Given a Lagrangian (field) theory, the second variation of the Lagrangian density integrated on a hypersurface $\Sigma$ yields a symplectic form $\omega_{\Sigma}$. This makes the space of solutions $L_{\Sigma}$ associated to the hypersurface $\Sigma$ into a \emph{symplectic manifold}. Note that $\omega_{\Sigma}$ changes sign under change of orientation of the hypersurface $\Sigma$. We suppose here that (as true in the simplest cases) this symplectic form is non-degenerate. (The complications arising in the contrary case are conceptually unimportant here.) Remarkably, given a region $M$ the image under $r_M$ of the solution space $L_M$ in $L_{\partial M}$ is generically a \emph{Lagrangian submanifold}. The underlying theory was developed in the ground breaking work of Kijowski and Tulczyjew \cite{KiTu:symplectic} to which we refer the interested reader.

\subsection{Axiomatization}
\label{sec:caxioms}

The structures and relations that arise in the assignments of solution spaces to hypersurfaces and regions coming from a classical theory can be codified into an axiomatic system. Interestingly, this suggests a new way to \emph{define} a classical theory. Instead of particles, fields and equations of motions a classical theory would be defined as a system of assignments of abstract solution spaces, satisfying the axioms. We present in the following one version of such an axiomatic system, roughly based on the above considerations. Note that as in the case of the positive formalism (Subsection~\ref{sec:paxioms}) we codify composition of regions in terms of two axioms, one for disjoint composition, (C6) and one for self-composition, (C7). Given a hypersurface $\Sigma$, we denote the same hypersurface equipped with the opposite orientation by $\overline{\Sigma}$.

\begin{itemize}
\item[(C1)] Associated to each hypersurface $\Sigma$ is a manifold $L_{\Sigma}$. $L_{\Sigma}$ is equipped with a non-degenerate symplectic form $\omega_{\Sigma}$, making it into a symplectic manifold.
\item[(C2)] Associated to each hypersurface $\Sigma$ there is an (implicit) involution $L_\Sigma\to L_{\overline\Sigma}$, such that $\omega_{\overline{\Sigma}}=-\omega_{\Sigma}$.
\item[(C3)] Suppose the hypersurface $\Sigma$ decomposes into a union of hypersurfaces $\Sigma=\Sigma_1\cup\cdots\cup\Sigma_n$. Then, there is an (implicit) isomorphism $L_{\Sigma_1}\times\cdots\times L_{\Sigma_n}\to L_\Sigma$. The isomorphism preserves the symplectic form.
\item[(C4)] Associated to each region $M$ is a manifold $L_M$.
\item[(C5)] Associated to each region $M$ there is a map $r_M:L_M\to L_{\partial M}$. The image $L_{\tilde{M}}$ of $r_M$ is a Lagrangian submanifold of $L_{\partial M}$.
\item[(C6)] Let $M_1$ and $M_2$ be regions and $M= M_1\sqcup M_2$ be their disjoint union. Then $L_M$ is the product $L_{M}=L_{M_1}\times L_{M_2}$. Moreover, $r_M=r_{M_1}\times r_{M_2}$.
\item[(C7)] Let $M$ be a region with its boundary decomposing as a union $\partial M=\Sigma_1\cup\Sigma\cup \overline{\Sigma'}$, where $\Sigma'$ is a copy of $\Sigma$. Let $M_1$ denote the gluing of $M$ to itself along $\Sigma,\overline{\Sigma'}$ and suppose that $M_1$ is a region. Then, there is an injective map $r_{M;\Sigma,\overline{\Sigma'}}:L_{M_1}\toi L_{M}$ such that
\begin{equation}
 L_{M_1}\toi L_{M}\rightrightarrows L_\Sigma
\label{eq:xsbdy}
\end{equation}
is an exact sequence. Here the arrows on the right hand side are compositions of the map $r_M$ with the projections of $L_{\partial M}$ to $L_\Sigma$ and $L_{\overline{\Sigma'}}$ respectively (the latter identified with $L_\Sigma$). Moreover, the following diagram commutes, where the bottom arrow is the projection.
\begin{equation}
\xymatrix{
  L_{M_1} \ar[rr]^{r_{M;\Sigma,\overline{\Sigma'}}} \ar[d]_{r_{M_1}} & & L_{M} \ar[d]^{r_{M}}\\
  L_{\partial M_1}  & & L_{\partial M} \ar[ll]}
\end{equation}
\end{itemize}

For more details we refer the reader to the articles \cite{Oe:holomorphic} and \cite{Oe:affine} where similar axiomatic systems are exhibited for the case of linear and affine field theory respectively. One may also relate this to the multisymplectic approach to field theory. For a discrete spacetime version, see \cite{ArZa:msymeffgbft}.

\subsection{Observables}
\label{sec:cobs}

An \emph{observable} is the standard tool to model an ideal measurement or observation in classical physics. In the present context a measurement is to take place in a spacetime region $M$. Thus, an observable is a real valued function $L_M\to\R$ on the space of solutions in $M$. To formalize this, we associate to any region $M$ its space of observables $\cobs_M$. This space inherits a lot of properties from the real numbers. In particular, it is a real vector space by addition of functions, even an algebra by multiplication of functions and a partially ordered vector space by partial ordering of functions. For the latter see Definition~\ref{dfn:povs} and Proposition~\ref{prop:cpovexa} in the appendix. An important class of observables are given by characteristic functions, i.e., functions that only take the values $0$ and $1$. We shall refer to these as \emph{binary observables}. We can think of these as providing \lbl{YES}/\lbl{NO} answers as to whether a particular solution has a certain property. For those observables the multiplication realizes the logical AND operation for properties. We axiomatize the observables as follows.

\begin{itemize}
\item[(CO1)] Associated to each spacetime region $M$ is a partially ordered unital commutative algebra $\cobs_M$, called \emph{observable algebra}.
\item[(CO2a)] Let $M_1$ and $M_2$ be regions and $M=M_1\sqcup M_2$ be their disjoint union. Then, there is an injective positive algebra homomorphism $\cocomp:\cobs_{M_1}\times\cobs_{M_2}\toi\cobs_M$. This operation is required to be associative in the obvious way.
\item[(CO2b)] Let $M$ be a region with its boundary decomposing as $\partial M=\Sigma_1\cup\Sigma\cup\overline{\Sigma'}$, where $\Sigma'$ is a copy of $\Sigma$. Let $M_1$ denote the gluing of $M$ to itself along $\Sigma,\overline{\Sigma'}$ and suppose it is a region. Then, there is a positive unital algebra homomorphism $\cocomp_{\Sigma}:\cobs_{M}\to\cobs_{M_1}$. This operation is required to commute with itself and with (CO2a) in the obvious way.
\end{itemize}

The operation of Axiom (CO2a) can be thought of as extending the observables on $L_{M_1}$ and on $L_{M_2}$ each to $L_M=L_{M_1}\times L_{M_2}$ and then multiplying them. Alternatively, we might have axiomatized the extension itself of the two observables and recovered the operation of (CO2a) by subsequently using the algebra structure from Axiom (CO1). (Indeed this choice was made in the article \cite{Oe:feynobs}.) However, we can also recover the extension of the observables from Axiom (CO2a) by composing with the unit observable. So both axiomatizations are equivalent. However, the attentive reader will notice that the present choice is more in line with the axiomatization of the positive formalism (Subsection~\ref{sec:paxioms}). In addition, there might be reasons (mathematical and physical ones) to encode observables in a different way for which the present version of Axiom (CO2a) is preferable. See also the discussion in Subsection~\ref{sec:cgenprobes} below.

We can use the present framework to answer questions about the physics in regions subject to \emph{boundary conditions}. A boundary condition for a region $M$ is precisely a (germ of a) solution on the boundary $\partial M$. That is, for a region $M$ the space of boundary conditions is $L_{\partial M}$. The simplest question to ask would be as to whether given a region $M$, a boundary condition $\varphi\in L_{\partial M}$ is admissible, i.e., physically realizable or not. $\varphi$ is admissible if there exists a solution $\phi\in L_M$ such that $\varphi$ arises from $\phi$ by restriction to the boundary, i.e., $\varphi=r_M(\phi)$. A slightly more complex question would be as to whether given a boundary solution $\varphi$, a certain property $F$ is realized in $M$, where $F\in\cobs_M$ is a binary observable. Of course, this question can only make sense if $\varphi$ is admissible. Then, the answer would be $F(\phi)$ (with $1$ signifying \lbl{YES} and $0$ signifying \lbl{NO}), where as before, $\varphi=r(\phi)$. For this answer to be well defined we need either $\phi$ to be unique given $\varphi$, or $F$ to give the same value for all solutions in $M$ that reduce to $\varphi$ on the boundary. For simplicity, we shall assume the latter to hold for all observables.\footnote{ This condition is less restrictive than what one would at first suspect. In a suitable context the condition means that the observables are ``gauge invariant''.} In an analogous way we can consider observables $F$ that take arbitrary real values, where $F(\phi)$ would then represent a measurement value, given the boundary condition $\varphi$. Again, this makes sense only if $\varphi$ is admissible.

\subsection{Statistical theory}
\label{sec:cstat}

We proceed to consider statistical classical physics. Instead of solutions we are thus interested in \emph{statistical distributions} of solutions. For simplicity we shall model such statistical distributions as functions and assume the existence and choice of certain measures on solution spaces as needed. In particular, we associate with any hypersurface $\Sigma$ the space $\bcg_{\Sigma}$ of real valued functions on hypersurface solutions $L_{\Sigma}$. The positive functions $\bc_{\Sigma}\subseteq \bcg_{\Sigma}$ are those that can be interpreted as statistical distributions (without normalization) on $L_{\Sigma}$. It is clear that $\bcg_{\Sigma}$ is a partially ordered vector space with $\bc_{\Sigma}$ forming the cone of positive elements. Again, see Definition~\ref{dfn:povs} and Proposition~\ref{prop:cpovexa} in the appendix. In fact $\bcg_{\Sigma}$ has more structure such as that of an algebra, but we do not need this here. In line with a statistical perspective we change our nomenclature and refer to statistical distributions on hypersurface solutions rather than to individual solutions themselves as \emph{boundary conditions}.

The question of the admissibility of boundary conditions for a region becomes more complicated in the statistical setting. Given a region $M$ and a boundary condition $b\in\bc_{\partial M}$ viewed as an ensemble, various solutions in the ensemble might be admissible while others might not be. This motivates the introduction of a notion of \emph{compatibility}, quantified in terms of a positive real number,
\begin{equation}
 \int_{L_M} b(r_M(\phi))\, \xd\mu_M(\phi) .
\label{eq:ccomp}
\end{equation}
Here $\mu_M$ is a measure on the space $L_M$ of solutions in $M$. (We assume that we have a corresponding measure for any region $M$, in a compatible way.) Note that only that part of the boundary statistical distribution $b$ contributes to the integral that consists of solutions induced from the interior of $M$.

We define a real valued \emph{pairing} $\cobs_M\times\bcg_{\partial M}\to\R$ between observables $F$ in a region $M$ and (generalized) boundary conditions $b$ on its boundary $\partial M$ as follows,
\begin{equation}
 \lv F,b\rv_M\defeq\int_{L_M} F(\phi) b(r_M(\phi))\, \xd\mu_M(\phi) .
\label{eq:cpairing}
\end{equation}
Note that $F$ is a positive function on $L_M$ if and only if the pairing yields a positive real number for any boundary condition $b\in\bc_{\partial M}$ (assuming reasonable regularity properties). That is, the partial order structure of $\cobs_M$ as a space of functions on $L_M$ is precisely the same as that of it as a space of functions on $\bc_{\partial M}$ via the pairing. In particular, the pairing is a positive bilinear map. To be more precise, the pairing might not always be defined, depending on the integrability of the product of the observable with the generalized boundary condition. However, due to positivity of the integral we can always define it for products of positive functions if we include positive infinity, $\infty$, as a possible value. This is precisely formalized by writing it as a map $\cobs_M^+\times\bc_{\partial M}\to [0,\infty]$ that is \emph{unbounded positive bilinear}, compare Definition~\ref{dfn:wposmap} of the appendix.

If we denote the observable that is the function with value $1$ by $\np$, then the compatibility (\ref{eq:ccomp}) of a boundary condition $b\in\bc_{\partial M}$ arises as the value of the pairing $\lv\np,b\rv_M$. We call $\np$ the \emph{null observable}. It is an \emph{order unit} of the partially ordered vector space $\cobs_M$ if we restrict to bounded observables, see Definition~\ref{dfn:ou} and Proposition~\ref{prop:cpovexa}.
We proceed to the question of the realization of a certain property $F$ in $M$, where $F\in\cobs_M$ is a binary observable, given a boundary condition $b\in\bc_{\partial M}$. Viewing $b$ as an ensemble, the answer will be a \emph{probability}. Indeed, it is easy to see that this probability $\Pi$ can be calculated as follows, using the pairing (\ref{eq:cpairing}) we have just defined,
\begin{equation}
 \Pi=\frac{\lv F,b\rv_M}{\lv\np,b\rv_M} .
\label{eq:cprob}
\end{equation}
We can also interpret this as the \emph{fraction} of the admissible boundary solutions in the distribution that satisfies property $F$. Note that for a binary observable $F$ we have in particular $0\le F\le \np$ so that $0\le\Pi\le 1$. (Except if $\lv\np,b\rv_M=0$ in which case the question is ill defined or if $\lv\np,b\rv_M=\infty$ in which case the boundary condition is not normalizable.) We still obtain probabilities if we let $F$ be a probabilistic, i.e., convex combination of binary observables. We call observables obtained in this way \emph{normalized primitive} observables. It is easy to see that these are (possibly up to suitable completion in the infinite dimensional case) all observables $F\in\cobs_M$ that satisfy $0\le F\le \np$. For general observables $F\in\cobs_M$ the formula (\ref{eq:cprob}) yields its expectation value $\Pi$ subject to the boundary condition $b\in\bc_{\partial M}$.

\subsection{Classical statistical theory as a version of the positive formalism}
\label{sec:cpos}

It is becoming clear at this point that the emerging framework for classical statistical physics fits into the positive formalism introduced in Section~\ref{sec:first}. We can complete the development of the relevant mathematical structures following steps laid out in Section~\ref{sec:first}. We start with the inner product on slice regions, see Subsection~\ref{sec:sregip}. Recall that for any hypersurface $\Sigma$ we have a corresponding slice region $\hat{\Sigma}$. By definition, the boundary $\partial\hat{\Sigma}$ of the slice region decomposes into two copies of the original hypersurface $\Sigma$. We now consider the bilinear map $\bcg_{\Sigma}\times\bcg_{\Sigma}\to\R$ induced from the pairing (\ref{eq:cpairing}) of the space of boundary conditions $\bcg_{\partial \hat{\Sigma}}=\bcg_{\Sigma}\tens\bcg_{\Sigma}$ with the null probe (i.e., null observable) in $\hat{\Sigma}$ according to definition (\ref{eq:ipfrompairing}),
\begin{equation}
  \lb b,c \rb_{\Sigma} = \lv\np, b\tens c \rv_{\hat{\Sigma}}
  = \int_{L_{\hat{\Sigma}}} (b\tens c)(r_{\hat{\Sigma}}(\phi))\, \xd\mu_{\hat{\Sigma}}(\phi)
  = \int_{L_{\Sigma}} b(\phi) c(\phi)\, \xd\mu_{\Sigma}(\phi) .
\label{eq:cip}
\end{equation}
We explain the equality on the right-hand side. The space $L_{\hat{\Sigma}}$ of solutions in $\hat{\Sigma}$ is by definition the same as the space $L_{\Sigma}$ of germs of solutions on $\Sigma$. The map $r_{\hat{\Sigma}}: L_{\hat{\Sigma}}\to L_{\partial\hat{\Sigma}}$ is really a map $L_{\Sigma}\to L_{\Sigma}\times L_{\Sigma}$ given by $\phi\mapsto(\phi,\phi)$. We also use the notation $\mu_{\Sigma}$ rather than $\mu_{\hat{\Sigma}}$ for the measure on $L_{\hat{\Sigma}}=L_{\Sigma}$. The so defined bilinear form is seen to be positive-definite (assuming non-degeneracy). Again, we have been sloppy by not taking into account that the inner product might be ill-defined for certain pairs due to non-integrability. And again, we can remedy this by restricting it to positive elements (i.e., boundary conditions) and including positive infinity as a possible value. In this way we obtain a \emph{positive-definite unbounded sharply positive symmetric bilinear form} $\bc_{\Sigma}\times\bc_{\Sigma}\to\R$, see Definitions~\ref{dfn:wposbmap}, \ref{dfn:wsp} and \ref{dfn:wpd} in the appendix. Thus, the space $\bcg_{\Sigma}$ becomes a \emph{partially ordered unbounded inner product space}, see Definition~\ref{dfn:wpoip}. The present example is further elaborated on in Proposition~\ref{prop:cwpoipexa} and Remark~\ref{rem:cwpoipexa}. In particular, the subspace of \emph{square-integrable functions} is a maximal partially ordered subspace. We denote it by $\bcg^2_{\Sigma}\subseteq\bcg_{\Sigma}$. We also denote the subspace of \emph{integrable functions} by $\bcg^1_{\Sigma}\subseteq\bcg_{\Sigma}$. This is the maximal partially ordered subspace that can be paired with constant functions.

Let $\{\xi_k\}_{k\in I}$ be an orthonormal basis of $\bcg^2_{\Sigma}$ and $b_1,b_2\in\bcg^2_{\Sigma}$. The completeness relation (\ref{eq:complip}) takes the form
\begin{equation}
 \int_{L_{\Sigma}} b_1(\phi) b_2(\phi)\, \xd\mu_{\Sigma}
 = \sum_{k\in I} \int_{L_{\Sigma}\times L_{\Sigma}} b_1(\phi_1) \xi_k(\phi_1) \xi_k(\phi_2) b_2(\phi_2)\, \xd\mu_{\Sigma\times\Sigma} .
\end{equation}
Here we denote by $\mu_{\Sigma\times\Sigma}$ the product measure on $L_\Sigma\times L_{\Sigma}$. This means that we can consider
\begin{equation}
  \delta(\phi_1,\phi_2)\defeq\sum_{k\in I} \xi_k(\phi_1) \xi_k(\phi_2)
\label{eq:cdist}
\end{equation}
as a ``delta''-distribution $L_{\Sigma}\times L_{\Sigma}\to\R$ concentrated on the diagonal.

Consider now the geometric setup of Axiom (CO2) of Subsection~\ref{sec:cobs}. That is, we have region $M$ with boundary decomposing as $\partial M=\Sigma_1\cup\Sigma\cup\overline{\Sigma'}$ such that $\Sigma'$ is a copy of $\Sigma$. We suppose the gluing of $M$ to itself along $\Sigma,\overline{\Sigma'}$ is the admissible region $M_1$. Let $F:L_M\to\R$ be an observable in $M$ and recall that the observable it induces in $M_1$ by restricting from $L_M$ to $L_{M_1}$ is called $\cocomp_{\Sigma} F$, compare Axiom (CO2). The distribution (\ref{eq:cdist}) then allows to relate $F$ and $\cocomp_{\Sigma} F$ as follows. For $b\in\bcg_{\Sigma_1}$ we have the identity,
\begin{equation}
 \int_{L_{M_1}} \cocomp_{\Sigma} F(\phi) b(r_{M_1}(\phi))\,\xd\mu_{M_1}
 = \int_{L_M} F(\phi) b(\phi_1) \delta(\phi_{\Sigma},\phi_{\Sigma'})\,\xd\mu_M .
\label{eq:cobscompid}
\end{equation}
(Strictly speaking we should restrict $b$ and $F$ to be positive to correctly handle the case of an infinite value.)
Here we have used the notation $r_M(\phi)=(\phi_1,\phi_{\Sigma},\phi_{\Sigma'})$, making explicit the components of the hypersurface solution space $L_{\partial M}=L_{\Sigma_1}\times L_{\Sigma}\times L_{\Sigma'}$. In words, the identity is obtained by restricting the integral on the right hand side via the distribution $\delta$ to those solutions in $M$ that have coinciding restrictions to germs on $\Sigma$ and $\Sigma'$. By Axiom (C7) of Subsection~\ref{sec:caxioms} these are in correspondence to solutions in $M_1$ yielding the left-hand side. This requires that the measures $\mu_M$ and $\mu_{M_1}$ are compatible. As previously mentioned, we assume such compatibility throughout. Conversely, to make this concept of compatibility precise we could define it precisely through identities such as (\ref{eq:cobscompid}).

With spaces of (generalized) boundary conditions $\bcg_{\Sigma}$ as defined and spaces of probes $\pr_M$ given by spaces of observables $\cobs_M$, all axioms of the positive formalism (Subsection~\ref{sec:paxioms}) are satisfied (with the modifications indicated in Subsection~\ref{sec:infinidim}). In particular, recalling the definition of the pairing (\ref{eq:cpairing}) and of the distribution $\delta$, (\ref{eq:cdist}) it is easy to recognize the identity (\ref{eq:cobscompid}) as the composition identity for probes of the positive formalism in Axiom (P5b), setting $\pcomp=\cocomp$.
What we see is how the axioms of the positive formalism arise here as a mere consequence of the ``more fundamental'' axioms of classical theory exhibited in Subsections~\ref{sec:caxioms} and \ref{sec:cobs}.

More crucial than the reproduction of the mathematical structures and axiomatic system, however, is the coincidence of the formulas for measurable physical quantities as well as their operational meaning. As a consequence, the discussion from Section~\ref{sec:first} of different scenarios for extracting measurable quantities applies here essentially in full, complementing the much more limited discussion we have provided in the present section so far.

\subsection{Generalized classical probes}
\label{sec:cgenprobes}

There is one sense in which the scope of the framework for classical theory introduced in this section so far is more limited than that of the positive formalism of Section~\ref{sec:first}. This comes from the fact that the notion of observable is more restrictive than that of probe. An observable is here understood as arising from an ideal measurement that does not disturb the classical equations of motion. A probe on the other hand does not need to satisfy such a restriction. Indeed, we have discussed at length in Section~\ref{sec:first} that for characterizing a typical measurement we would generically need two probes $P$ and $Q$. The probe $Q$ represents the measurement apparatus alone, while $P$ represents the apparatus with the measurement outcome superimposed. For the classical observables defined here the apparatus is always given by the null probe, i.e., $Q=\np$. However, in general an apparatus can have the purpose of deliberately altering the physics in a region. Such an apparatus cannot be modeled by the classical observables as considered. (An example is the switch in Subsection~\ref{sec:pf_example}.)

It is not difficult to remedy this shortcoming of the classical framework by introducing a suitable notion of probe. For any region $M$ an apparatus in $M$ would modify the equations of motion in $M$. The space of modified solutions in $M$, however, should still be a Lagrangian submanifold of the boundary solution space $L_{\partial M}$. The simplest choice here is to admit any Lagrangian submanifold of $L_{\partial M}$ as arising from some ``apparatus''. Combining such a Lagrangian submanifold $A_M^P$ with an observable $f^P:A_M^P\to\R$ would then define a probe $P$ via modification of the pairing (\ref{eq:cpairing}),
\begin{equation}
 \lv P,b\rv_M\defeq\int_{A_M^P} f^P(\phi) b(r_M^P(\phi))\, \xd\mu_{A_M^P}(\phi) .
\label{eq:cppairing}
\end{equation}
Note that we also need a measure $\mu_{A_M^P}$ for the Lagrangian submanifold $A_M^P$.

What we have considered so far would not be the most general version of a probe as convex or even linear combinations of probes should be probes. To distinguish it, we might call it a generating probe.
This also suggests a different possibility for mathematically encoding the notion of such a probe. Namely, the notions of submanifold, measure and (positive) observable can be compactly encoded in terms of a (positive) measure on the boundary solution space. That is, a (primitive) generating probe in $M$ would simply be a (positive) measure on $L_{\partial M}$ that is supported on a Lagrangian submanifold. An arbitrary (primitive) probe would then be an arbitrary (positive) measure on $L_{\partial M}$ (or perhaps something slightly more restrictive). Since we are interested here merely in feasibility in principle it is beyond the scope of this paper to discuss the most suitable notion of classical probe or the most convenient mathematical realization of solution spaces, statistical distributions, observables or probes. In any case, the resulting structures should satisfy the axioms of the positive formalism (Subsection~\ref{sec:paxioms}) and given a suitably general notion of probe, the example of Subsection~\ref{sec:pf_example} can in principle be realized in classical physics.

\subsection{What is special about classical physics?}
\label{sec:cspecial}

Unsurprisingly, the framework developed in this section, being built on known properties of classical physics exhibits additional structure as compared to the positive formalism of Section~\ref{sec:first} alone.
One feature of classical physics, often emphasized in contradistinction to quantum physics, is the fact that the observables form a commutative algebra. This is incorporated into Axiom (CO1) in Subsection~\ref{sec:cobs} above. The operational significance of this algebra structure is limited, however. It enters into the composition of observables in disjoint regions, Axiom (CO2) which in turn underlies Axiom (P5a) of the positive formalism for this classical setting. On the other hand, a multiplicative composition of two observables in the same region is not contemplated here as its operational meaning is unclear. What is more, when replacing observables with more general probes the algebra structure is lost altogether. However, if we model classical probes on measures on boundary solution spaces as sketched above, an analogue of Axiom (CO2) is still valid. The axiom arises then from the product of measures rather than from the product of functions.

An operationally more significant ingredient of the spaces of observables or of probes is their partial order structure (compare Subsection~\ref{sec:hierarchies}). Not all partially ordered vector spaces are alike and the axioms of the positive formalism leave considerable room here. Indeed, spaces of real valued functions that satisfy the reasonable condition that for any admissible function also its positive and negative parts are admissible are partially ordered vector spaces of a special kind: They are \emph{lattices}. That is, given two elements $a$, $b$ of such a space there is a unique element $\min(a,b)$ which is maximal among those elements that are smaller or equal to both $a$ and $b$. Similarly, given $a$, $b$ there is a unique element $\max(a,b)$ that is minimal among those elements that are larger or equal to both $a$ and $b$. (Note that in the present vector space setting the existence of $\min$ and $\max$ does not need to be required separately as one implies the other, $\max(a,b)=-\min(-a,-b)$.) For functions, $\min$ and $\max$ are simply obtained by taking the pointwise minimum or maximum. See also Definition~\ref{dfn:lattice} and Proposition~\ref{prop:cpovexa} in the appendix.
The spaces of classical observables are thus lattices. Suppose now that we consider more general classical probes, modeled as sketched above as measures on the space of boundary solutions $L_{\partial M}$. Assume that there is a reference measure on $L_{\partial M}$ with all measures being absolutely continuous with respect to the reference measure. We can then write the pairing of a probe $P$ with a (generalized) boundary condition similarly to expression (\ref{eq:cpairing}) as,
\begin{equation}
 \lv P,b\rv_M\defeq\int_{L_{\partial M}} f^P(\phi) b(\phi)\, \xd\mu_{\partial M}(\phi) .
\label{eq:cmpairing}
\end{equation}
Here $\mu_{\partial M}$ is the reference measure on $L_{\partial M}$ and $f^P$ is the Radon-Nikodym derivative with respect to the reference measure of the measure corresponding to the probe $P$. The space of measures becomes a space of functions (given by the Radon-Nikodym derivatives) on $L_{\partial M}$ in this way. What is more, as for observables, the partial order as functions on $L_{\partial M}$ coincides with that derived from the pairing with $\bc_{\partial M}$. Note that this partial order structure is independent of the choice of reference measure. Thus, also this space of more general classical probes forms a lattice. We do not exclude here, however, the possibility that some other suitable space of classical probes may not have the structure of a lattice.

Of similar operational significance is the partial order structure on the space of boundary conditions. These are statistical distributions on boundary solution spaces and modeled as functions. In particular, these spaces also form lattices. Note that even if we modeled them in a different way, say as measure spaces, these would be lattices. Conversely, by the Stone-Krein-Kakutani-Toshia vector lattice theorem \cite{Kad:repcomalg} essentially any partially ordered vector space forming a lattice can be identified as the space of (continuous) functions on a (topological) space. That is, given the lattice property we are naturally led to consider the boundary conditions as statistical distributions and thus to interpret them as data of a classical theory.

\subsection{Time evolution}
\label{sec:cevol}

We aim to connect in this subsection with a more traditional description of classical physics in terms of \emph{states} of a \emph{system} evolving in time. This is to be seen as parallel to the Section~\ref{sec:evolution} where a time evolution perspective was considered for the positive formalism in general. When convenient we will freely use notations, conventions and results from that section without introducing them again. For simplicity and definiteness we specialize as in that section to Minkowski spacetime in a fixed inertial reference frame, viewed as a direct product $\R\times\R^3$.

We denote the space of (germs of) solutions at time $t$, i.e., on the equal-time hypersurface at time $t$ by $L_t$. This space is the \emph{(instantaneous) phase space} at time $t$ or the space of \emph{initial data} at time $t$. Recall from Subsection~\ref{sec:ceom} or from Axiom (C1) of Subsection~\ref{sec:caxioms} that $L_t$ carries a symplectic structure $\omega_t$, making it into a symplectic manifold. We proceed to describe evolution of the system in time. To this end consider an initial time $t_1$ and a final time $t_2$. By Axiom (C4) there is a space of solutions $L_{[t_1,t_2]}$ that describes the physically realizable configurations of the system in the time interval $[t_1,t_2]$. Restricting to the boundary is implemented by a map $r_{[t_1,t_2]}:L_{[t_1,t_2]}\to L_{t_1}\times L_{t_2}$ according to Axiom (C5). We restrict our considerations to the case that the system has a well defined initial value problem. Consequently, the spaces $L_{[t_1,t_2]}$, $L_{t_1}$ and $L_{t_2}$ are all in one-to-one correspondence. Moreover, the restriction of $r_{[t_1,t_2]}$ to either $L_{t_1}$ or $L_{t_2}$ implements this correspondence. We denote the induced \emph{time evolution map} by $v_{[t_1,t_2]}: L_{t_1}\to L_{t_2}$.
Let $\phi\in L_{[t_1,t_2]}$ be a solution in the time interval $[t_1,t_2]$. Denote $r_{[t_1,t_2]}(\phi)=(\varphi_1,\varphi_2)$ where $\varphi_1\in L_{t_1}$ and $\varphi_2\in L_{t_2}$. By definition, $v_{[t_1,t_2]}(\varphi_1)=\varphi_2$. Given $t_1< t_2 < t_3$ we obviously have the composition property,
\begin{equation}
 v_{[t_1,t_3]}= v_{[t_2,t_3]}\circ v_{[t_1,t_2]} .
\end{equation}

Recall that the sign of the symplectic structure $\omega_t$ depends on the orientation of the hypersurface at time $t$, as codified in Axiom (C2). Here we orient all hypersurfaces the same way with respect to the flow of time. More specifically we give them the orientation induced as initial boundaries of time interval regions. That implies, however, that they have opposite orientation with respect to that induced as final boundaries of time interval regions. We consider the symplectic structure on the boundary of a time interval region, taking into the account the orientations.
 Let $\Delta$ and $\Delta'$ be infinitesimal solutions near $\phi$, i.e., $\Delta,\Delta'\in T_{\phi} L_{[t_1,t_2]}$. Denote the induced boundary map for infinitesimal solutions by $\xd r_{[t_1,t_2],\phi}:T_{\phi} L_{[t_1,t_2]}\to T_{\varphi_1} L_{t_1}\times T_{\varphi_2} L_{t_2}$. Denote $\xd r_{[t_1,t_2],\phi}(\Delta)=(\delta_1,\delta_2)$ and $\xd r_{[t_1,t_2],\phi}(\Delta')=(\delta_1',\delta_2')$. With $\xd v_{[t_1,t_2],\varphi_1}:T_{\varphi_1} L_{t_1}\to T_{\varphi_2} L_{t_2}$ the induced time evolution map for infinitesimal solutions we get $\xd v_{[t_1,t_2],\varphi_1}(\delta_1)=\delta_2$ and $\xd v_{[t_1,t_2],\varphi_1}(\delta_1')=\delta_2'$. We obtain,
\begin{equation}
\omega_{\partial[t_1,t_2],\phi}(\xd r_{[t_1,t_2]}(\Delta),\xd r_{[t_1,t_2]}(\Delta'))
=\omega_{t_1, \varphi_1}(\delta_1,\delta_1') - \omega_{t_2, \varphi_2}(\delta_2,\delta_2') .
\label{eq:ctisymp}
\end{equation}
By Axiom (C5), the image of $L_{[t_1,t_2]}$ in $L_{\partial [t_1,t_2]}$ is a Lagrangian submanifold. In particular for infinitesimal solutions coming from $L_{[t_1,t_2]}$ the boundary symplectic structure, expression (\ref{eq:ctisymp}), must vanish. That is (with $v=v_{[t_1,t_2]}$),
\begin{equation}
 \omega_{t_1, \varphi_1}(\delta_1,\delta_1') = \omega_{t_2, \varphi_2}(\delta_2,\delta_2')
 = \omega_{t_2, v(\varphi_1)}(\xd v_{\varphi_1}(\delta_1),\xd v_{\varphi_1}(\delta_1'))
\end{equation}
In other words, the symplectic structure $\omega_t$ is \emph{conserved} under time-evolution. Stated equivalently, the time-evolution maps $v_{[t_1,t_2]}$ are \emph{symplectomorphisms} or \emph{canonical transformations}.

It is common to identify all the solution spaces $L_t$ with a single copy $L$, which is then simply called \emph{the phase space} of the system. We shall do so in the following. If the system is time-translation symmetric, the evolution map $v$ depends on the duration only and we write $v_{\Delta}=v_{[t,t+\Delta]}$. These satisfy
\begin{equation}
v_{\Delta_1+\Delta_2}=v_{\Delta_2}\circ v_{\Delta_1},
\end{equation}
and thus form a one-parameter group of symplectomorphisms. The vector field $X$ generating this group describes the infinitesimal time-evolution in phase space. In \emph{Hamiltonian mechanics} the vector field arises from a \emph{Hamiltonian function} $H:L\to\R$ on phase space via the relation
\begin{equation}
\xd H(Y)=\omega(X,Y)
\end{equation}
for all vector fields $Y$.

We switch to a statistical description, adapting Subsections~\ref{sec:cstat} and \ref{sec:cpos} to the time-evolution context. A state is thus a \emph{statistical distribution} on phase space. Mathematically it is a positive element in the space $\bcg$ of (suitable) real valued functions on phase space. We call $\bcg$ the space of \emph{generalized states} with the set $\bc$ of states forming the cone of positive elements. Recall that we consider the phase space $L$ to carry a measure $\mu$ (coming from the measure associated to an equal-time hypersurface as a slice region). In fact, if $L$ is a finite-dimensional manifold of dimension $2n$ a well behaved measure is given by the $2n$-form $\omega^n$, also called \emph{Liouville form}. We recall that the space $\bcg$ carries a (generally unbounded) inner product defined in terms of this measure, see expression (\ref{eq:cip}),
\begin{equation}
  \lb b,c \rb  = \int_{L} b(\phi) c(\phi)\, \xd\mu(\phi) .
\label{eq:cipp}
\end{equation}

Time-evolution of (statistical) states is induced from the time-evolution on phase space. Using the notation $\npe_{[t_1,t_2]}:\bcg\to \bcg$ for the time-evolution map that maps states at time $t_1$ to states at time $t_2$ we have,
\begin{equation}
 \npe_{[t_1,t_2]}(b)=b\circ v_{[t_1,t_2]}^{-1} .
\end{equation}
By construction $\npe_{[t_1,t_2]}$ is \emph{positive}. Since the generalized state space $\bcg$ is a \emph{lattice}, this is the same as \emph{boundary positive}, compare Subsection~\ref{sec:statesevol}. There, the map $\npe_{[t_1,t_2]}$ was given in terms of the null probe via formula (\ref{eq:npedef}). It is easily verified that this yields the same result,
\begin{align}
\npe_{[t_1,t_2]}(b) & =\sum_{k\in I} \lv \np,b\tens\xi_k\rv_{[t_1,t_2]} \xi_k
 =\sum_{k\in I} \int_{L_{[t_1,t_2]}} (b\tens\xi_k)(r_{[t_1,t_2]}(\phi))\xd\mu_{[t_1,t_2]}(\phi)\, \xi_k \nonumber\\
 & =\sum_{k\in I} \int_L b(\varphi) \xi_k(v_{[t_1,t_2]}(\varphi)) \xd\mu(\varphi)\, \xi_k
 =\sum_{k\in I} \int_L b(v_{[t_1,t_2]}^{-1}(\varphi)) \xi_k(\varphi) \xd\mu(\varphi)\, \xi_k \nonumber\\
 & =\sum_{k\in I} \lb b\circ v_{[t_1,t_2]}^{-1}, \xi_k \rb\, \xi_k
 =b\circ v_{[t_1,t_2]}^{-1} .
\end{align}
Here, we have taken $\{\xi_k\}_{k\in I}$ to be an orthonormal basis of $\bcg^2\subseteq\bcg$.

We consider observables. For simplicity we restrict to \emph{instantaneous observables}. These are observables associated to instantaneous slice regions, i.e., they are effectively functions on phase space. Let $F:L\to\R$ be a function that we want to interpret as an observable. Given initial and final states $b_1,b_2\in\bcg$ the probe associated to $F$ (and also denoted by $F$) yields the value,
\begin{equation}
\lb \tilde{F}(b_1),b_2\rb =
\lv F, b_1\tens b_2\rv = \int_L F(\varphi) b_1(\varphi) b_2(\varphi) \xd\mu(\varphi) .
\end{equation}
This is the specialization of expression (\ref{eq:cpairing}) to instantaneous slice regions. We also rewrite the probe as a map $\tilde{F}:\bcg\to\bcg$, see expressions (\ref{eq:pmapdef}) and (\ref{eq:pmap}) in Subsection~\ref{sec:statesevol}. By comparison to the inner product (\ref{eq:cipp}) we can read off that the probe map $\tilde{F}$ acts by multiplication of functions,
\begin{equation}
\tilde{F}(b)=F\cdot b .
\end{equation}

The natural choice for the \emph{state of maximal uncertainty} $\ou\in\bc$ (compare Subsection~\ref{sec:pou}) is the constant function with value $1$, i.e., $\ou(\varphi)=1$ for all $\varphi\in L$. (We assume $\bcg$ here to consist of bounded functions only.) We can use this to \emph{normalize} states (compare Subsection~\ref{sec:causnorm}). A state $b\in\bc$ is \emph{normalized} if $\lb b,\ou\rb =1$. Explicitly, this condition takes the form,
\begin{equation}
 \int_L b(\varphi)\xd\mu(\varphi) = 1 .
\end{equation}
Unsurprisingly, this is just the standard normalization condition for a probability distribution. Note that for a state to be \emph{normalizable} it must live in the subspace $\bcg^1\subseteq\bcg$.
Consider the measurement of an observable with initial normalized state $b\in\bc$ and final state $\ou$, i.e., disregarding the fate of the system after the measurement. The expectation value $\Pi$ of the observable is given by the quotient (\ref{eq:cprob}), compare Subsection~\ref{sec:cstat}. This is here,
\begin{equation}
\Pi=\frac{\lb\tilde{F}(b),\ou\rb}{\lb b,\ou\rb}
 =\lb\tilde{F}(b),\ou\rb = \int_L F(\varphi) b(\varphi) \xd\mu(\varphi) .
\end{equation}
As expected, this recovers the standard notion of expectation value of an (instantaneous) observable $F$ in state $b$.

Recall that we called an observable that takes only values $0$ and $1$ a \emph{binary observable}. Such an observable can be thought of as encoding a property corresponding here to the subset of the phase space where it takes the value $1$. Since for such an observable $P$ we have $0\le P\le\np$ the corresponding probe map as well as its dual (time reversed version) are \emph{normalization decreasing}. That is, viewed as \emph{selective probes} those observables satisfy \emph{forward} as well as \emph{backward causality}.

As in Subsection~\ref{sec:cgenprobes} we may consider probes that are not observables. The simplest possibility here is a temporary modification of the dynamics, i.e., of the equations of motion in a time interval $[t_1,t_2]$. Since the spaces of solutions before and after the alteration are unchanged, so will be the (initial and final) phase space. We can thus represent this as a modified time-evolution map $w_{[t_1,t_2]}:L\to L$. We shall expect this also to be a symplectomorphism. If the measure $\mu$ on phase space is invariant under any symplectomorphism as it ideally should be, this implies that the modified dynamics (as well as its time-reversed version) is \emph{normalization preserving},
\begin{equation}
 \lb b\circ w_{[t_1,t_2]}^{-1},\ou\rb=\int_L b(w_{[t_1,t_2]}^{-1}(\varphi))\xd\mu(\varphi) =  \int_L b(\varphi)\xd\mu(\varphi)=\lb b,\ou\rb .
\end{equation}
That is, the probe encoding the modified dynamics viewed as a \emph{non-selective probe} satisfies both forward and backward causality.

\section{Quantum theory}
\label{sec:quantum}

In the present section we sketch the development of a framework for quantum theory that implements the principles of locality and operationalism as outlined at the beginning of Section~\ref{sec:first} for the positive formalism. In contrast to Section~\ref{sec:first} and in analogy to the development for classical physics in Section~\ref{sec:classical} we proceed in a constructive manner. That is, our starting point will be quantum theory as we know it. Any further development of it will be based upon mere reformulation or motivated strongly by inference from its known structure. Locality and operationalism will be merely guiding principles. The perspective we are adopting here is outlined in the paper \cite{Oe:reveng}. The presented development largely retraces the steps of the program for foundations of quantum theory known as the \emph{general boundary formulation} \cite{Oe:gbqft,Oe:feynobs,Oe:dmf}. We shall see that it leads back to the positive formalism. In fact, it was in this way (and not in the way as presented in Section~\ref{sec:first}) that the positive formalism was originally encountered.

\subsection{Spacetime}
\label{sec:qsts}

Our first goal will be to implement a notion of locality as outlined in Subsection~\ref{sec:spacetime}. As for classical theories, a spacetime background is an important ingredient of quantum theories. In so far, the discussion about suitable spacetime systems for classical theory (Subsection~\ref{sec:csts}) is fully applicable to quantum theory as well and does not need to be repeated. A crucial difference arises, however, in the much more prominent role that the spacetime background plays in the quantum theory in its standard formulation. By \emph{standard formulation} we shall understand here quantum theory formulated in terms of a complex Hilbert space\footnote{Hilbert spaces are always \emph{separable} and \emph{complex} if not specified otherwise.} per system with an algebra of observables acting as operators on it.\footnote{Although we emphasize the standard formulation for simplicity, the argument applies equally to certain other formulations such as that of \emph{algebraic quantum field theory}.} In the standard formulation, the predicted outcome of a composite measurement depends crucially on the \emph{temporal order} of the component measurements. This temporal order must be supplied by a (space)time background structure. This excludes in particular settings with merely topological or differential spacetime structure as discussed for classical general relativity in Subsection~\ref{sec:csts}. To proceed, we assume provisionally that we are given a global Minkowski spacetime with regions and hypersurfaces arising as submanifolds.

\subsection{Locality and the path integral}
\label{sec:qlocpath}

\begin{figure}
\centering
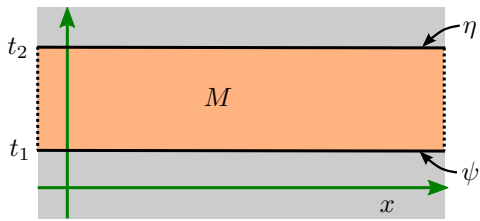
\caption{Spacetime representation of a transition amplitude with initial state $\psi$ and final state $\eta$.}
\label{fig:tampl}
\end{figure}

Given a spacetime system, the next step is to identify a structure that would allow to encode the physics in a spacetime region $M$. The standard formulation provides a tool for this, the \emph{transition amplitude}. Unfortunately, it applies only to very special regions, namely those determined by time intervals. Given a time interval $[t_1,t_2]$ consider the spacetime region $M=[t_1,t_2]\times\R^3$ obtained by extending it over all of $3$-dimensional space $\R^3$. Let $\cH$ be the Hilbert space of our quantum theory and $U_{[t_1,t_2]}:\cH\to\cH$ the unitary operator that encodes \emph{time-evolution} from time $t_1$ to time $t_2$. Then, the physics in $M$ is encoded in the transition amplitudes
\begin{equation}
 \langle\eta, U_{[t_1,t_2]}\psi\rangle
\label{eq:tampl}
\end{equation}
for $\psi\in\cH$ an \emph{initial state} and $\eta\in\cH$ a \emph{final state}, see Figure~\ref{fig:tampl}. Physically, the modulus square of this transition amplitude yields the probability $\Pi$ for a measurement at time $t_2$ to find the state $\eta$ given that the state $\psi$ was prepared at time $t_1$,
\begin{equation}
 \Pi=|\langle\eta, U_{[t_1,t_2]}\psi\rangle|^2
\label{eq:probtampl}
\end{equation}
(We suppose $\psi$ and $\eta$ to be normalized.)

A suggestion that these transition amplitudes might be generalized to apply to generic spacetime regions was first articulated by Dirac in the last section of a visionary paper on the Lagrangian in quantum theory, published in 1933 \cite{Dir:lagqm}. (Dirac called these objects ``generalized transformation functions''.) However, this idea appears to have been immediately abandoned by the mainstream, including by Dirac himself. It lived on in Japan, however, (see the introductory remarks in a paper by Watanabe \cite{Wat:condprobphys}) where it influenced the Tomonaga-Schwinger formulation of quantum field theory \cite{Tom:relwave,Sch:qed1}. It was not until the 1980s that Dirac's vision of the ``generalized transformation functions'' resurfaced in an unexpected form. In Witten's work on a geometric understanding of quantum field theory (and vice versa) he generalized the notion of ``quantum field theory'' considerably. A part of this work was the deeper understanding and exploitation of the locality properties of quantum field theory, manifest in the path integral \cite{Wit:physgeom,Wit:qftjones}. In the following we give a self-contained account of the relevant aspects.

In classical theory localization is achieved by localizing spaces of solutions of the equations of motion (Subsection~\ref{sec:ceom}). There is no quantum analog of this. However, there is a well known tool to calculate transition amplitudes that does suggest localizability: the \emph{Feynman path integral}. (Incidentally, the paper of Dirac mentioned above was cited by Feynman in his paper on the path integral \cite{Fey:stnrqm} as a basis for its development.) We recall that quantum theories are most often obtained by \emph{quantization} of a classical theory. The path integral provides just such a prescription of building a quantum theory based on the data of a classical theory. It is ubiquitous in modern quantum theory, particularly so in quantum field theory. To express the transition amplitude in terms of a path integral we have to set up the classical theory first. Suppose the configuration space of our classical theory at an instant $t$ of time is $K_t$. Then, quantizing via the Schrödinger representation, the Hilbert space $\cH_t$ at time $t$ is a space of square integrable complex valued functions on $K_t$. (We allow here for the possibility that the Hilbert space may be different for different times.) Let $S_{[t_1,t_2]}$ denote the classical \emph{action} in the time interval $[t_1,t_2]$. The transition amplitude (\ref{eq:tampl}) may then be written as a formal integral over the space $K_{[t_1,t_2]}$ of configurations in the time interval $[t_1,t_2]$,
\begin{equation}
 \langle\eta, U_{[t_1,t_2]}\psi\rangle =
 \int_{K_{[t_1,t_2]}} \psi(\phi|_{t_1}) \overline{\eta(\phi|_{t_2})} e^{\im S_{[t_1,t_2]}(\phi)} \xd\mu(\phi) ,
 \label{eq:tamplpi}
\end{equation}
where the measure $\mu$ is translation invariant.

From this point onward we take a strictly spacetime point of view, restricting attention to quantum field theory. Thus, $K_{[t_1,t_2]}$ is really the space of configurations in the region $[t_1,t_2]\times\R^3$. It can be easily defined for a generic spacetime region $M$ for which we denote it by $K_M$. Similarly, the action $S_{[t_1,t_2]}$ arises as an integral over the spacetime region $[t_1,t_2]\times\R^3$. We write $S_M$ for the action in a region $M$. Thus, the ingredients of the path integral expression (\ref{eq:tamplpi}) for the transition amplitude generalize straightforwardly to a generic region $M$. The only remaining obstacle is that there are two wave functions in the integral, one for the initial state and one for the final state. These are associated with the two boundary components (initial and final) of the region $[t_1,t_2]\times\R^3$. The boundary $\partial M$ of a generic region $M$ does not necessarily admit an analogous decomposition into two connected components. It is thus convenient to combine the two states into a single object. Indeed, the space $K_{\partial [t_1,t_2]\times\R^3}$ of field configurations on the boundary of the region $[t_1,t_2]\times\R^3$ decomposes as the direct product of the configuration spaces on the initial ($t_1$) and final ($t_2$) hypersurfaces, $K_{\partial [t_1,t_2]\times\R^3}=K_{t_1}\times K_{t_2}$. Quantizing this product space à la Schrödinger yields a Hilbert space $\cH_{\partial [t_1,t_2]\times\R^3}$ which is the (completed) tensor product of the initial and final Hilbert spaces, $\cH_{\partial [t_1,t_2]\times\R^3}=\cH_{t_1}\tens\cH_{t_2}^*$. Note that instead of $\cH_{t_2}$ its dual Hilbert space $\cH_{t_2}^*$ appears in the tensor product. This is due to the opposite orientation of the final hypersurface at $t_2$ compared to the initial one at $t_1$ as boundary components of the region $[t_1,t_2]\times\R^3$. Thus, we can view the transition amplitude as a map $\rho_{[t_1,t_2]\times\R^3}:\cH_{t_1}\tens\cH_{t_2}^*\to\bC$ via
\begin{equation}
  \rho_{[t_1,t_2]\times\R^3}(\psi\tens\eta^*)=\langle\eta, U_{[t_1,t_2]}\psi\rangle .
 \label{eq:rhoampl}
\end{equation}
With this notation the \emph{(generalized) amplitude} for a region $M$ is a map $\rho_M:\cH_{\partial M}\to\bC$ where $\cH_{\partial M}$ is a Hilbert space of \emph{(generalized) states} associated with the boundary hypersurface $\partial M$. In terms of the path integral, the amplitude for a state $\Psi\in\cH_{\partial M}$ is given by,
\begin{equation}
 \rho_M(\Psi) =
 \int_{K_{M}} \Psi(\phi|_{\partial M}) e^{\im S_{M}(\phi)} \xd\mu(\phi) .
 \label{eq:amplpi}
\end{equation}
This reduces precisely to the expression (\ref{eq:tamplpi}) for $\Psi=\psi\tens\eta^*$. Note that the wave function of the dual state $\eta^*$ is precisely the complex conjugate of the wave function for $\eta$.

If the Hilbert spaces of states are infinite-dimensional, the amplitude map $\rho_{[t_1,t_2]\times\R^3}$ given by expression (\ref{eq:rhoampl}) is necessarily \emph{unbounded} and well-defined only on a dense subspace $\cH_{\partial M}^\circ$ of the boundary state space $\cH_{\partial M}$. This is also generically true for arbitrary regions $M$. For simplicity, we continue to pretend in the following that the amplitude map $\rho_M$ is defined on all of $\cH_{\partial M}$. The necessary modifications for a correct treatment are minor and quite straightforward. We refer the interested reader to the already mentioned reference \cite{Oe:feynobs} for such a treatment.

\subsection{Composition and the path integral}
\label{sec:qcomppi}

Transition amplitudes satisfy a crucial temporal composition property, inherited from the composition property of time-evolution operators. Given times $t_1<t_2<t_3$ we must have $U_{[t_1,t_3]}=U_{[t_2,t_3]}\circ U_{[t_1,t_2]}$. With an orthonormal basis $\{\zeta_k\}_{k\in I}$ of the Hilbert space $\cH$ this translates for transition amplitudes to 
\begin{equation}
  \langle\eta, U_{[t_1,t_3]}\psi\rangle =
  \sum_{k\in I} \langle\eta, U_{[t_2,t_3]}\zeta_k\rangle
  \langle\zeta_k, U_{[t_1,t_2]} \psi\rangle .
\label{eq:tacomp}
\end{equation}
To see this property in the path integral description we note first that the sum over a complete basis of states yields a ``delta''-distribution in configuration space, concentrated on the diagonal,
\begin{equation}
  \delta(\phi_1,\phi_2)\defeq\sum_{k\in I} \overline{\zeta_k(\phi_1)} \zeta_k(\phi_2) .
\label{eq:qdist}
\end{equation}
(Mathematically this is analogous to the $\delta$ in expression (\ref{eq:cdist}), except for the fact that we work over the complex numbers here.) We then have the composition identity,
\begin{align*}
 & \int_{K_{[t_1,t_3]}} \psi(\phi|_{t_1}) \overline{\eta(\phi|_{t_3})} e^{\im S_{[t_1,t_3]}(\phi)} \xd\mu(\phi) \\
 & =\int_{\phi\in K_{[t_1,t_2]}} \int_{\phi'\in K_{[t_2,t_3]}} \psi(\phi|_{t_1}) \delta(\phi|_{t_2},\phi'|_{t_2}) \overline{\eta(\phi'|_{t_3})}  e^{\im S_{[t_1,t_2]}(\phi)} e^{\im S_{[t_2,t_3]}(\phi')} \xd\mu(\phi)\xd\mu(\phi') \\
 & =\sum_{k\in I}\int_{\phi\in K_{[t_1,t_2]}} \psi(\phi|_{t_1}) \overline{\zeta_k(\phi|_{t_2})} e^{\im S_{[t_1,t_2]}(\phi)} \xd\mu(\phi) \int_{\phi'\in K_{[t_2,t_3]}} \zeta_k(\phi'|_{t_2}) \overline{\eta(\phi'|_{t_3})} e^{\im S_{[t_2,t_3]}(\phi')} \xd\mu(\phi') ,
\end{align*}
exactly reproducing expression (\ref{eq:tacomp}). This identity relies crucially on the locality of the configuration spaces and on the additivity of the action under composition of regions.

\begin{figure}
\centering
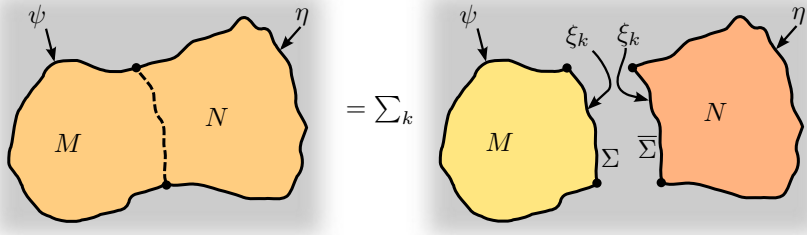
\caption{Illustration of the composition identity for generalized amplitudes in quantum field theory.}
\label{fig:comp_tqft}
\end{figure}

Given the path integral expressions it is straightforward to generalize the temporal composition property to a much more powerful spacetime composition property. What is more, we can express this composition property on the level of generalized state spaces and generalized amplitudes, forgetting the underlying path integral.
Consider regions $M$ and $N$ with decomposable boundaries $\partial M=\Sigma_1\cup\Sigma$ and $\partial N=\Sigma_2\cup\overline{\Sigma}$, that can be glued along the common hypersurface $\Sigma$. Given an orthonormal basis $\{\zeta_k\}_{k\in I}$ of the Hilbert space $\cH_{\Sigma}$ of states associated to the hypersurface $\Sigma$ we obtain the \emph{composition identity},
\begin{equation}
  \rho_{M\cup N}(\psi\tens\eta)=\sum_{k\in I} \rho_M(\psi\tens\zeta_k) \rho_N(\zeta_k^*\tens\eta) .
\label{eq:qabincompid}
\end{equation}
This is illustrated in Figure~\ref{fig:comp_tqft}.

\subsection{Topological Quantum Field Theory}
\label{sec:tqft}

The generalized notions of amplitude and state space together with their properties such as the composition identity (\ref{eq:qabincompid}) may be formalized into an axiomatic system. This requires a slight refinement of the notion of spacetime system introduced in Subsection~\ref{sec:spacetime}. All manifolds (regions and hypersurfaces) need to carry an \emph{orientation}. In particular, given a region $M$ its boundary $\partial M$ inherits an orientation from $M$. Given a hypersurface $\Sigma$ we denote the same hypersurface with opposite orientation by $\overline{\Sigma}$. We present in the following such an axiomatic system without justifying it in detail. For purposes of the present paper these precise details are unimportant. For a more in depth explanation we refer the reader to the paper \cite{Oe:gbqft}.

\begin{itemize}
\item[\textbf{(T1)}] Associated to each hypersurface $\Sigma$ there is a Hilbert space $\cH_{\Sigma}$, called the \emph{state space} of $\Sigma$. We denote its inner product by $\langle\cdot,\cdot\rangle_{\Sigma}$. For $\Sigma$ the empty set, $\cH_{\Sigma}=\bC$.
\item[\textbf{(T1b)}] Associated to each hypersurface $\Sigma$ is a conjugate linear isometry $\iota_{\Sigma}:\cH_{\Sigma}\to\cH_{\overline{\Sigma}}$. This map is an involution in the sense that $\iota_{\overline{\Sigma}}\circ\iota_{\Sigma}$ is the identity on $\cH_{\Sigma}$.
\item[\textbf{(T2)}] Given a hypersurface $\Sigma$ decomposing into a union $\Sigma_1\cup\cdots\cup\Sigma_n$, there is an isometric isomorphism of Hilbert spaces $\tau:\cH_{\Sigma_1}\tens\cdots\tens\cH_{\Sigma_n}\to\cH_{\Sigma}$. This is required to be associative in the obvious way. We often omit writing $\tau$ explicitly.
\item[\textbf{(T2b)}] The involution $\iota$ is compatible with the above decomposition. That is, $\tau \circ(\iota_{\Sigma_1}\tens\cdots\tens\iota_{\Sigma_n}) =\iota_\Sigma\circ\tau$.
\item[\textbf{(T4)}] Associated to each region $M$ there is a linear map $\rho_M:\cH_{\partial M}\to\bC$, called the \emph{amplitude (map)}.
\item[\textbf{(T3x)}] Let $\Sigma$ be a hypersurface and $\hat{\Sigma}$ the associated slice region. The boundary $\partial\hat{\Sigma}$ decomposes into the union $\partial\hat{\Sigma}=\overline{\Sigma}\cup\Sigma'$, where $\Sigma'$ denotes a second copy of $\Sigma$. Then, the bilinear map $(\cdot,\cdot)_\Sigma:\cH_{\overline{\Sigma}}\times\cH_{\Sigma'}\to\bC$ given by $\rho_{\hat{\Sigma}}\circ\tau$ is related to the inner product on $\cH_{\Sigma}$ via $\langle\cdot,\cdot\rangle_\Sigma=(\iota_\Sigma(\cdot),\cdot)_\Sigma$.
\item[\textbf{(T5a)}] Let $M_1$ and $M_2$ be regions and $M=M_1\sqcup M_2$ their disjoint union. Then, for all $\psi_1\in\cH_{\partial M_1}$ and $\psi_2\in\cH_{\partial M_2}$,
\begin{equation}
 \rho_{M}(\psi_1\tens\psi_2)= \rho_{M_1}(\psi_1)\rho_{M_2}(\psi_2) .
\label{eq:qglueaxa}
\end{equation}
\item[\textbf{(T5b)}] Let $M$ be a region with its boundary decomposing as $\partial M=\Sigma_1\cup\Sigma\cup\overline{\Sigma'}$, where $\Sigma'$ is a copy of $\Sigma$. Let $M_1$ denote the gluing of $M$ to itself along $\Sigma,\overline{\Sigma'}$ and suppose it is a region. Then, for any orthonormal basis $\{\zeta_k\}_{k\in I}$ of $\cH_\Sigma$, we have for all $\psi\in\cH_{\partial M_1}$,
\begin{equation}
 \rho_{M_1}(\psi) =\sum_{k\in I}\rho_M\left(\psi\tens\zeta_k\tens\iota_\Sigma(\zeta_k)\right) .
\label{eq:qglueaxb}
\end{equation}
\end{itemize}

We make one technical remark about the axiomatic system. Rather than considering the composition property (\ref{eq:qabincompid}) directly as an axiom we have two axioms dealing with spacetime composition: (T5a) and (T5b). The first covers the case of a disjoint composition while the second codifies a self-composition. Combining these we can recover the binary composition of the identity (\ref{eq:qabincompid}). However, the presented form allows for more flexibility and has technical advantages.

The presented axiomatic system is but a particular version of an axiomatic framework proposed by Segal and Atiyah at the end of the 1980s \cite{Seg:cftproc,Ati:tqft}. This was inspired by Witten's work at the time and called \emph{topological quantum field theory}. The adjective ``topological'' comes from the fact that originally it was applied mainly to \emph{topological field theories}. These are theories which are defined on topological manifolds or only depend on the topology of underlying manifolds even if these are equipped with more structure. The reason of interest in these theories is that they often possess only finitely many degrees of freedom, allowing for their rigorous treatment. Many of these theories do not represent physical quantum field theories but are motivated mathematically. In particular, they might be used to construct invariants of manifolds or invariants of knots \cite{Tur:qinv}. In this way topological quantum field theory lead to a revolution of various fields of mathematics, including algebraic topology, low dimensional topology and knot theory. It has also connections to quantum groups and category theory. (In fact the latter connects back to categorical diagrammatics as used in Section~\ref{sec:catdiag}, but further explaining this would go far beyond this paper's scope.)

It is worth mentioning one particular way in which the axioms as we have presented them differ from common definitions of topological quantum field theory (such as Atiyah's original one \cite{Ati:tqft}). In these definitions there is an additional datum on boundaries of regions: Each connected component of the boundary is labeled either as ``in'' or ``out''. The amplitude map for a region is then presented as a linear map between the state spaces associated to these two boundary parts, i.e., as a map $\cH_{\text{in}}\to\cH_{\text{out}}^*$ rather than $\cH_{\text{in}}\tens\cH_{\text{out}}\to\bC$ as corresponds to our axiomatization. This has the advantage that a composition of amplitudes can be implemented as a simple composition of maps, gluing an ``in'' to an ``out'' hypersurface. Also this means that one can view the spacetime system as forming a category with objects being the hypersurfaces and morphisms being the regions understood as \emph{cobordisms}. The axiomatic system implements then a \emph{functor} from this cobordism category to a category of vector spaces (in our case complex Hilbert spaces). There are two important reasons why we do not follow this tradition, one technical and one conceptual. The technical reason is that if the vector spaces are infinite dimensional (the generic case in quantum field theory) dualization becomes problematic. Relabeling an ``in'' part as an ``out'' part to convert an amplitude into a map as indicated above is then generically not well defined. The conceptual reason is that the ``in'' vs.\ ``out'' choice does in general not have any physical meaning. In the early literature it is sometimes motivated from a direction of time, i.e., the ``in'' part corresponding to an initial spacelike hypersurface and the ``out'' part to a final one. However, taking this seriously would severely restrict the types of region that could be considered, throwing us back to essentially just the standard formulation with ordinary transition amplitudes. A more adequate mathematical home for the presented version of the axioms could be the recently developed ``blob complex'' of Morrison and Walker \cite{MoWa:blobhomology}.

While topological quantum field theory arose as a formalization of properties of quantum field theory, Segal in particular proposed to take it as a \emph{definition} \cite{Seg:cftproc}. Since physically realistic quantum field theories involve infinitely many degrees of freedom, making this precise poses considerable technical challenges. Nevertheless, considering $2$-dimensional conformal field theory, Segal was able to advance this program considerably \cite{Seg:cftdef}. He is also developing the application to quantum field theory in a $4$-dimensional Euclidean setting \cite{Seg:fklectures}.
On the other hand, without abandoning the Lorentzian signature of physical spacetime, some success has been achieved by considering the simplest class of ``realistic'' theories, linear and affine field theories \cite{Oe:holomorphic,Oe:affine}.
An ambitious programme for a TQFT-like encoding of quantum field theory based on the BV-BFV formalism was also launched by Cattaneo, Mnev and Reshetikhin \cite{CaMnRe:pertqgaugebdy}. Elements of the TQFT approach are also present in ``mainstream'' conformal field theory, for example in the notion of \emph{radial quantization} \cite{DFMaSe:cft}.

\subsection{Towards operationalism -- boundary measurements}
\label{sec:qbdymeasure}

What we have considered so far is essentially a spacetime localized description of unitary time-evolution in quantum field theory. This does not include any spacetime localized description of measurement processes. Indeed, if we are merely interested in a new (and potentially better in some ways) mathematical description of standard quantum field theory there might not be any need for this. After all, in standard quantum field theory, the measurements of interest are idealized to happen at past and future infinity in time. They are encoded via the \emph{S-matrix} which, from a conceptual point of view is nothing but an asymptotic transition amplitude. That is, in the spacetime diagram of Figure~\ref{fig:tampl} we send the initial time $t_1\to-\infty$ and the future time $t_2\to\infty$. The probability (density) used to calculate scattering cross sections is obtained (in the limit) from the usual modulus square expression (\ref{eq:probtampl}).

Recall that in Subsection~\ref{sec:qsts} we insisted on the necessity for a temporal background structure to spacetime in order to ensure the applicability of the standard formulation of quantum theory. However, for the axiomatic system of topological quantum field theory as outlined in Subsection~\ref{sec:tqft} we have apparently not had any need for such a background structure. But it is precisely in the moment we consider actual measurements (via the S-matrix) that the need for this temporal structure comes back. In contrast, the examples of topological theories we mentioned earlier are not quantum theories in the physical sense. The mathematical quantities of interest in these theories do not have an interpretation as probabilities or expectation values of measurements.

Can we do better? It turns out that the probability rule (\ref{eq:probtampl}) for transition amplitudes admits a considerable generalization, valid for general spacetime regions and not requiring any explicit temporal structure \cite{Oe:gbqft}. To explain this, we first consider the more limited question of time-reversal symmetry in the description of measurements associated with transition amplitudes. 
Often measurement processes in quantum theory are presented using the picture of a ``collapse'' or ``reduction''. A measurement associated with a transition amplitude (Figure~\ref{fig:tampl}) would be described as follows: At time $t_1$ an initial state $\psi$ is prepared which then evolves unitarily until time $t_2$. At that time the measurement occurs, causing the state to suddenly collapse to the final state. In the case at hand the final state is $\eta$ with probability given by expression (\ref{eq:probtampl}). Otherwise it is the normalized projection of $\psi$ onto the orthogonal complement of $\eta$. This description distinguishes in an essential way past and future and seems to preclude any time-symmetric formulation of the measurement process. However, the collapse picture, while sometimes a convenient mental image, should not be taken as a physical description of the measurement process. Indeed, it has long been understood that the formulas for the prediction of measurement outcomes (including composite ones) in quantum theory are perfectly time-reversal symmetric. This is explained in particular in the seminal work of Aharonov, Bergmann and Lebowitz \cite{AhBeLe:timesymqm}. We illustrate this with a retrodictive interpretation of the probability (\ref{eq:probtampl}) associated to a transition amplitude. Suppose that the system is prepared randomly in states of an orthonormal basis of the Hilbert space $\cH$, one of which is $\psi$. Then, for the ensemble of measurements with outcome $\eta$ the probability (or frequency) that $\psi$ was prepared is given by formula (\ref{eq:probtampl}). (Watanabe called this kind of retrodiction ``blind retrodiction'' \cite{Wat:symphysiii}.)

The lesson here is that rather than the temporal relation between the elements of the measurement process what is essential for its description is their conditional relation. It just so happens that we are usually more interested in conditioning future events on past ones rather than the other way round. Usually we want to predict, not retrodict. In the following, for definiteness, we nevertheless use a language that is mostly guided by the traditional temporal connotation of conditioning in quantum measurements. We use the terms \emph{preparation} or \emph{knowledge} for what is conditioned upon and \emph{observation} or \emph{question} for what is conditioned.
For a general spacetime region the boundary does not necessarily decompose into two components. Consequently, it is not generally possible to ascribe a prepared state to one boundary component and an observed one to another. Instead we need to view the boundary as a whole and ascribe the elements of preparation and observation to the same boundary. It is also convenient at this point to remember that rather than the state vectors it is only their rays, i.e., the one-dimensional subspaces of the Hilbert space they generate, that are relevant for probabilities. From this it is a small step to generalize from one-dimensional to arbitrary subspaces.

The generalized probability rule for a spacetime region $M$ works as follows. Associated to the boundary $\partial M$ we have a Hilbert space $\cH_{\partial M}$ of generalized states. We need two closed subspaces $\cS$ and $\cA$ of this Hilbert space such that $\cA\subseteq\cS$. The subspace $\cS$ represents the preparation or knowledge while the subspace $\cA$ represents the observation or question. The inclusion $\cA\subseteq\cS$ represents that in posing a question we take into account what we already know.\footnote{The inclusion condition may be dropped at the expense of slightly more complicated formulas and a certain care with handling the probabilities \cite{Oe:probgbf}. However, it is unclear if anything is gained by doing so.} Let $\{\zeta_k\}_{k\in I}$ be an orthonormal basis of the Hilbert space $\cH_{\partial M}$ that restricts for the subset $J\subseteq I$ to a basis of $\cS$ and for the subset $K\subseteq J$ to a basis of $\cA$. The probability to observe $\cA$ given $\cS$ was prepared (or is known) is then,
\begin{equation}
 \Pi(\cA|\cS)=\frac{\sum_{k\in K}|\rho_M(\zeta_k)|^2}{\sum_{k\in J}|\rho_M(\zeta_k)|^2} .
\label{eq:probbdy}
\end{equation}
Note that the summands in numerator and denominator are positive and the sum in the denominator includes the sum in the numerator. Thus, if the expression is well defined we have as expected for probabilities,
\begin{equation}
 0\le \Pi(\cA|\cS)\le 1 .
\end{equation}
There are different cases when the quotient (\ref{eq:probbdy}) is not well defined. These are interpreted as corresponding to situations where the measurement is physically not well posed. If the denominator vanishes this means that the knowledge encoded in $\cS$ does not correspond to any physically realizable condition on the boundary. If the denominator is infinite this signals that the condition encoded by $\cS$ is insufficient to determine a probability.
For later use we also note an alternative form of expression (\ref{eq:probbdy}). Denote by $\po_{\cS},\po_{\cA}$ orthogonal projection operators onto the subspaces $\cS,\cA$ respectively. The probability (\ref{eq:probbdy}) may also be written as,
\begin{equation}
 \Pi(\cA|\cS)=\frac{\sum_{k\in I} \rho_M(\po_{\cA}\zeta_k)\overline{\rho_M(\zeta_k)}}{\sum_{k\in I}\rho_M(\po_{\cS}\zeta_k)\overline{\rho_M(\zeta_k)}} .
\label{eq:probbdyp}
\end{equation}

Although expression (\ref{eq:probbdy}) or (\ref{eq:probbdyp}) look rather different from the probability rule (\ref{eq:probtampl}) for transition amplitudes, the latter arises as a special case of the former as follows. In the preparation we fix the state $\psi\in\cH_{t_1}$ while we leave undetermined the final state in $\cH_{t_2}$. This leads to the subspace
\begin{equation}
 \cS = \{\psi\tens\zeta^* : \zeta^*\in\cH_{t_2}^*\}\subseteq \cH_{t_1}\tens\cH_{t_2}^*=\cH_{\partial M} .
\label{eq:qcondinit}
\end{equation}
For the observation we already know that we have $\psi\in\cH_{t_1}$ but additionally ask for $\eta\in\cH_{t_2}$. This yields the one-dimensional subspace
\begin{equation}
 \cA = \{\lambda\psi\tens\eta^* : \lambda\in\bC\}\subseteq \cS .
\label{eq:qcondoned}
\end{equation}
It is then easy to check that the denominator in the quotient (\ref{eq:probbdy}) or (\ref{eq:probbdyp}) is equal to $1$ while the numerator reproduces the usual probability expression (\ref{eq:probtampl}). We refer the reader interested in a more in-depth development and justification of the generalized probability rule to the original papers \cite{Oe:gbqft,Oe:probgbf}.

At this point we have at our disposal a manifestly local and timeless description of an important class of measurement processes (namely those localized on boundaries of spacetime regions). Moreover, this is compatible with and integrates into the local description of quantum field theory in terms of topological quantum field theory (Subsection~\ref{sec:tqft}). The boundary measurements just described have already lead to a corresponding generalization of the S-matrix. Such generalized S-matrices arise as asymptotic amplitudes for spacetime regions that do not take the special shape of an extended time interval. A simple case is to consider a hypercylinder in Minkowski spacetime that is a sphere in space, extended over all of time. The asymptotic amplitude arises when the radius of the sphere is sent to infinity \cite{CoOe:smatrixgbf}. Of particular interest are applications in curved spacetime \cite{Col:desitterpaper,CoDo:smatrixcsp}, especially when there are obstructions to constructing the standard S-matrix such as in Anti-de~Sitter spacetime \cite{DoOe:complexads}. This has also lead to new investigations of the Unruh effect \cite{CoRa:unruh,BiHaRo:bdymixed}.

\subsection{Observables}
\label{sec:qobs}

We proceed to add a further important ingredient to our local and axiomatic description of quantum field theory: \emph{observables}. In contrast to non-relativistic quantum mechanics, in quantum field theory these are localized not only in time, but also in space. In fact, the simplest ones are localized at spacetime points and equipped with a \emph{label} indicating this point. The typical example is a \emph{field operator} $\phi(t,x)$, where $(t,x)$ indicates a spacetime point. (For simplicity we use a notation that suggests a scalar field.) The most important composition of observables in quantum field theory is via the \emph{time-ordered product}. Irrespective of the order in which observables are written down, they are composed as operators in the temporal order corresponding to the labels. The earliest is applied first etc. That this ordering is invariant under Poincaré transformations is ensured by the \emph{relativistic causality} condition. This states that the commutator between observables must vanish if they are relatively spacelike localized (in which case Poincaré transformations may change their temporal order). That is (in the Heisenberg picture),
\begin{equation}
 [\phi(t,x),\phi'(t',x')]=0\qquad\text{if $(t-t',x-x')$ is spacelike}. 
\end{equation}

\begin{figure}
\centering
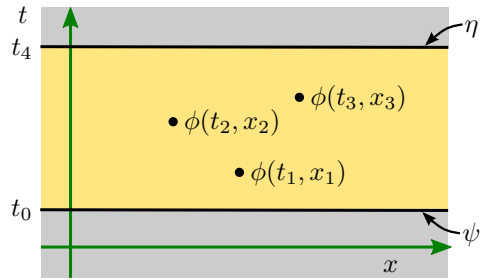
\caption{Illustration of a transition amplitude with field operators inserted.}
\label{fig:tampl_np}
\end{figure}

In path integral quantization the matrix elements of time-ordered products take a simple form. We consider the example of a product of three field operators illustrated in Figure~\ref{fig:tampl_np}. Let $t_0<t_1<t_2<t_3<t_4$ and $x_1,x_2,x_3\in\R^3$ and $\psi,\eta\in\cH$. Then,
\begin{align}
  & \langle\eta, U_{[t_3,t_4]} \phi(t_3,x_3) U_{[t_2,t_3]} \phi(t_2,x_2) U_{[t_1,t_2]} \phi(t_1,x_1) U_{[t_0,t_1]}\psi\rangle \nonumber \\ 
 & = \int_{K_{[t_0,t_4]}} \psi(\phi|_{t_0}) \overline{\eta(\phi|_{t_4})}
 \phi(t_1,x_1) \phi(t_2,x_2) \phi(t_3,x_3) e^{\im S_{[t_0,t_4]}(\phi)} \xd\mu(\phi) .
\label{eq:npointpi}
\end{align}
In text books on quantum field theory the Heisenberg picture is mostly used and the left hand side would be written as
\begin{equation}
 \langle \eta, \mathrm{T} \phi(t_3,x_3) \phi(t_2,x_2)\phi(t_1,x_1)\psi\rangle,
\end{equation}
with $\mathrm{T}$ explicitly indicating time-ordering.

Comparing to the expression (\ref{eq:tamplpi}) for a simple transition amplitude we see that the difference consists of merely inserting the product of the classical observables into the path integral. Since these depend locally on field configurations in accordance with the labels we can immediately write down the generalization for arbitrary spacetime regions. Let $M$ be a spacetime region, $F:K_M\to\R$ a classical observable\footnote{In contrast to the notion of classical observable introduced in Subsection~\ref{sec:cobs} the domain is here all of configuration space rather than only the space of solutions. This is a notable feature of path integral quantization.} and $\Psi\in\cH_{\partial M}$ a generalized state on the boundary. The amplitude with inserted observable will be denoted by $\rho_M^F$ and is a map $\cH_{\partial M}\to\bC$ just as the amplitude without observable. We get,
\begin{equation}
 \rho_M^F(\psi)=  \int_{K_{M}} \Psi(\phi|_{\partial M}) F(\phi) e^{\im S_{M}(\phi)} \xd\mu(\phi) ,
\end{equation}
compare expression (\ref{eq:amplpi}). We call $\rho_M^F$ an \emph{observable map}. Note that the classical observable $F$ may depend on field configurations in the whole region $M$, not merely at a point or a few points.

It is clear from the path integral expression (\ref{eq:npointpi}) that observable maps compose exactly in the same way as amplitudes. Moreover, amplitudes arise as special cases of observable maps with the observable being the constant function with value $1$. Proceeding as for the amplitudes we may forget the path integral and just keep the properties of the observable maps that were induced by the path integral. We cast these into axioms, extending in this way the axiomatic system of topological quantum field theory (Subsection~\ref{sec:tqft}) with a notion of observable \cite{Oe:obsgbf,Oe:feynobs}. As in Subsection~\ref{sec:tqft} we may drop at this point the restriction that the spacetime system contain a temporal background structure.
\begin{itemize}
\item[\textbf{(O1)}] Associated to each spacetime region $M$ there is a real vector space $\qobs_M$ of linear maps $\cH_{\partial M}\to\bC$, called \emph{observable maps}. In particular, $\rho_M\in\qobs_M$.
\item[\textbf{(O2a)}] Let $M_1$ and $M_2$ be regions and $M=M_1\cup M_2$ be their disjoint union. Then, there is an injective bilinear map $\qcomp:\qobs_{M_1}\times\qobs_{M_2}\toi\qobs_{M}$ such that for all $O_1\in\qobs_{M_1}$ and $O_2\in\qobs_{M_2}$ and $\psi_1\in\cH_{\partial M_1}$ and $\psi_2\in\cH_{\partial M_2}$,
\begin{equation}
 O_1\qcomp O_2(\psi_1\tens\psi_2)= O_{1}(\psi_1)O_{2}(\psi_2) .
\end{equation}
This operation is required to be associative in the obvious way.
\item[\textbf{(O2b)}] Let $M$ be a region with its boundary decomposing as a disjoint union $\partial M=\Sigma_1\cup\Sigma\cup \overline{\Sigma'}$ and $M_1$ given as in (T5b). Then, there is a linear map $\qcomp_{\Sigma}:\qobs_{M}\to\qobs_{M_1}$ such that for all $O\in\qobs_{M}$ and any orthonormal basis $\{\zeta_i\}_{i\in I}$ of $\cH_\Sigma$ and for all $\psi\in\cH_{\partial M_1}$,
\begin{equation}
 \qcomp_{\Sigma}(O)(\psi)
 =\sum_{i\in I}O(\psi\tens\zeta_i\tens\iota_\Sigma(\zeta_i)) .
\end{equation}
This operation is required to commute with itself and with (O2a) in the obvious way.
\end{itemize}

\begin{figure}
\centering
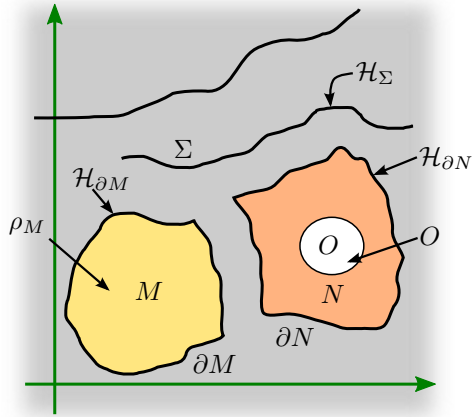
\caption{Assignment of quantum data to regions and hypersurface in topological quantum field theory with observables: $\cH_{\Sigma}$, $\cH_{\partial M}$, $\cH_{\partial N}$ are complex Hilbert spaces associated to the hypersurfaces $\Sigma$, $\partial M$ and $\partial N$ respectively. $\rho_M:\cH_{\partial M}\to\bC$ is the amplitude map for region $M$. $O:\cH_{\partial N}\to\bC$ is an observable map associated to region $N$.}
\label{fig:st_quant_af}
\end{figure}

The assignment of quantum data to regions and hypersurfaces according to the combined axiomatic system is illustrated in Figure~\ref{fig:st_quant_af}.
Rather than merely adding the observable axioms to the axiomatic system exhibited in Subsection~\ref{sec:tqft} we may merge them into a more integrated axiomatic system  as follows. We replace axioms (T4), (T5a) and (T5b) by axioms (O1), (O2a) and (O2b) and then add the features of the amplitude map as a special kind of observable map back in. In particular, equation (\ref{eq:qglueaxa}) of axiom (T5a) would appear as $\rho_M=\rho_{M_1}\qcomp\rho_{M_2}$. Similarly, equation (\ref{eq:qglueaxb}) in axiom (T5b) would appear as $\rho_{M_1}=\qcomp_{\Sigma}(\rho_M)$. This results in a striking resemblance between the present axiomatic system and that of the positive formalism (Subsection~\ref{sec:paxioms}). For hypersurfaces we have complex Hilbert spaces of states vs.\ partially ordered inner product spaces of boundary conditions, which are in particular real Hilbert spaces. For regions we have complex linear observable maps vs.\ real linear (and positive) probes. The amplitude map corresponds to the null probe. In the present case we need manifolds with orientation while the positive formalism gets along without orientation. Apart form this detail composition axioms for hypersurfaces and for regions are completely analogous. This is all the more surprising as they seem to have a completely different origin. While the composition property for quantum field theory comes from the path integral, that of the positive formalism is deduced from physical principles.

At this point it is an open problem whether quantum field theories of physical relevance (such as those that form the Standard Model of Elementary Particle Physics) can be encoded in a non-trivial way in an axiomatic system similar to the one presented here. (Restricting to time-interval regions this can be trivially done.) However, the obstacles that currently impede this appear to be more related to the general difficulties of making quantum field theory mathematically rigorous than to any conceptual problems. In particular, in the simplest case of linear quantum field theory with observables (which is the basis of perturbation theory) and assuming the availability of certain additional structure on hypersurfaces (complex structures in the sense of geometric quantization) this problem is solved \cite{Oe:feynobs}. The solution in this case also includes a rigorous quantization functor that takes in classical field theory as axiomatically presented in Subsections~\ref{sec:caxioms} and \ref{sec:cobs}. We also remind the reader of the efforts in this direction already mentioned at the end of Subsection~\ref{sec:tqft}.

\subsection{Expectation values}
\label{sec:qexpval}

As the name suggests, observables in quantum theory were conceived as mathematical objects encoding measurement processes. However, observables as used in quantum field theory fulfill this role only in a very limited sense. The objects of interest there are mostly vacuum expectation values of time-ordered products of observables, called \emph{$n$-point functions}. (E.g., as depicted in Figure~\ref{fig:tampl_np} but with initial and final states given by the vacuum state and initial and final time sent to infinity.) These serve two principal purposes. Firstly, in perturbation theory, the $n$-point functions of an interacting theory can be expanded into sums over integrals over $n$-point functions of the corresponding linear theory (see e.g.\ \cite{ItZu:qft}). In this role there is no direct relation to measurement. Secondly, the S-matrix for a scattering process of $k$ particles can be recovered from the $n$-point functions with $n\le k$. This is due to the famous reduction formula by Lehmann, Symanzik and Zimmermann \cite{LSZ:reduction}. In this case the S-matrix is closely related to measurable quantities, but the measurement in question is idealized to take place at asymptotic infinity. In particular, the spacetime localization of the observables is not related to the spacetime localization of the measurement.

\begin{figure}
\centering
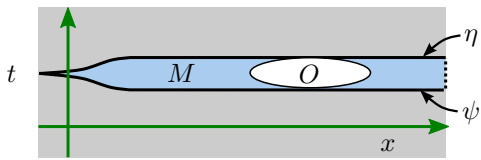
\caption{An observable map associated to a spacelike hypersurface corresponds to an operator.}
\label{fig:qobsslice}
\end{figure}

The notion of observables as operators in the standard formulation is recovered for observable maps that are associated to spacelike slice regions. Consider such a slice region, say for the equal-time hypersurface $\Sigma_t$ at time $t$ in Minkowski spacetime, see Figure~\ref{fig:qobsslice}. The relation between an observable map $O\in\qobs_{\hat{\Sigma_t}}:\cH_t\tens\cH_t^*\to\bC$ and the corresponding operator $\hat{O}:\cH_{t}\to\cH_{t}$ is via matrix elements,
\begin{equation}
  O(\psi\tens\eta^*)=\langle\eta, \hat{O}\psi\rangle .
 \label{eq:qobsop}
\end{equation}
This is closely analogous to equation (\ref{eq:rhoampl}) for transition amplitudes.
The \emph{expectation value} of an operator playing the role of an observable is of course a measurable quantity in the standard formulation. Thus, setting $\eta=\psi$ in relation (\ref{eq:qobsop}) we get the expectation value of the observable encoded by the operator $\hat{O}$ in the state $\psi$. Note that for the interpretation of the operator $\hat{O}$ as an observable in this sense it must be \emph{hermitian}. There is, however, no corresponding condition in the presented axioms that would ensure this. In any case, at least for spacelike slice observable maps that satisfy this additional condition we have an interpretation in terms of a measurement with corresponding (temporal) localization on the slice hypersurface.

The standard formula for the expectation value looks quite unnatural and special in the present axiomatic context. Formula (\ref{eq:probbdyp}) is suggestive of a more natural candidate for a \emph{generalized expectation value} \cite{Oe:obsgbf}. Let $M$ be a region and $O\in\qobs_{M}$ an observable map in $M$. Let $\cS\subseteq\cH_{\partial M}$ be a closed subspace of the boundary state space as in Subsection~\ref{sec:qbdymeasure}. Define the generalized expectation value of $O$ given $\cS$ as,
\begin{equation}
  \langle O \rangle_{\cS}\defeq\frac{\sum_{k\in I} O(\po_{\cS}\zeta_k) \overline{\rho_M(\zeta_k)}}{\sum_{k\in I}\rho_M(\po_{\cS}\zeta_k) \overline{\rho_M(\zeta_k)}} .
\label{eq:qgenev}
\end{equation}
There are several interesting special cases for this quantity. In particular, the probability (\ref{eq:probbdyp}) for a boundary measurement is recovered if we determine the observable through the subspace $\cA$ of the measurement via $O= \rho_M\circ\po_{\cA}$. It is also not difficult to see that the usual expectation value for slice observables can be recovered. To this end consider the previous context with a slice observable at time $t$ corresponding to the operator $\hat{O}$. If we set $\cS$ according to equation (\ref{eq:qcondinit}) as determined by the initial state $\psi\in\cH_t$ we obtain indeed,
\begin{equation}
 \langle O \rangle_{\cS}=\langle \psi,\hat{O}\psi\rangle .
\end{equation}
We mention another interesting special case. Still in the context of the slice observable, set $\cS$ to a one-dimensional subspace determined by a product state $\psi\tens\eta$ as given in equation (\ref{eq:qcondoned}). (Replace $\cA$ by $\cS$ in that expression.) It is easy to verify that the value (\ref{eq:qgenev}) then takes the form,
\begin{equation}
 \langle O \rangle_{\cS}=\frac{\langle \eta,\hat{O}\psi\rangle}{\langle \eta,\psi\rangle} .
\end{equation}
This is precisely the \emph{weak value} of the observable $O$ given pre-selection by $\psi$ and post-selection by $\eta$ \cite{AhAlVa:weakvalue}.
We emphasize, however, that also the generalized expectation value (\ref{eq:qgenev}) has the physical interpretation of a measurable probability or expectation value only in special circumstances that are not themselves codified in the axioms.

We shall refer to the axiomatic system obtained up to this point together with the rule (\ref{eq:probbdy}) for probabilities in boundary measurements and the generalized expectation value (\ref{eq:qgenev}) as the \emph{amplitude formalism} for quantum theory. We have already remarked the striking analogy between the axiomatic system of the amplitude formalism and that of the positive formalism as laid out in Subsection~\ref{sec:paxioms}. This analogy definitely breaks down, however, when examining the rules to extract measurable quantities. In the positive formalism the most important rules for obtaining probabilities are given by expressions (\ref{eq:elemprob}) and (\ref{eq:elembdy}). Replacing probes by observable maps and generalized boundary conditions by generalized states in these expressions does not yield quantities that can be interpreted as directly related to measurement outcomes in any way. Conversely, the analogues of the probability (\ref{eq:probbdy}) and generalized expectation value (\ref{eq:qgenev}) do not make any immediate sense in the positive formalism. What is more, there is a difference in scope. The positive formalism is meant to be capable in principle to codify any measurement that can be done in a given theory. In contrast we were able to codify some important classes of quantum measurements in the amplitude formalism, but these classes are certainly not meant to be exhaustive. In particular, this means that the amplitude formalism at this point would be insufficient as a \emph{definition} of quantum theory.

\subsection{Statistical quantum theory}
\label{sec:qstat}

A coherent and complete description of quantum measurement theory requires to work in the setting of statistical quantum theory. Before switching to this setting we introduce a few useful definitions and notations. For a complex Hilbert space $\cH$ we denote by $\rop(\cH)$ the real vector space of bounded \emph{self-adjoint operators} on $\cH$. We denote by $\pop(\cH)$ the cone of positive operators in $\rop(\cH)$ making the latter into a partially ordered vector space, compare Definition~\ref{dfn:povs} and Proposition~\ref{prop:qpovexa} of the appendix. There is an inner product in $\rop(\cH)$ given by,
\begin{equation}
\lb A, B\rb \defeq \tr(A^\dagger B),
\label{eq:qbip}
\end{equation}
where $\tr$ denotes the \emph{trace}. If $\cH$ is infinite-dimensional this inner product is not defined for every pair of operators. However, it can be defined for every pair of positive operators if we include positive infinity, $\infty$, as a possible value. In particular, it is thus a \emph{positive-definite unbounded sharply positive symmetric bilinear form}, compare Definitions~\ref{dfn:wposbmap}, \ref{dfn:wsp}, \ref{dfn:wpd} and Proposition~\ref{prop:qtrace}. $\rop(\cH)$ becomes in this way a \emph{partially ordered unbounded inner product space}. The inner product is well defined on the subspace of \emph{Hilbert-Schmidt operators} and makes this into a real Hilbert space. It is also well defined if either $A$ or $B$ is a \emph{trace-class operator}, i.e., has a well defined trace, see Remark~\ref{rem:qwpoipexa}.

In statistical quantum theory the concept of state is generalized to that of \emph{mixed state}. Mathematically a mixed state, also called \emph{density matrix} or \emph{density operator}, is represented by a positive operator $\sigma\in\pop(\cH)$ which is of trace-class and has unit trace. The notion of state that we have used so far is contained in the statistical notion of state as follows. An element of the Hilbert space $\cH$ determines a projection operator $\sigma\in\pop(\cH)$ onto the one-dimensional subspace spanned by this element. Such a state is then called a \emph{pure state}. Sometimes the term ``mixed state'' is reserved for those density operators that are not pure states. Any mixed state can be written as a (possibly infinite) linear combination of pure states with positive coefficients. This decomposition is generically not unique, however. Thus, the interpretation of a mixed state as an \emph{ensemble} of pure states (as suggested by the word ``statistical'') has to be taken with care.

The time evolution of mixed states is easily inferred from that of pure states via the correspondence of the latter to $1$-dimensional projectors. Thus, if the evolution of a state as an element in $\cH$ from time $t_1$ to time $t_2$ is implemented via the application of the unitary operator $U_{[t_1,t_2]}$ the corresponding evolution of a density operator is implemented by conjugation with $U_{[t_1,t_2]}$ as follows,
\begin{equation}
  \tilde{U}_{[t_1,t_2]}:\rop(\cH)\to\rop(\cH),\qquad \sigma\mapsto U_{[t_1,t_2]}\sigma U_{[t_1,t_2]}^\dagger .
\end{equation}
The operator $\tilde{U}_{[t_1,t_2]}$ is also called a \emph{super-operator} to emphasize that it does not act on the original Hilbert space $\cH$ but on the space of operators on it. It is immediate to see that the super-operators of time evolution compose just as the ordinary time evolution operators. That is, for $t_1< t_2 < t_3$ we have $\tilde{U}_{[t_1,t_3]}=\tilde{U}_{[t_2,t_3]}\circ \tilde{U}_{[t_1,t_2]}$.

Analogous to the Hilbert space formulation (compare Subsection~\ref{sec:qlocpath}), we can encode the time evolution super-operator $\tilde{U}$ in terms of its matrix elements. This leads to an analogue of the notion of transition amplitude. For an initial state $\sigma\in\pop(\cH)$ and a final state $\tau\in\pop(\cH)$ this is,
\begin{equation}
  \lb \tau, \tilde{U}_{[t_1,t_2]}\sigma\rb =\tr(\tau U_{[t_1,t_2]} \sigma U_{[t_1,t_2]}^\dagger) .
\label{eq:mtransprob}
\end{equation}
Note that this quantity is positive and also less or equal to $1$ (supposing that $\sigma$ and $\tau$ are normalized to have unit trace). Indeed, if $\sigma$ and $\tau$ are one-dimensional projectors onto spaces spanned by normalized vectors $\psi$ and $\eta$ respectively it is immediate to check that expression (\ref{eq:mtransprob}) coincides with the transition probability (\ref{eq:probtampl}).

Clearly the transition probabilities (\ref{eq:mtransprob}) compose in time just like the transition amplitudes (see expression (\ref{eq:tacomp})), by inserting a complete basis. That is, we have,
\begin{equation}
  \lb\tau, \tilde{U}_{[t_1,t_3]}\sigma\rb =
  \sum_{k\in I} \lb\tau, \tilde{U}_{[t_2,t_3]}\xi_k\rb
  \lb\xi_k, \tilde{U}_{[t_1,t_2]} \sigma\rb .
\label{eq:tpcomp}
\end{equation}
Here $\{\xi_k\}_{k\in H}$ is a real orthonormal basis of the subspace of $\rop(\cH_{t_2})$ formed by the self-adjoint Hilbert-Schmidt operators with respect to the inner product (\ref{eq:qbip}). Note in particular, that the basis elements $\xi_k$ are neither in general positive nor trace-class. So the matrix elements appearing in the sum on the right-hand side of expression (\ref{eq:tpcomp}) are not in general positive, let alone probabilities.

We now switch to the context of the amplitude formalism with generalized state spaces associated to hypersurfaces and amplitude maps associated to regions. Our goal is to work out the corresponding objects of statistical quantum theory and their properties. The remainder of this subsection is an abbreviated account of some of the contents of the paper \cite{Oe:dmf}.
An obvious choice for the structure to be associated to a hypersurface $\Sigma$ is the space $\rop(\cH_{\Sigma})$ of self-adjoint operators over the Hilbert space $\cH_{\Sigma}$. For a region $M$ we need a generalization of the transition probability (\ref{eq:mtransprob}) in analogy to the generalization from transition amplitudes to amplitude maps described in Subsection~\ref{sec:qlocpath}. The right choice turns out to be the (unbounded) linear map $A_M:\rop(\cH_{\partial M})\to\R$ given as follows in terms of the amplitude map for $M$,
\begin{equation}
A_M(\sigma)\defeq\sum_{k\in I} \overline{\rho_M(\zeta_k)}\rho_M(\sigma\zeta_k) ,
\label{eq:probmap}
\end{equation}
where $\{\zeta_k\}_{k\in I}$ is an orthonormal basis of $\cH_{\partial M}$.
Since the amplitude map $\rho_M$ takes complex values it is a non-trivial fact that $A_M$ takes only real values. To see this consider the case of $\sigma$ being a projection operator. We can then use a basis adapted to $\sigma$ so that the summands on the right hand side of expression (\ref{eq:probmap}) are all positive. Thus, $A_M(\sigma)$ is positive in this case. By approximating self-adjoint operators with real linear combinations of projection operators we obtain the desired result. What is more, if $\sigma$ is a positive operator it can be approximated by positive linear combinations of projection operators so that $A_M(\sigma)$ is positive in this case. That is, $A_M$ is an \emph{unbounded positive linear map} from the partially ordered vector space $\rop(\cH_{\partial M})$ to the real numbers, see Definition~\ref{dfn:wposmap}.  We call $A_M$ the \emph{probability map} associated to the region $M$.

In the amplitude formalism the assignments of Hilbert spaces to hypersurfaces satisfy important properties, relating to orientation change of the hypersurface and hypersurface decompositions. These are codified in axioms (T1), (T1b), (T2) and (T2b) of Subsection~\ref{sec:tqft}. To see the implications of the hypersurface assignments in the present statistical setting recall the following facts. Consider complex Hilbert spaces $\cH_1$ and $\cH_2$ as well as their (completed) tensor product $\cH_1\tens\cH_2$. Then, the real (completed) tensor product of the corresponding spaces of self-adjoint operators $\rop(\cH_1)\tens\rop(\cH_2)$ is naturally isomorphic to the space of self-adjoint operators over the tensor product, $\rop(\cH_1\tens\cH_2)$. What is more, this isomorphism is positive, i.e., a tensor product of positive operators is mapped to a positive operator, see Proposition~\ref{prop:qhdec}. We also mention that the space of self-adjoint operators over a Hilbert space $\cH$ and over its dual $\cH^*$ are naturally isomorphic. All in all this means the following. Axiom (T1) is replaced by the assignment of the space $\rop(\cH_{\Sigma})$ to the hypersurface, which is a partially ordered vector space. We denote this space from now on by $\bcg_{\Sigma}$. Also, if $\Sigma$ is the empty set we have $\bcg_{\Sigma}=\R$. Due to the canonical isomorphism between $\rop(\cH_{\Sigma})$ and $\rop(\cH_{\overline{\Sigma}})$ we do not need to care about the orientation of $\Sigma$ and there is no need for an analogue of Axiom (T1b). Rather, $\bcg_{\overline{\Sigma}}=\bcg_{\Sigma}$. Axiom (T2) is replaced by a very similar axiom, also in terms of an isomorphism of (now real) vector spaces involving the tensor product, but which is also positive. Axiom (T2b) again has no analogue. As the attentive reader will surely have noticed at this point, the new axioms we have obtained are precisely the Axioms (P1) and (P2) of the positive formalism, see Subsection~\ref{sec:paxioms}. There is one more structure on the spaces $\bcg_{\Sigma}$ that we have mentioned, but not yet codified. This is the inner product (\ref{eq:qbip}). We could also add this to the Axiom (P1) and to (P2) as it is compatible with the isomorphism associated to hypersurface decompositions. However, we have left it out deliberately, as it will follow from further axioms that we are to consider in a moment.

We turn to the probability maps assigned to regions. As a first step we show that they indeed specialize to the transition probabilities for regions that have the shape of a time-interval, $M=[t_1,t_2]\times\R^3$. As before, we identify the corresponding Hilbert spaces $\cH=\cH_{t_1}=\cH_{t_2}$ and therefore also the spaces of generalized mixed states $\bcg=\bcg_{t_1}=\bcg_{t_2}$. Let $\{\zeta_k\}_{k\in I}$ be an orthonormal basis of $\cH$. Then, $\{\zeta_k\tens\zeta_l^*\}_{(k,l)\in I\times I}$ is an orthonormal basis of $\cH_{\partial M}=\cH\tens\cH^*$. Given states $\sigma,\tau\in\bc$ we have $\sigma\tens\tau\in\bc_{\partial M}$ a state. We evaluate the probability map for $M$ on the state $\sigma\tens\tau$ according to definition (\ref{eq:probmap}),
\begin{align}
A_M(\sigma\tens\tau) & = \sum_{(k,l)\in I\times I} \overline{\rho_M(\zeta_k\tens\zeta_l^*)}\rho_M(\sigma\zeta_k \tens \tau\zeta_l^*) \nonumber\\
& = \sum_{(k,l)\in I\times I} \overline{\langle\zeta_l,U_{[t_1,t_2]}\zeta_k\rangle}
   \langle\tau\zeta_l,U_{[t_1,t_2]}\sigma\zeta_k\rangle \nonumber\\
& = \tr(\tau U_{[t_1,t_2]} \sigma U_{[t_1,t_2]}^\dagger) .
\end{align}
We recover expression (\ref{eq:mtransprob}) as claimed.
A very similar derivation yields the inner product (\ref{eq:qbip}) from the probability map for slice regions using Axiom (T3x) of the amplitude formalism. Thus, consider a hypersurface $\Sigma$ and the associated slice region $\hat{\Sigma}$. Given generalized states $\sigma,\tau\in\bcg_{\Sigma}$ and an orthonormal basis $\{\zeta_k\}_{k\in I}$ of $\cH_{\Sigma}$ we have,
\begin{align}
  A_{\hat{\Sigma}}(\sigma\tens\tau) & = \sum_{(k,l)\in I\times I} \overline{\rho_{\hat{\Sigma}}(\zeta_k\tens\zeta_l^*)}\rho_{\hat{\Sigma}}(\sigma\zeta_k \tens \tau\zeta_l^*) \nonumber\\
& = \sum_{(k,l)\in I\times I} \overline{\langle\zeta_l,\zeta_k\rangle_{\Sigma}}
   \langle\tau\zeta_l,\sigma\zeta_k\rangle_{\Sigma} \nonumber\\
& = \tr(\tau\sigma)=\lb\tau,\sigma\rb_{\Sigma} .
\end{align}

We obtain properties analogous to Axioms (T4) and (T3x) of the amplitude formalism. Formulating these as axioms themselves can be done as follows:
\begin{itemize}
\item[\textbf{(P4')}] Associated to each region $M$ there is an unbounded positive linear map $A_M:\bcg_{\partial M}\to\R$ called the \emph{probability map}.
\item[\textbf{(P3x')}] Let $\Sigma$ be a hypersurface and $\hat{\Sigma}$ the associated slice region. Then, the map $\lb\cdot,\cdot\rb_{\Sigma}:\bcg_{\Sigma}\times\bcg_{\Sigma}\to\R$ given by $\lb b_1,b_2\rb_{\Sigma}\defeq A_{\hat{\Sigma}}(b_1\tens b_2)$ is a positive-definite unbounded sharply positive symmetric bilinear form and makes $\bcg_{\Sigma}$ into a partially ordered unbounded inner product space.
\end{itemize}
In this way we have included the inner product on $\bcg_{\Sigma}$ axiomatically, although we have not yet codified its compatibility with the hypersurface decompositions of Axiom (P2).

The most remarkable analogy of the statistical setting with the amplitude formalism arises in the composition properties of the probability maps. These turn out to be completely analogous to the composition properties of the amplitude maps, codified in Axioms (T5) and (T5b) of Subsection~\ref{sec:tqft}, and can be derived directly from the latter. We include the explicit demonstration here only for the case of composition in the context of Axiom (T5b). Thus, consider a region $M$ with boundary decomposing as $\partial M=\Sigma_1\cup \Sigma\cup\overline{\Sigma'}$ such that $\Sigma'$ is a copy of $\Sigma$. Denote the gluing of $M$ to itself along $\Sigma,\overline{\Sigma'}$ by $M_1$ and suppose this is an admissible region. Suppose $\sigma\in\bcg_{\Sigma_1}$ and let $\{\zeta_{1,k}\}_{k\in I}$ and $\{\zeta_{k}\}_{k\in J}$ denote orthonormal basis of the Hilbert spaces $\cH_{\Sigma_1}$ and $\cH_{\Sigma}$ respectively. Define $\xi_{k l}$ to be the operator on $\cH_{\Sigma}$ defined by $\xi_{k l}(\zeta_m)=\delta_{l,m}\zeta_k$. Then, using Definition (\ref{eq:probmap}) and Axiom (T5b) we have,
\begin{align}
A_{M_1}(\sigma) & =\sum_{k\in I} \overline{\rho_{M_1}(\zeta_{1,k})}\rho_{M_1}(\sigma\zeta_{1,k}) \nonumber \\
 & = \sum_{k\in I\; l,m\in J} \overline{\rho_{M}(\zeta_{1,k}\tens\zeta_l\tens\zeta_l^*)}\rho_{M}(\sigma\zeta_{1,k}\tens\zeta_m\tens\zeta_m^*) \nonumber \\
 & = \sum_{k\in I\; l,m,n\in J} \overline{\rho_{M}(\zeta_{1,k}\tens\zeta_l\tens\zeta_l^*)}\rho_{M}(\sigma\zeta_{1,k}\tens\xi_{m,n}(\zeta_l)\tens\xi_{m,n}(\zeta_l)^*) \nonumber \\
 & = \sum_{k\in I\; l,m,n,j\in J} \overline{\rho_{M}(\zeta_{1,k}\tens\zeta_l\tens\zeta_j^*)}\rho_{M}(\sigma\zeta_{1,k}\tens\xi_{m,n}(\zeta_l)\tens\xi_{m,n}(\zeta_j)^*) \nonumber \\
 & = \sum_{k\in I\; m,n\in J} A_M(\sigma\tens\xi_{m,n}\tens\xi_{m,n}^\dagger) .
\label{eq:probcomp5}
\end{align}
Note that $\{\xi_{k l}\}_{(k,l)\in J\times J}$ is an orthonormal basis of the complex Hilbert space of Hilbert-Schmidt operators on $\cH_{\Sigma}$ with the Hilbert-Schmidt inner product (\ref{eq:qbip}). In expression (\ref{eq:probcomp5}) we can replace this orthonormal basis by any other one. In particular we may choose a basis that consists of self-adjoint operators only. This is then also an orthonormal basis of the real Hilbert space of self-adjoint Hilbert-Schmidt operators.
The composition properties of the probability maps can be codified into axioms as follows,
\begin{itemize}
\item[\textbf{(P5a')}] Let $M_1$ and $M_2$ be regions and $M=M_1\cup M_2$ their disjoint union. Then, for all $\sigma_1\in\bcg_{\partial M_1}$ and $\sigma_2\in\bcg_{\partial M_2}$,
\begin{equation}
 A_{M}(\sigma_1\tens\sigma_2)= A_{M_1}(\sigma_1) A_{M_2}(\sigma_2) .
\label{eq:qpglueaxa}
\end{equation}
\item[\textbf{(P5b')}] Let $M$ be a region with its boundary decomposing as $\partial M=\Sigma_1\cup\Sigma\cup\overline{\Sigma'}$, where $\Sigma'$ is a copy of $\Sigma$. Let $M_1$ denote the gluing of $M$ to itself along $\Sigma,\overline{\Sigma'}$ and suppose it is a region. Then, for any orthonormal basis $\{\xi_k\}_{k\in I}$ of the subspace of Hilbert-Schmidt operators of $\bcg_\Sigma$, we have for all $\sigma\in\bcg_{\partial M_1}$,
\begin{equation}
 A_{M_1}(\sigma) =\sum_{k\in I} A_M\left(\sigma\tens\xi_k\tens \xi_k\right) .
\label{eq:qpglueaxb}
\end{equation}
\end{itemize}

We arrive at an axiomatic formulation of properties of quantum statistical field theory analogous to the properties of quantum field theory encoded into the axioms of topological quantum field theory and the amplitude formalism of Subsection~\ref{sec:tqft}. More precisely Axioms (P1), (P2), (P4'), (P3x'), (P5a'), (P5b') are analogs to Axioms (T1), (T2), (T4), (T3x), (T5a), (T5b) respectively. As explained previously, there are no analogs of Axioms (T1b) and (T2b) as all structures in the statistical setting are invariant under orientation change of hypersurfaces. At this time the attentive reader will have noticed that the Axioms (P1), (P2), (P4'), (P3x'), (P5a'), (P5b') can be obtained as a simplification of the axioms of the positive formalism (Subsection~\ref{sec:paxioms}) arrived at by taking away the spaces of probes, merely retaining the null probes, i.e., probability maps. This also involves a slight notational change by setting $A_M(\sigma)=\lv\np,\sigma\rv_M$. Indeed, it was by following the steps outlined in the present subsection that the positive formalism was originally discovered in the paper \cite{Oe:dmf}. The derivation from first principles outlined in Section~\ref{sec:first} as well as the realization that the positive formalism can also describe classical theories came later \cite{Oe:firstproc}.

The fact that the axioms of the present setting describe a more general class of theories than statistical quantum field theory reflects a loss of structure as compared to the axioms of Subsection~\ref{sec:tqft}. To see this more clearly it is convenient to momentarily redefine the spaces $\bcg_{\Sigma}$ to be the complex vector spaces $\op(\cH_{\Sigma})$ of bounded operators on the corresponding Hilbert spaces $\cH_{\Sigma}$. It is easy to see that all constructions work with these spaces as well and we basically obtain a complexification of the presented setting. In particular, the probability maps are complex valued maps that yield real or positive values on the subsets of self-adjoint and positive operators respectively. In fact, it is natural to start working in this way and then realize that everything restricts accordingly when restricting to the self-adjoint operators. An advantage of the complex setting is that we see more structure: The spaces associated to hypersurfaces carry not merely a partial order and an inner product but also an algebra structure. This algebra structure is something we (deliberately) loose when transiting to the positive formalism. The real spaces of self-adjoint operators do in fact still carry a shadow of this structure, namely they form \emph{Jordan algebras}. We have not axiomatized this either, however, as there is apparently no need for this structure in the present framework.

The construction outlined in the present subsection is also interesting from a purely mathematical point of view. It can be understood as a \emph{modulus-square functor} that takes an ordinary topological quantum field theory (based on vector spaces with inner products) to a new type of ``positive'' topological quantum field theory based on partially ordered vector spaces and positive maps. More specifically, it replaces vector spaces on hypersurfaces by tensor products of these with their duals associated to the orientation reversed hypersurfaces. For regions it replaces amplitude maps with a product of an amplitude map with its version for the orientation reversed region. All in all we are ``gluing'' or ``tensoring'' the original theory to an orientation reversed copy of itself. Then, the inner product is used to restrict to the self-adjoint real subspaces and extract the partial order structure. After the process, the orientations of the manifolds may be ``forgotten''. To see how this description corresponds to the one outlined above, note that the complex Hilbert space of Hilbert-Schmidt operators on a complex Hilbert space $\cH$ is canonically isomorphic to the (completed) tensor product $\cH\tens\cH^*$. Moreover, under this isomorphism the definition of the probability map (\ref{eq:probmap}) in terms of the amplitude map takes the simple form, for $\psi\in\cH_{\partial M}$ and $\eta\in\cH_{\overline{\partial M}}= \cH_{\partial M}^*$,
\begin{equation}
 A_M(\psi\tens\eta)=\rho_M(\psi)\rho_{\overline{M}}(\eta)=\rho_M(\psi)\overline{\rho_M(\iota_{\overline{\partial M}}(\eta))}.
\end{equation}

\subsection{Measurement and quantum operations}
\label{sec:qmesop}

We are now turning to the question of measurement in the statistical setting. The natural first step is to reconsider the notions of measurement that we have already incorporated into the amplitude formalism. We begin with the notion of boundary measurement as described in Subsection~\ref{sec:qbdymeasure}. Thus consider a region $M$ with subspaces $\cA\subseteq\cS\subseteq\cH_{\partial M}$ of the boundary Hilbert spaces determining preparation or knowledge ($\cS$) and the observation or question ($\cA$). As before, we encode $\cS$ and $\cA$ in terms of their corresponding projection operators $\po_{\cS}$ and $\po_{\cA}$. Comparing the definition (\ref{eq:probmap}) of the probability map $A_M$ with formula (\ref{eq:probbdyp}) for the probability we obtain the extremely simple and compelling result,
\begin{equation}
 \Pi(\cA|\cS)=\frac{A_M(\po_{\cA})}{A_M(\po_{\cS})} .
\end{equation}
This gives further justification for calling $A_M$ \emph{probability map}. What is more, adapting notation (identifying the probability map with the null probe) we recover precisely the probability (\ref{eq:elembdy}) for boundary measurements in the positive formalism (special case $Q=\np$). Note also that $0\le\po_{\cA}\le\po_{\cS}$ are positive operators, i.e., represent generalized mixed states in $\bc_{\partial M}$. We find therefore complete agreement between the notion of boundary measurement introduced in Subsection~\ref{sec:qbdymeasure} and that of the positive formalism introduced in Subsection~\ref{sec:hierarchies}. In fact this allows us to adapt the former to the latter and drop the restriction that the entries need to be projection operators in favor of allowing arbitrary positive operators.

We proceed to consider observable maps and the generalized expectation values of Subsection~\ref{sec:qexpval} in the statistical setting. The role of the probability map (\ref{eq:probmap}) in the boundary measurement probability (\ref{eq:probbdyp}) suggests an analogous definition for observables with a corresponding role in generalized expectation values (\ref{eq:qgenev}). Given a region $M$ and an observable map $\rho_M^O:\cH_{\partial M}\to\bC$ we define the \emph{expectation map} $A_M^O:\bcg_{\partial M}\to\bC$ as,
\begin{equation}
A_M^O(\sigma)\defeq \sum_{k\in I} \overline{\rho_M(\zeta_k)}\rho_M^O(\sigma\zeta_k) ,
\label{eq:qevmap}
\end{equation}
where $\{\zeta_k\}_{k\in I}$ is an orthonormal basis of $\cH_{\partial M}$. Given a closed subspace $\cS\subseteq\cH_{\partial M}$ and the corresponding projection operator $\po_{\cS}$ this simplifies the formula for the generalized expectation value (\ref{eq:qgenev}) to the following expression,
\begin{equation}
  \langle O \rangle_{\cS}\defeq\frac{A_M^O(\po_{\cS})}{A_M(\po_{\cS})} .
\label{eq:qgenevs}
\end{equation}
Expectation maps compose just like observable maps, satisfying properties analogous to Axioms (O1), (O2a), (O2b) of Subsection~\ref{sec:qobs}. This can be demonstrated in a way very similar to the corresponding demonstration for amplitude maps in Subsection~\ref{sec:qstat}. Also note that the probability map can be viewed as a special expectation map. In this way we can completely encode the quantum field theoretic notion of observable in the statistical setting. In fact integrating an axiomatization of the expectation map with the axioms discussed in Subsection~\ref{sec:qstat} yields an axiomatic system almost exactly like the positive formalism (Subsection~\ref{sec:paxioms}). In this we consider the expectation maps to be analogous to probes by equating $A_M^P(b)$ with $\lv P,b \rv_M$. This then brings into formal coincidence the formula (\ref{eq:qgenevs}) for the generalized expectation value with the expectation value (\ref{eq:elemprob}) for a measurement induced by probes. However, crucial discrepancies remain. For one, the expectation maps (\ref{eq:qevmap}) generically yield complex and not real values, even applied to a self-adjoint or positive operator. Related to this, they lead to actual probabilities or expectation values only in special cases, as already discussed in Subsection~\ref{sec:qexpval}. On the other hand, the expectation maps are clearly not sufficient for expressing all possible measurements, in contrast to the role of the probes in the positive formalism.

We proceed to review relevant aspects of the general theory of measurement in quantum theory. To this end we return to the context of the standard formulation. That is, the spacetime regions of interest are time intervals and the Hilbert spaces $\cH_t$ for all equal-time hypersurfaces are identified with ``the'' Hilbert space $\cH$ of the ``system''. We start by considering a measurement encoded through an observable given in terms of a self-adjoint operator $\hat{O}:\cH\to\cH$. Let the spectral decomposition of $\hat{O}$ be as follows,\footnote{For simplicity we consider here only the case that $\hat{O}$ has a pure point spectrum.}
\begin{equation}
 \hat{O}=\sum_{k\in I}\lambda_k \po_k,
\end{equation}
where $\lambda_k$ are the pairwise distinct real eigenvalues and $\po_k$ the orthogonal projectors on the corresponding eigenspaces with $\sum_{k\in I} \po_k = \one$. According to the Lüders rule \cite{Lue:messprozess} an initial state $\sigma\in\bc$ is transformed through the measurement as follows,
\begin{equation}
 \sigma\mapsto \sum_{k\in I} \po_k \sigma \po_k .
\label{eq:projmes}
\end{equation}
The probability $\Pi$ of finding the value $\lambda_n$ for the measurement outcome is,
\begin{equation}
 \Pi=\tr(\po_n\sigma\po_n) .
\end{equation}
This supposes that the state $\sigma$ is normalized to have unit trace, $\tr(\sigma)=1$. Otherwise we would have to divide by the value of the trace to obtain the probability.

The super-operators $\bcg\to\bcg$ that describe any state change that may be effected through a measurement or other physical process by coupling the system to another system are called \emph{quantum operations}. The type of measurement just discussed constitutes just one special class of measurements that can be performed in quantum theory. Consequently, the super-operators of the form of formula (\ref{eq:projmes}) represent only a special class of quantum operations. Clearly, a minimal requirement for a super-operator to describe a transformation of states is that it needs to be \emph{positive}, i.e., map positive operators to positive operators. It turns out that the requirement for a quantum operation is only slightly more stringent as was understood by Kraus around 1970 \cite{Kra:statechanges}: A quantum operation is described by a super-operator that is \emph{completely positive}. Such a super-operator $S$ can be specified in terms of a set $\{K_k\}_{k\in I}$ of \emph{Kraus operators} on $\cH$ as acting on a state $\sigma$ by,
\begin{equation}
 S:\sigma\mapsto \sum_{k\in I} K_k \sigma K_k^\dagger .
\end{equation}
In addition, quantum operations are usually also required not to increase the normalization of states, i.e., satisfy $\tr(S(\sigma))\le\tr(\sigma)$. This translates for the Kraus operators to the inequality $\sum_{k\in I} K_k^\dagger K_k\le\one$. The special case $\tr(S(\sigma))=\tr(\sigma)$ for all states $\sigma$ is equivalent to the condition $\sum_{k\in I} K_k^\dagger K_k=\one$. A quantum operation is then called \emph{non-selective} or a \emph{quantum dynamical map}. This is used to model a measurement process where no selection of the final state according to measurement outcome has taken place. It also models physical processes that are induced from the unitary evolution of a larger system that contains the system of interest.

Consider now a binary measurement with possible outcomes \lbl{YES}/\lbl{NO}. We are interested in the probability for an affirmative outcome. There are two quantum operations associated to the measurement. One is a non-selective quantum operation $Q$ that describes merely the influence of the measurement apparatus, i.e., it alters the dynamics due to the presence of the apparatus. The other is a \emph{selective} quantum operation $P$ that filters out the final states corresponding to a negative outcome. We can describe this with a set of Kraus operators $\{K_i\}_{i\in I}$ such that $\sum_{i\in I} K_i^\dagger K_i=\one$. These yield the quantum operation $Q$. The quantum operation $P$ is obtained by eliminating those operators that collectively correspond to the negative result. We can describe this in terms of a subset $J\subset I$ of the index set. Note also that this implies $P\le Q$. The probability $\Pi$ for an affirmative outcome for an initial normalized state $\sigma$ is,
\begin{equation}
 \Pi = \tr\left(P(\sigma)\right)= \tr\left(\sum_{i\in J} K_i \sigma K_i^\dagger\right) .
\end{equation}
The quantum operation $Q$ does not explicitly appear in this expression. For this reason its significance is not necessarily emphasized in the literature. However, if we were to relax the normalization conditions on the Kraus operators and on the states, the probability would take the more complicated form,
\begin{equation}
 \Pi = \frac{\tr\left(P(\sigma)\right)}{\tr\left(Q(\sigma)\right)}
 = \frac{\tr\left(\sum_{i\in J} K_i \sigma K_i^\dagger\right)}{\tr\left(\sum_{i\in I} K_i \sigma K_i^\dagger\right)} .
\label{eq:probqon}
\end{equation}

As a first step in returning to the spacetime setting developed in the present section we switch to a view where the boundary of the time-interval region $M=[t_1,t_2]\times\R^3$ which was the arena for the measurement processes just discussed is viewed as one hypersurface. The generalized boundary state space is the tensor product $\bcg_{\partial M}=\bcg_{t_1}\tens\bcg_{t_2}$. (Recall that this tensor product is defined as the space of self-adjoint operators on the (completed) tensor product $\cH_{t_1}\tens\cH_{t_2}^*$.) Now, consider a linear map $S:\bcg_{t_1}\to\bcg_{t_2}$. This induces an unbounded linear map $S':\bcg_{t_1}\tens\bcg_{t_2}\to\R$ by dualization of $\bcg_{t_2}$ (which is self-dual). Concretely, we have $S'(\sigma\tens\tau)=\tr(\tau\, S(\sigma))$. Now, we have the following remarkable fact: The map $S$ is \emph{completely positive} if and only if the map $S'$ is \emph{unbounded positive}, see Proposition~\ref{prop:cpwpos} of the appendix. In the terminology of Subsection~\ref{sec:statesevol}, complete positivity turns out to be precisely the same as \emph{boundary positivity}. In other words, an (unbounded) linear map $\bcg_{\partial M}\to\R$ corresponds to a quantum operation (without the normalization condition) if and only if it is positive. This prompts us to declare the concept of a \emph{(generalized) quantum operation} for any spacetime region $M$ (not just time intervals) to be given by an unbounded positive map $\bcg_{\partial M}\to\R$. From here onward we take this, rather than the previously defined expectation maps as the basic concept for describing measurements in the statistical setting. Note also that the non-factorizing structure of quantum operations makes it clear that an adequately complete notion of quantum measurement cannot be encoded into the amplitude formalism.

The normalization conditions for states and quantum operations in the standard formulation are dependent on the restriction to time-interval regions. Thus, we need to drop them in the setting of a more general spacetime system. That this is not necessarily a problem is illustrated by the probability expression (\ref{eq:probqon}) which, as we have stated, is valid without the normalization conditions imposed. Translating the quantum operations $Q$ and $P$ in that expression to their corresponding maps $Q'$ and $P'$ on the boundary state space yields,
\begin{equation}
 \Pi = \frac{P'(\sigma\tens \one)}{Q'(\sigma\tens \one)} .
\label{eq:probqob}
\end{equation}
Here $\one$ is the identity operator on (the final copy of) the Hilbert space $\cH$. It encodes the state of \emph{maximal uncertainty}, also called \emph{maximally entangled state}. Its physical meaning here is that we do not assume any knowledge about the system or what happens to it after the measurement. Of course, we could also assume such knowledge, i.e., impose \emph{post-selection} in which case the boundary state $b\in\bc_{\partial M}$ will not take the simple form $b=\sigma\tens\one$. (See also Subsections~\ref{sec:pou}, \ref{sec:causnorm} and \ref{sec:qevol}.)

At this point it becomes plainly clear that up to a minor adjustment of notation we recover precisely formula (\ref{eq:elemprob}) for measurement probabilities in the positive formalism. Not only that, but the notion of generalized quantum operation is precisely coincident with that of a \emph{primitive probe} in the positive formalism. More general (not necessarily probabilistic) expectation values are recovered precisely as in the positive formalism via linear combinations of generalized quantum operations respectively primitive probes. This leads to the notion of non-primitive probes also in the quantum setting. By comparison with the probability maps of Subsection~\ref{sec:qstat} it is clear that the composition properties of quantum operations in time intervals generalize precisely to the spacetime composition properties embodied in the Axioms (P5a) and (P5b) of the positive formalism, see Subsection~\ref{sec:paxioms}. Finally, identifying the probability maps as the null probes leaves us precisely with the axiomatic system of the positive formalism.

We have thus succeeded in bringing quantum theory into a form that coincides precisely with a special version of the positive formalism, including crucially its notions of measurement, probability and expectation values.

\subsection{Towards a characterization of quantum theory}
\label{sec:qchar}

As already mentioned in Subsection~\ref{sec:qstat} structure is lost in the transition from the amplitude formalism to the positive formalism. The primary structure in the standard formulation of quantum theory is the complex Hilbert space of states. In the amplitude formalism such a complex Hilbert space $\cH_{\Sigma}$ is associated to any hypersurface $\Sigma$. Each complex Hilbert space gives rise to the corresponding algebra of bounded operators $\op(\cH_{\Sigma})$. For the positive formalism we forget the algebra structure and retain only the real subspace $\bcg_{\Sigma}=\rop(\cH_{\Sigma})$ of self-adjoint operators with its partial order structure. Obviously, not every partially ordered vector space can be constructed in this way. The question is thus as how to characterize the partially ordered vector spaces that originate in this way. A partial answer was given by Kadison in 1951 \cite{Kad:orderpropbsaop}, introducing the notion of \emph{anti-lattice}. An anti-lattice is a partially ordered set such that for any two elements $a$, $b$ the following is true. If there is a unique element $\min(a,b)$ which is maximal among those elements that are smaller or equal to both $a$ and $b$ then $a\le b$ or $b\le a$. In other words, $\min(a,b)$ only exists if it is forced to exist due to $a\le b$ (in which case $\min(a,b)=a$) or $b\le a$ (in which case $\min(a,b)=b$). Recall that in contrast, in a \emph{lattice} $\min(a,b)$ always exists (compare Subsection~\ref{sec:cspecial}). This justifies the name ``anti-lattice''. Kadison showed that the partially ordered vector space of self-adjoint operators on a complex Hilbert space is an anti-lattice, compare Definition~\ref{dfn:lattice} and Proposition~\ref{prop:qpovexa} of the appendix.

Recalling corresponding comments on classical theory in Subsection~\ref{sec:cspecial} this yields a compelling picture of classical and quantum theories as opposite extremes in a potential spectrum of theories that fit into the positive formalism. The distinction emphasized here is through the properties of the partially ordered vector spaces forming the spaces of generalized boundary conditions. One extreme, a lattice structure, would correspond to classical theories. The other extreme, an anti-lattice structure, would correspond to quantum theories.

One should take this picture with caution, however, especially on the quantum side. There are partially ordered vector spaces that are anti-lattices, but that do not arise exactly in the way described from a complex Hilbert space. Kadison himself showed in the mentioned article that the self-adjoint elements of any \emph{von~Neumann algebra} with trivial center, called \emph{factor}, form an anti-lattice. A \emph{type I} factor is the algebra of all bounded operators on a complex Hilbert space, corresponding to the situation discussed, but there exist other types of factors (called type II and type III). Kadison's result was generalized by Archbold in 1972, showing that any \emph{$C^*$-algebra} that is \emph{prime} yields an anti-lattice in this way \cite{Arc:primecsal}. (This includes the factors.) Not much seems to be known about the existence of vector anti-lattices that cannot be obtained in this way. So, it is unclear to which extent the anti-lattice property may be sufficient for qualifying a theory as quantum. On the other hand, as the algebraic approach to quantum field theory shows \cite{HaKa:aqft}, there are good reasons for admitting at least certain spaces that do not arise from type I factors as state spaces of quantum theories.

\subsection{Time evolution}
\label{sec:qevol}

The natural home of both the amplitude formalism and of the positive formalism as developed for quantum theory is a spacetime setting. Nevertheless, in developing these formalisms we have taken recourse many times to a time-evolution setting as this is the home of the standard formulation of quantum theory. In the present subsection we complement this by indicating how the time-evolution perspective developed in Section~\ref{sec:evolution} for the positive formalism in general specializes to quantum theory in so far as this may not have become clear yet.

For definiteness and simplicity we restrict to taking spacetime to be Minkowski spacetime in a fixed inertial frame, written as $\R\times\R^3$.
Associated to the equal-time hypersurface at time $t$ we have a complex Hilbert space $\cH_t$ of (pure) states by Axiom (T1), see Subsection~\ref{sec:tqft}. As is usual in the standard formulation of quantum theory we assume a natural identification of all these Hilbert spaces with a single copy $\cH$, \emph{the} Hilbert space of (pure states) of the system. In Subsection~\ref{sec:qlocpath} we have described how the amplitude map generalizes transition amplitudes. We indicate here how one could go the opposite way. Given a time interval $[t_1,t_2]$ we define the \emph{time-evolution operator} $U_{[t_1,t_2]}:\cH\to\cH$ via equation (\ref{eq:rhoampl}) from the amplitude map given by Axiom (T4). With an orthonormal basis $\{\zeta_k\}_{k\in I}$ of $\cH$ we can also write this as,
\begin{equation}
 U_{[t_1,t_2]}\psi=\sum_{k\in I} \rho_{[t_1,t_2]}(\psi\tens\zeta_k^*) \zeta_k .
\end{equation}
We shall assume as usual that the time-evolution operator is an automorphism of the state space, i.e., that it is \emph{unitary}.

Transiting to the statistical setting, we have a state space $\bcg$ which is a partially ordered unbounded inner product space. This is the space $\rop(\cH)$ of self-adjoint operators on the complex Hilbert space $\cH$. Its inner product is the (unbounded) Hilbert-Schmidt inner product, compare Subsection~\ref{sec:qstat}. The time-evolution operator $\npe_{[t_1,t_2]}:\bcg\to\bcg$ characterized by relation (\ref{eq:npechar}) is the super-operator that was denoted by $\tilde{U}_{[t_1,t_2]}$ in Subsection~\ref{sec:qstat}, compare expression (\ref{eq:mtransprob}).

A super-operator is the quantum version of what was called a \emph{probe map} in Subsection~\ref{sec:statesevol}. As already mentioned in Subsection~\ref{sec:qmesop} the \emph{complete positivity} of a super-operator is precisely equivalent to the \emph{boundary positivity} of Subsection~\ref{sec:statesevol}, i.e., to the primitivity of the associated probe. That is, the notion of primitive probe coincides precisely with the standard notion of \emph{quantum operation} (but without the normalization condition). The \emph{state of maximal uncertainty} $\ou\in\bc$ of Subsection~\ref{sec:pou} is precisely the state of maximal uncertainty of Subsection~\ref{sec:qmesop}, i.e., the identity operator $\one$ on the Hilbert space $\cH$. The \emph{normalization} of states and \emph{forward causality} are implemented in the standard formulation of quantum theory precisely in the way described in Subsection~\ref{sec:causnorm}. Note in particular, that since the inner product on $\bcg$ is expressible in terms of the trace (\ref{eq:qbip}), so is the normalization condition for states, taking for $b\in\bc$ the familiar form,
\begin{equation}
 1=\lb \ou, b\rb = \tr(\one b)=\tr(b) .
\end{equation}
Thus, the property of a probe map to be \emph{normalization preserving} or \emph{normalization decreasing} is in quantum theory usually called \emph{trace preserving} or \emph{trace decreasing}. By the implementation of forward causality a quantum operation is usually required to be trace decreasing. If it is even trace preserving it is also called a \emph{quantum dynamical map} or a \emph{quantum channel}. Essentially all constructions of Section~\ref{sec:evolution} apply to quantum theory in its standard formulation. This includes the diagrammatic tools and the categorical viewpoint.
This is not coincidental. The version of the positive formalism of Section~\ref{sec:evolution} is in large parts inspired by and modeled on the standard formulation of quantum theory.

We add a remark on causality and observables. The non-selective probe map associated to a measurement determined by an observable, expression (\ref{eq:projmes}), takes the same form under time-reversing dualization. In particular, both this map and its dual are normalization preserving. Correspondingly, the associated selective probe maps and their duals are both normalization decreasing. That is, a measurement determined by an observable is automatically compatible both with forward and with backward causality.

\subsection{Further related work}
\label{sec:qcomp}

We have presented in this section the development of the (quantum) positive formalism as originating mainly from a formalization of quantum field theory. As we have seen, this development builds on a lot of previous work, and especially heavily on the framework of \emph{topological quantum field theory}. However, it interweaves with and builds on previous work in many other areas as well. In the following we expose relations to other work that is specific to the quantum setting and has not been mentioned previously. For related work not specific to the quantum setting see Subsection~\ref{sec:pcomp}.

We start with the categorical viewpoint. As already mentioned, the formalization of quantum field theory in categorical language is at the core of the origins of topological quantum field theory and dates from the end of the 1980s. The elaboration of a related diagrammatic language dates from the same time. This was not appreciated at the time, however, in the quantum foundations community. This changed only with the rediscovery of the categorical formalization and associated diagrammatics in the time-evolution setting by Abramsky and Coecke \cite{AbCo:catsemquant}. They used a Hilbert space formulation corresponding to what we have called the (time-evolution version of the) amplitude formalism. They proposed to deal with measurements by using direct sums, with one summand for each possible measurement outcome. The more natural setting for including measurement processes, the statistical setting with density operators was formalized in categorical language by Selinger \cite{Sel:cccandcpm}, see also comments in Subsection~\ref{sec:pcomp}. This corresponds to the time-evolution version of the positive formalism. Remarkably, Selinger also introduced a functor from the Hilbert space category to the density operator category, called the \emph{CPM construction}. This is quite similar to a time-evolution version of the \emph{modulus-square functor} we have discussed in Subsection~\ref{sec:qstat}.

Hardy's \emph{operator tensor formulation} of quantum theory \cite{Har:optensqt} can also be understood from the categorical point of view, even though that is not made explicit in his paper. However, a version of the corresponding diagrammatics is fully developed and used as a central tool for performing calculations. Hardy also brings in a notion of locality and emphasizes that diagrams need not be (temporally) foliated to be evaluated, thus moving in the direction of deemphasizing time. However, a fixed causal structure on processes remains imprinted into this formalism due to the normalization conditions implementing forward causality, compare corresponding remarks at the end of Subsection~\ref{sec:evolcompsys}.

A central feature of the general boundary formulation compared to the standard formulation of quantum theory is that it removes a fixed notion of time as an essential ingredient. There are many works in the literature that can be seen as steps in this same direction. A first step is to remove any asymmetry under time reversal which the quantum theoretical measurement process superficially seems to possess. As already mentioned in Subsection~\ref{sec:qbdymeasure}, Aharonov, Bergmann and Lebowitz showed that the rules for calculating probabilities for outcomes of composite measurements are really symmetric under time reversal once pre- and post-selection are properly taken into account \cite{AhBeLe:timesymqm}. This work provided the starting point for the \emph{two-time formalism} of Aharonov and collaborators, where an initial and final state are treated as a combined entity to achieve a time-symmetric quantum formalism, see \cite{AhVa:twostateuprev} for a review. This foreshadows considering the state space associated to the boundary of a time-interval region as a single entity in the amplitude and positive formalisms. This idea was also recently (and independently) explored by Popescu and collaborators \cite{SGBLSP:prepost}. It is interesting to note in this context that states that would normally be described as mixed can appear pure when considered in the full boundary state space. In the context of the Unruh effect this is true for example for the Minkowski vacuum state as seen in Rindler spacetime as shown by Colosi and Rätzel \cite{CoRa:unruh}.

The \emph{quantum conditional state formalism} proposed by Leifer and Spekkens \cite{LeiSpe:causneut} aims to isolate correlation from causation in quantum theory. In particular, it seeks a unified treatment for ``experiments involving two systems at a single time and those involving a single system at two times''. In contrast to how this is achieved in the positive formalism this involves a wide ranging ``quantum'' generalization of Bayesian inference. Also, it assigns Hilbert spaces to ``elementary'' spacetime regions rather than to boundaries of such regions as in TQFT and the amplitude formalism. At the moment this approach remains unfinished and its precise relation to the standard formulation unclear.

Oreshkov and Cerf have recently sketched a formalism for quantum theory without predefined time \cite{OrCe:opquantwotime}. It is more complicated than the positive formalism in that it distinguishes states and effects. Also states and effects are conceived of as pairs of objects analogous to the pair of selective and non-selective quantum operations for processes. Nevertheless, the resulting formulas for probabilities in a closed diagram are essentially equivalent to those of the quantum version of the positive formalism. Crucially, their formalism effectively implements the composition identity (\ref{eq:bincompid}). Also, the trace normalization properties for quantum operations are abandoned as they must be, since they would fix a causal structure (compare Subsections~\ref{sec:causnorm} and \ref{sec:evolcompsys}). Interestingly, the authors go on to propose a line of investigation to probe causal structure as emerging from (generalized) quantum field theories through the dynamics as imprinted on what we would call the null probe and boundary state spaces of small compact regions. It appears that a natural home for such an investigation could be the positive formalism with its explicit spacetime locality properties.

\section{Examples}
\label{sec:examples}

The positive formalism is rooted via the preceding amplitude formalism in quantum field theory, with already a rich history of development and applications, compare Section~\ref{sec:quantum} and also remarks on quantum gravity in Section~\ref{sec:discussion}. Via the modulus-square functor (Subsection~\ref{sec:qstat}) all these applications carry over to the positive formalism. In contrast, applications in the foundations of quantum theory and in quantum information theory are just beginning to be explored. We present in this section two initial examples.

\subsection{A hybrid: quantum teleportation}
\label{sec:qteleport}

In the following we consider the famous \emph{quantum teleportation protocol} \cite{BBCJPW:qteleport} as formalized within the time-evolution version of the positive formalism (Section~\ref{sec:evolution}). This formalization is not a feat unique to the positive formalism. Rather, it can be accomplished by any sufficiently general version of the convex operational framework, recall Subsection~\ref{sec:pcomp}.
We include it here to demonstrate explicitly the applicability of the positive formalism to \emph{quantum information theory} and in particular to models that are \emph{hybrid}, treating classical and quantum components in a unified way. To do so we naturally incorporate notions of \emph{classical-quantum} and \emph{quantum-classical channels} \cite{Hol:statquant}.

\begin{figure}
  \centering
  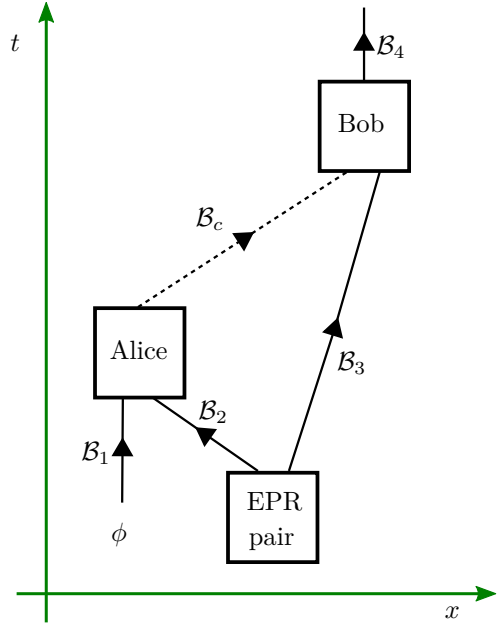
  \caption{Directed circuit diagram for the quantum teleportation protocol. Solid lines represent quantum systems, the dashed line a classical system. The quantum state $\phi$ is teleported from Alice to Bob using an entangled EPR pair as a quantum resource and classical communication from Alice to Bob.}
  \label{fig:quantum_teleportation}
\end{figure}

We use the time-evolution version of the positive formalism as we also wish to implement full forward causality via normalization conditions on probes, see Section~\ref{sec:evolution}. It is thus convenient to use the directed version of the circuit diagrams. The circuit diagram for the quantum teleportation protocol is shown in Figure~\ref{fig:quantum_teleportation}. As usual, the boxes represent probes and the links state spaces. We consider quantum systems given in terms of an $N$-dimensional Hilbert space $\cH$. Thus the generalized state spaces $\bcg_1,\bcg_2,\bcg_3,\bcg_4$ (see figure) are all copies of the partially ordered inner product space $\rop(\cH)$ of self-adjoint operators on $\cH$. Recall that the positive elements (proper states) in $\rop(\cH)$ are the positive operators. Moreover, the inner product in $\rop(\cH)$ is the trace of the operator product, compare formula (\ref{eq:qbip}). We also recall that the state $\ou$ of maximal uncertainty is the identity operator in $\rop(\cH)$.

In contrast, $\bcg_c$ is a space of generalized states of a classical system. Since we want to teleport states from a Hilbert space of dimension $N$, we need to transmit classical information equivalent to the choice of one element out of $N^2$ elements. That is, the relevant classical solution space $L$ has $N^2$ elements which we label $x_{i j}$ with $i,j\in\{1,\dots,N\}$. The space of classical generalized states $\bcg_c$ is the partially ordered inner product space of real valued functions on $L$. Recall that the positive elements (proper states) are the positive functions. Also, the inner product is the integral of the product, compare formula (\ref{eq:cip}). The natural measure here is the \emph{counting measure} since $L$ is a finite set. Thus, integrals over $L$ reduce to sums. Concretely, for $b,c\in\bcg_c$,
\begin{equation}
\lb c,b\rb_c=\int_L b(x) c(x)\xd\mu(x)=\sum_{i,j=1}^N b(x_{ij}) c(x_{ij}) .
\end{equation}
Recall also that the state of maximal uncertainty $\ou_c$ is the constant function with value $1$. It is convenient to chose an orthonormal basis in $\bcg_c$ of characteristic functions for the points in $L$. That is, we define $\chi_{i j}:\bcg_c\to\R$ via
\begin{equation}
\chi_{n m}(x_{i j})=\delta_{n,i}\delta_{m,j} .
\end{equation}
Note that these are normalized since $\lb\ou_c,\chi_{n m}\rb=1$.

Fix an orthonormal basis $\{\zeta_i\}_{i\in\{0,\dots,N-1\}}$ of $\cH$. Define the state $\psi_{n m}\in\cH\tens\cH$ for $n,m\in\{0,\dots,N-1\}$ by,
\begin{equation}
\psi_{n m}\defeq \frac{1}{\sqrt{N}}\sum_{j=0}^{N-1} e^{2\pi\im j n/N} \zeta_j\tens\zeta_{j+m} .
\end{equation}
Here $j+m$ is understood modulo $N$. Note that these states form an orthonormal basis of $\cH\tens\cH$. We denote the corresponding orthogonal projectors by $\po_{n m}\in\rop(\cH\tens\cH)$. We use throughout the natural isomorphism $\rop(\cH\tens\cH)=\rop(\cH)\tens\rop(\cH)$.
We also define the unitary operators $U_{n m}:\cH\to\cH$ for $n,m\in\{0,\dots,N-1\}$ by,
\begin{equation}
 U_{n m}:\eta\to \sum_{j=0}^{N-1} e^{2\pi\im j n/N} \zeta_j \langle \zeta_{j+m},\eta\rangle_{\cH} .
\end{equation}

The quantum teleportation protocol works as follows (see Figure~\ref{fig:quantum_teleportation}): An EPR pair of entangled quantum states is created. This is the state $\po_{0 0}\in\bcg_{2}\tens\bcg_{3}$. Note that this state is normalized, i.e., $\lb \ou_2\tens\ou_3,\po_{0 0}\rb_{2,3}=\tr_{2,3}(\po_{0 0})=1$, since the projector is one-dimensional. The state is sent to Alice and Bob, with Alice receiving the $\bcg_2$-component and Bob receiving the $\bcg_3$-component. Now, Alice receives an unknown state $\phi\in\bcg_1$. Alice performs a measurement on the tensor product state in $\bcg_1\tens\bcg_2$. She discards the resulting quantum state, but sends the measurement outcome through a classical system $\bcg_c$ to Bob. That is, Alice performs an operation $A:\bcg_1\tens\bcg_2\to\bcg_c$. Bob, who might be at a considerable distance, receives the classical state from Alice in $\bcg_c$. Depending on the information contained in it, he transforms the quantum state in $\bcg_3$ resulting in an outgoing state in $\bcg_4$. In particular, he performs an operation $B:\bcg_c\tens\bcg_3\to\bcg_4$. If Alice and Bob perform the correct operations, the resulting state is identical to $\phi$.

We turn to define the operations $A$ and $B$. From Alice's perspective what she does can be described as follows: She performs a projective measurement of the joint state in $\bcg_1\tens\bcg_2$ with respect to the complete set $\{\po_{n m}\}_{n,m\in\{0,\dots,N-1\}}$ of orthogonal one-dimensional projection operators. If the outcome is along $\po_{i j}$ she encodes this by sending the element $x_{i j}$ in the classical ``solution space'' $L$. However, since the outcome of the measurement is not certain, we need to describe it and Alice's behavior statistically. Thus, instead of outputting an element of $L$ we will get a classical statistical distribution, i.e., an element of $\bcg_c$. If the outcome along $\po_{i j}$ was certain, this would be the normalized distribution supported on the single element $x_{i j}$, that is the pure state $\chi_{i j}$. In general we get a mixed state. It is now straightforward to write down the operation $A$ for any input $\sigma\in\bcg_1\tens\bcg_2$,
\begin{equation}
 A(\sigma)= \sum_{n,m=0}^{N-1} \lb \ou_1\tens\ou_2, \po_{n m} \sigma\rb_{1,2} \chi_{n m}
 = \sum_{n,m=0}^{N-1} \tr_{1,2}(\po_{n m} \sigma) \chi_{n m} .
\end{equation}
Note that $A$ is normalization preserving,
\begin{equation}
 \lb\ou_c,\sum_{n,m=0}^{N-1} \tr_{1,2}(\po_{n m} \sigma) \chi_{n m}\rb_c
 =\sum_{n,m=0}^{N-1} \tr_{1,2}(\po_{n m} \sigma)=\tr_{1,2}(\sigma)=\lb \ou_1\tens\ou_2,\sigma\rb_{1,2} .
\end{equation}
In the language of quantum information theory, $A$ is an example of a \emph{quantum-classical channel}.

Bob's perspective is the following: He receives from Alice an element $x_{i j}$ of $L$ that indicates that he should apply the unitary transformation $U_{i j}$ on the quantum state in $\bcg_3$, outputting to $\bcg_4$. Again we need to model this statistically with $B$ transforming an input $b\tens\sigma\in\bcg_c\tens\bcg_3$ as,
\begin{equation}
 B(b\tens\sigma)= \sum_{n,m=0}^{N-1} b(x_{n m})\, U_{n m}\sigma U_{n m}^\dagger .
\end{equation}
$B$ is also normalization preserving,
\begin{equation}
 \lb\ou_4,\sum_{n,m=0}^{N-1} b(x_{n m})\, U_{n m}\sigma U_{n m}^\dagger\rb_4
 =\lb\ou_c,b\rb_c \lb\ou_3,\sigma\rb_3
\end{equation}
In the language of quantum information theory, $B$ could be called a \emph{classical/quantum-quantum channel}.

The teleportation protocol thus results in the transformation $\bcg_1\to\bcg_4$ given by the composite that can be read off from the diagram of Figure~\ref{fig:quantum_teleportation} as a directed circuit diagram (according to the rules in Section~\ref{sec:catdiag}):
\begin{equation}
 \phi\mapsto B ((A\tens\id_3) (\phi\tens\po_{0 0})) .
\end{equation}
The validity of the protocol means that the right-hand side is equal to $\phi$ again. This is a straightforward calculation for which we refer the reader to the original paper.

Rather, we emphasize the applicability of the positive formalism to the present example from quantum information theory, in particular, with quantum and classical components interacting. More specifically we note that in the present unified perspective, the operations of Alice and Bob are each described in terms of a single \emph{non-selective} probe map that is \emph{normalization preserving}.

\subsection{Indefinite causal structure}
\label{sec:indefcausal}

In this subsection we consider an example with \emph{indefinite causal structure}. Again, this example is primarily presented to demonstrate the range of applicability of the positive formalism beyond the standard formulation of quantum theory. The example was introduced in a ground-breaking paper by Oreshkov, Costa and Brukner \cite{OrCoBr:quantcorrcaus}.

\begin{figure}
  \centering
  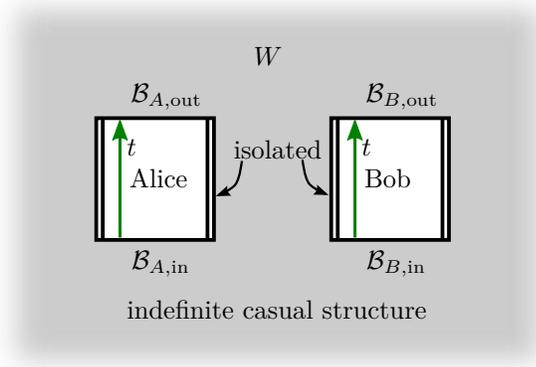
  \caption{Example with indefinite causal structure. Alice and Bob are in local spacetime regions with definite time and causal structure. However, they interact with the exterior only at an initial and a final time. They are surrounded by a region $W$ without definite temporal or causal structure.}
  \label{fig:indef_causal}
\end{figure}

Consider two parties, Alice and Bob, in their respective laboratories, see Figure~\ref{fig:indef_causal}. The two laboratories are isolated, except for an initial and a final time (in the respective
laboratories reference frame, not necessarily a global time) at which they may receive or send out a quantum state. Alice receives a state through the system $\bcg_{A,\text{in}}$ and later sends a state through the system $\bcg_{A,\text{out}}$. The respective systems for Bob are $\bcg_{B,\text{in}}$ and $\bcg_{B,\text{out}}$. Within each laboratory and for the duration of the experiment quantum theory is valid as prescribed by the standard formulation. Now, instead of assuming that Alice and Bob are timelike separated so that one can signal to the other, but not the other way round, or that they are spacelike separated so that neither can signal to the other, we want to leave any possibilities for interaction through the exterior spacetime region $W$ (gray in Figure~\ref{fig:indef_causal}) open. In particular, we do not equip $W$ with any definite temporal or causal structure.

While this situation is outside of the convex operational framework and outside of the standard formulation of quantum theory in particular, it fits straightforwardly into the positive formalism. Consider a set of \emph{selective} primitive probes $\{M_{A,i}\}_{i=\{1,\dots,n_A\}}$ for Alice that represent alternative outcomes for some experiment that Alice performs. The sum of these is the \emph{non-selective} primitive probe $M_A= \sum_{i=1}^{n_A} M_{A,i}$ that represents the experiment without any assumption about the outcome. Since we describe Alice's laboratory in complete accordance with the standard formulation, the probe map $\tilde{M}_A:\bcg_{A,\text{in}}\to\bcg_{A,\text{out}}$ corresponding to the probe $M_A$ must be \emph{normalization preserving}. (The probe maps $\tilde{M}_{A,i}:\bcg_{A,\text{in}}\to\bcg_{A,\text{out}}$ are then automatically \emph{normalization decreasing}.) We consider an analogous situation for Bob with selective primitive probes $\{M_{B,j}\}_{j=\{1,\dots,n_B\}}$ summing to a non-selective primitive probe $M_B= \sum_{j=1}^{n_B} M_{B,j}$ with normalization preserving probe map. The exterior region carries a single probe $W$. For this we assume no normalization condition as there is no notion of causality we want to impose. Since this probe does not carry any outcome we consider it \emph{non-selective}, although we do not associate any condition with that except of course that $W$ is primitive, i.e., positive.

The probability $\Pi(i,j)$ for a measurement outcome $i\in\{1,\dots,n_A\}$ for Alice and $j\in\{1,\dots,n_B\}$ for Bob is given by expression (\ref{eq:elemprob}) of Section~\ref{sec:first}. In this case, $P$ and $Q$ are composite probes and there is no explicit appearance of a boundary condition since combining the regions of Alice, Bob and the exterior does not leave any external boundary, compare Subsection~\ref{sec:pfcd}. Thus we get,
\begin{equation}
\Pi(i,j)=\frac{W\pcomp M_{A,i}\pcomp M_{B,j}}{W\pcomp M_{A}\pcomp M_{B}} .
\label{eq:prob_indef_caus}
\end{equation}
Choosing an orthonormal basis $\{\xi_{A,\text{in}}^k\}_{k\in I_{A,\text{in}}}$ of the state space $\bcg_{A,\text{in}}$ and correspondingly for the other states spaces, we can write all probes in terms of (multi-)matrices. For the probe $W$ the corresponding matrix can be written as,
\begin{equation}
\hat{W}^{k l m n}=\lv W,\xi_{A,\text{in}}^k\tens\xi_{A,\text{out}}^l\tens\xi_{B,\text{in}}^m\tens\xi_{B,\text{out}}^n\rv ,
\end{equation}
and similarly for the other probes. Expression (\ref{eq:prob_indef_caus}) for the probability $\Pi(i,j)$ can then be written as a contraction of matrices, recall the example in Subsection~\ref{sec:pf_example}. $\hat{W}^{k l m n}$ recovers precisely the \emph{process matrix} that Oreshkov, Costa and Brukner invented to describe the present example outside of the standard formulation. (They also imposed an extra condition on $\hat{W}^{k l m n}$ that amounts to the denominator in expression (\ref{eq:prob_indef_caus}) being equal to $1$ for any possible normalization preserving primitive probes $M_A$ and $M_B$.) The main aim of their paper was to start a classification of possible process matrices with respect to their causal behavior. In particular, they derived an inequality that could be violated only if Alice and Bob would be neither spacelike nor timelike separated in the sense of the standard formulation. Remarkably, they found a process matrix violating the inequality, possibly even in a maximal way. We refer the interested reader to the original paper.

In the context of fixed global causal structure we can of course also have bidirectional signaling between two adjacent regions if these are separated by a \emph{timelike hypersurface}. A state space for a timelike hypersurface was introduced in quantum field theory for the first time in \cite{Oe:timelike}. It plays a prominent role in a novel version of the S-matrix, where particles (and thus quantum information) flow into and out of the interior of spacetime through an asymptotic timelike hypercylinder \cite{CoOe:smatrixgbf}. All this is done in the amplitude formalism described in Section~\ref{sec:quantum} of the present work, see in particular the remarks at the end of Subsection~\ref{sec:qbdymeasure}. It carries over to the positive formalism via the modulus-square functor, compare Subsection~\ref{sec:qstat}.

\section{Refinements}
\label{sec:refine}

\subsection{Instruments and expectation values}
\label{sec:instruments}

In the development of the positive formalism as well as in examples we have mostly focused on probes that implement measurements with two or a finite number of possible outcomes. However, outcomes of realistic measurements often lie in a continuum. We have covered this case to some extent with the non-primitive probes that give rise to expectation values, compare Subsection~\ref{sec:peval}. However, this represents only one rather limited way to implement such measurements. Also, our discussion lacked precision. We shall remedy these defects in the present subsection.

In the standard formulation of quantum theory, there is a notion of \emph{instrument} that formalizes measurements with outcomes in arbitrary measurable sets \cite{DaLe:opapquantprob}. It is quite straightforward to adapt this notion to the positive formalism. Let $M$ be a region, $X$ be a set and $\cN$ a $\sigma$-algebra of subsets of $X$.

\begin{dfn}
An \emph{instrument} is a map $\cI:\cN\to\prp_M$ with the following properties:
\begin{itemize}
\item $\cI(\emptyset)=0$.
\item Given a sequence $\{E_n\}_{n\in\N}$ of disjoint elements of $\cN$ and any $b\in\bc_{\partial M}$ we have:
\begin{equation}
\lv\cI(\bigcup_{n\in\N} E_n),b\rv_M=\sum_{n\in\N} \lv\cI(E_n),b\rv_M
\label{eq:countadd}
\end{equation}
\end{itemize}
\end{dfn}

Note that both sides of the equation (\ref{eq:countadd}) are positive, but may be infinite, recall Subsection~\ref{sec:infinidim}. For fixed $b\in\bc_{\partial M}$ we define $\cI_b:\cN\to [0,\infty]$ by $\cI_b(E)\defeq\lv \cI(E),b\rv_M$. This is then an ordinary measure on $(X,\cN)$.

The notion of instrument also comes with the idea that $\cI(X)$ is \emph{non-selective}, while $\cI(E)$ for $E\subset X$ is \emph{selective}. In the general positive formalism we have no mathematical condition that imposes non-selectiveness. In the time-evolution version (Section~\ref{sec:evolution}) however, this can be implemented by requiring the probe map of $\cI(X)$ to be \emph{normalization preserving}. This is also how non-selectiveness is implemented in the conventional definition of instrument.

The probability $\Pi(E)$ for the outcome of a measurement to lie in the set $E\subseteq X$ given boundary condition $b\in\bc_{\partial M}$ is,
\begin{equation}
\Pi(E)=\frac{\lv\cI(E),b\rv_M}{\lv\cI(X),b\rv_M} .
\end{equation}
For the case $X=\R$ we can also give a precise construction of the non-primitive probe $P$ that yields the expectation value of the measurement. Given $b\in\bc_{\partial M}$, this is determined by the integral,
\begin{equation}
 \lv P,b\rv_M=\int_X x\, \xd\cI_b(x) .
\end{equation}

The previous considerations also suggest a hybrid, i.e., partly classical, partly quantum device. This generalizes both a quantum-classical and a classical-quantum channel \cite{Hol:statquant}.
Define  $\bcg_{c,\partial M}$ to be the space of real valued measurable functions on $X$. By adding a measure $\mu$ on $(X,\cN)$ this becomes a partially ordered (unbounded) inner product space, compare Proposition~\ref{prop:cwpoipexa} of the appendix. We enlarge the space of generalized boundary conditions of $M$ by this classical component. That is, the new space of generalized boundary conditions is $\bcg_{\partial M}\tens\bcg_{c,\partial M}$. We define the primitive probe $Q$ in $M$ by the equation, for any $b\tens c\in \bcg_{\partial M}\tens\bcg_{c,\partial M}$,
\begin{equation}
 \lv Q,b\tens c\rv_M=\int_X c(x)\, \xd\cI_b(x) .
\end{equation}
Note that $\cI$ can be recovered from $Q$ as follows. For a measurable set $E\subseteq X$ let $\chi_E$ denote its characteristic function. Then,
\begin{equation}
\lv \cI(E),b\rv_M=\lv Q,b\tens\chi_E\rv_M .
\end{equation}
It turns out that both hybrid devices employed in the quantum teleportation example of Subsection~\ref{sec:qteleport} are of this type. We leave the details to the reader.

\subsection{Quantum field theory}
\label{sec:refqft}

In the realm of quantum physics the most comprehensive and successful theories that we use to describe nature at a fundamental level are quantum field theories. For a version of the positive formalism to be taken seriously as a foundation of quantum theory it is thus crucial that it may serve as a basis for quantum field theory at least as well, but preferably better, than the standard formulation of quantum theory. While there is ample evidence for this (see Section~\ref{sec:quantum}), the amplitude formalism (and thus potentially also the positive formalism) as presented in Subsections~\ref{sec:tqft} and \ref{sec:qobs} does not yet provide a satisfactory axiomatization of quantum field theory. Some necessary refinements were left out for simplicity, others remain to be addressed in future work. We provide a brief discussion of some relevant issues.

The fact that amplitude maps are generically unbounded and defined on dense subspaces of state spaces only was mentioned at the end of Subsection~\ref{sec:qlocpath} and has been satisfactorily addressed in the literature \cite{Oe:holomorphic,Oe:affine,Oe:feynobs}. Another place where a weakening of the axioms is in order turns out to be the gluing rule (\ref{eq:qglueaxb}) of Axiom (T5b). Here, an additional scaling factor depending only on geometrical data needs to be introduced \cite{Oe:holomorphic,Oe:affine,Oe:feynobs}. This is called a \emph{gluing anomaly} \cite{Tur:qinv}. As a consequence, the positive formalism also acquires a gluing anomaly \cite{Oe:dmf}.

In standard quantum field theory the choice of \emph{vacuum} plays an important role in the construction of the Hilbert space of states. This vacuum encodes asymptotic boundary conditions of fields at infinity and is thus a global object that cannot be satisfactorily localized on pieces of hypersurfaces. Consequently, the standard construction of Hilbert spaces of states as well as the hypersurface decomposition rule of Axiom (T2) does not generalize to bounded hypersurfaces. One may avoid this issue by using \emph{local vacua} on hypersurfaces, recovering also Axiom (T2).\footnote{In geometric quantization the vacuum is encoded on a hypersurface in terms of a \emph{complex structure} \cite{AsMa:qfieldscurved}. Replacing a global vacuum with a local vacuum means then replacing a non-local complex structure with a local one.} However, the issue of relating the local vacua to a \emph{global vacuum} remains. A fully satisfactory solution of this issue may require replacing Hilbert spaces with a more suitable notion of state space. It might also be convenient to address this issue directly at the level of mixed state spaces, i.e., at the level of the positive formalism, see also Subsection~\ref{sec:fermions}.

Related to this, there is usually a notion of \emph{vacuum state} as a distinguished element of the Hilbert space of states in quantum field theory. With one Hilbert space per hypersurface in the amplitude formalism the natural question arises what plays the role of this vacuum state. It turns out to be natural to define one such vacuum state per hypersurface and relate these states in an axiomatic way through \emph{vacuum axioms} \cite{Oe:gbqft,Oe:holomorphic,Oe:affine}. These axioms can be straightforwardly translated into the positive formalism.

A more severe modification of the hypersurface decomposition rule (Axiom (T2) or (P2)) arises in the presence of \emph{gauge symmetries} in field theory. Classically, one associates there to hypersurfaces germs of solutions modulo gauge transformations. However, these gauge transformations need to be fixed on the boundaries of the hypersurface (also known as \emph{corners}). As a consequence, if the gauge symmetry is non-abelian, there are fewer gauge transformations on a hypersurface decomposed into pieces than on the undecomposed hypersurface. Upon quantization this results in the tensor product of the Hilbert spaces of the pieces to be larger than the Hilbert space of the undecomposed hypersurface. The map $\tau$ in Axiom (T2) is then no longer an isomorphism, but a \emph{partial isometry}. This is well understood in 2-dimensional quantum Yang-Mills theory \cite{Oe:2dqym}. The situation might be more complicated in higher spacetime dimensions. A corresponding modification of Axiom (P2) arises both in classical gauge theory as well as in quantum gauge theory. In the latter case this is induced by the functorial transition from the amplitude to the positive formalism.

Recall that in Subsection~\ref{sec:pfcd} we considered a version of the positive formalism where the notion of spacetime hypersurface is abstracted, replaced by a mere relation (represented by a \emph{link}) between \emph{processes} (which in turn abstract spacetime regions). It is at this point that we can clearly see that this abstraction, while completely compatible with the positive formalism as presented in Section~\ref{sec:first}, is untenable in general. It is precisely the refinements of Axioms (T2) and (P2) with their inclusion of lower dimensional topological information (in terms of \emph{corners}) that present an obstruction to this abstraction. It is thus only in sufficiently simplified models of physics that we may expect this abstracted version to hold. For more fundamental models, in particular involving quantum field theory or gauge symmetries we need the additional structure provided by the spacetime system.

\subsection{Fermions}
\label{sec:fermions}

In the quantum realm we have so far not made any distinction between bosonic and fermionic degrees of freedom. In fact, all of our treatment up to this point applies in the presented form only to bosonic quantum theory. In fermionic quantum theory the Hilbert spaces of states become \emph{graded}. They decompose into a direct sum of an even and an odd sector, depending on the number of fermions. Associated to this is a \emph{superselection rule} that imposes the preservation of these sectors \cite{WiWiWi:parity}. This is obeyed in particular by the unitary time-evolution operator. Correspondingly, the amplitude map becomes a \emph{graded} map. This simply means that it vanishes on the odd part of the state space. Implementing the grading in Axioms (T1) and (T4) of topological quantum field theory (Subsection~\ref{sec:tqft}) is straightforward.
The ``anti-commuting'' nature of fermions becomes evident when considering Axiom (T2). Here, we need to postulate that the $\tau$-maps for different orderings of the same decomposition are related by a minus sign if the permutation relating the orderings is odd.
It turns out that there is one more place where a consistent fermionic theory needs a grading. This is the map $\iota$ of Axiom (T1b). Rather than a conjugate linear isometry this becomes a conjugate linear \emph{graded isometry}. This means in particular that the map $\iota$ inverts the sign of the inner product on the odd sector of the state space. This has far reaching consequences. State spaces are no longer Hilbert spaces, but \emph{Krein spaces}. These are complete inner product spaces that can be decomposed into a direct sum of a positive-definite and a negative-definite part.

The reader might wonder why only Hilbert spaces, but not Krein space are mentioned in text book treatments of fermionic quantum (field) theory. The reason is that the hypersurfaces relevant for the construction of \emph{the} state space in the standard formulation of quantum theory are exclusively \emph{spacelike} hypersurfaces. On these the standard fermionic theories (such as the Dirac theory) turn out to lead to state spaces with definite signature. Moreover, there is need for only one global choice of time orientation for all spacelike hypersurfaces making the signatures on them equal. Positive signature is the conventional choice.

The amplitude formalism of fermionic quantum theory was developed and justified in detail in the paper \cite{Oe:freefermi}. This also includes a corresponding treatment of ``classical'' fermionic theory and a method of quantization. Moreover, the rule (\ref{eq:probbdy}) for extracting probabilities of boundary measurements is adapted to the fermionic case by enforcing superselection rules. The transition to the statistical setting is achieved in essentially the same way as in the bosonic setting, but suitably taking into account gradings and superselection rules. A first version of the positive formalism for fermionic theories was provided in the paper \cite{Oe:dmf}. However, a flaw of that version was that spaces of generalized boundary conditions had to be taken to be complex rather than real. A real and more definitive version was provided in the paper \cite{Oe:locqft}. The latter paper also proposes for fermionic theories a partial solution of the problem of global vacua (see Subsection~\ref{sec:refqft}) by making use of the positive formalism. A generalization of the notion of \emph{probe} to the fermionic setting has not been formally written down, but is straightforward.

\subsection{Abstracting boundaries}
\label{sec:absbdy}

From a strictly operationalist perspective the prominent role that boundary conditions play in the development of the positive formalism as laid out in Section~\ref{sec:first} might seem less than satisfactory. Rather than introducing boundary conditions early on it might be more desirable to derive them as auxiliary objects with the probes taking the primary role. For example, although the composition map for probes is mentioned already in Subsection~\ref{sec:probes}, just before the introduction of boundary conditions, a concrete form of this map is then derived using pairings with boundary conditions in Subsection~\ref{sec:probecomp}. Here one might try to take instead the composition maps of probes as fundamental and use these to derive the spaces of boundary conditions. As for the parings used to predict probabilities as in expression (\ref{eq:elemprob}), we can here replace boundary conditions by probes associated to all of the exterior as a region, compare Subsection~\ref{sec:pfcd}.

In the amplitude formalism for quantum field theory the analogue of taking the composition map for probes as fundamental would be taking the composition map for observables as fundamental (compare Subsection~\ref{sec:qobs}). This would bring the formalism closer to the \emph{factorization algebra} approach to quantum field theory, where this composition map is the central structure \cite{CoWi:facalgqft1}.

It is natural to wonder whether we can abstract or generalize the spacetime setting of the positive formalism. Spacetime regions serve to distinguish the physics in the interior from the exterior in order to be able to focus on the former. Are there other useful ways that we can distinguish part of the physics from the rest? As in Subsection~\ref{sec:pfcd} lets call such a part a \emph{process}, although \emph{subsystem} seems also somewhat appropriate. Correspondingly, hypersurfaces are the arena for communication or interaction between regions. Let us refer to possible generalizations as \emph{interfaces} between \emph{processes}. Depending on the context it is these interfaces rather than the processes that may be thought of as \emph{subsystems}. Thus, in the spacetime setting, regions are the carriers of processes while hypersurfaces are the carriers of interfaces. As already remarked in Subsection~\ref{sec:pfcd} it is straightforward to consider a version of the positive formalism where there is no mention of spacetime at all. The axioms (Subsection~\ref{sec:paxioms}) would be formulated purely in terms of processes and interfaces, with the tensor product rule of Axiom (P2) for combining interfaces. There would be no bounds on what the physical implementation of these notions could be. While this ``completely abstract'' positive formalism appears quite appealing at first sight it also has some serious limitations. As already mentioned in Subsection~\ref{sec:pfcd} the locality principle is very valuable in constraining the possible communications or interaction between processes. What is more, as discussed in Subsection~\ref{sec:refqft} we know this abstract setting to fail in quantum field theory (and probably also in classical gauge field theory). This is  because it misses lower dimensional data of the spacetime setting on the connectivity of the hypersurfaces, i.e., on \emph{corners}.

It is instructive to recall the place of the standard formulation of quantum theory in this light. This corresponds to the setting described in Subsection~\ref{sec:evolcompsys}. We start with a spacetime that has a global time variable and use the latter to foliate spacetime into spacelike hypersurfaces. We then restrict the spacetime system to these particular hypersurfaces and the regions that are bounded by pairs of them. After this ``time rigidification'' we abstract space and allowing interfaces (here also identified as subsystems) to be combined in ``space'' freely via the tensor product of Axiom (P2). The present perspective makes it particularly clear how an absolute notion of time is baked into the standard formulation of quantum theory and why it is necessary to go to spacetime in order to remove it.

The question remains as to the prospect of other useful arenas where processes and interfaces encode neither the spacetime system nor the absolute time/abstract space setting, but carry more structure than the purely abstract setting. An intriguing proposal in this direction was made recently by Hardy in the context of classical general relativity \cite{Har:2015infotheo,Har:opgenrel}. To this end a number of scalar fields are introduced that are local functions of the metric and possibly of additional fields. Then the \emph{WS-space} is introduced as the space of values of these fields, i.e., this is in general some $\R^n$. (``WS'' stands for Westman-Sonego \cite{WeSo:obsgeninv}.) The scalar fields could encode quantities that are directly measurable, making WS-space operationally much more accessible than spacetime. A solution of general relativity (and other fields) then gives rise to a submanifold in this WS-space. If the scalar fields are sufficiently regular, moving a small distance in spacetime should correspond to moving a small distance in this submanifold in WS-space. That is, a notion of spacetime locality should be inherited by WS-space. Also, WS-space is inherently invariant under gauge transformations. The core of the proposal is now to replace spacetime regions and hypersurfaces with regions and hypersurfaces in WS-space as the arena for processes and interfaces. This appears to be much more promising than setting up the positive formalism directly in spacetime for obtaining a manageable and useful description of classical statistical general relativity. This is also intended as a possible stepping stone towards an approach to quantum gravity.

\section{Conclusions and discussion}
\label{sec:discussion}

The \emph{positive formalism} presented in this paper as a novel framework for physical theories arises from a convergence of and acts as a unifier for different ideas and approaches in diverse fields. Thus, the present paper (or parts thereof) can be read from different perspectives and for various different purposes.

In one reading the emphasis is on the positive formalism as a first-principles approach to physics. The focus is thus on the derivation in Section~\ref{sec:first} of the positive formalism from the simple principles of \emph{locality} and \emph{operationalism} with only a minimum of empirical input. It is particularly remarkable that the \emph{composition law} as expressed in equation (\ref{eq:bincompid}) or Axioms (P5a) and (P5b) can be obtained in this way. This arose originally in a completely different way, as an abstraction of the path integral in quantum field theory, see Subsection~\ref{sec:qcomppi}. In fact, the full axiomatic system of the positive formalism (Subsection~\ref{sec:paxioms}) closely mirrors that of \emph{topological quantum field theory} (Subsection~\ref{sec:tqft}) which arose as an abstraction of quantum field theory.

Another reading focuses on the \emph{convex operational framework} or framework for \emph{generalized probabilistic theories}. This is recovered in Section~\ref{sec:evolution} from the positive formalism by specializing to a time-evolution setting. It requires as additional ingredients the notion of a \emph{state of maximal uncertainty} and \emph{normalization conditions} that implement \emph{causality}. This derivation from an originally spacetime local framework sheds new light on the convex operational framework. Thus a necessary self-duality identifying states and effects (final states) arises as well as a uniform treatment of composition, be that of different systems or of one system at different times, as anticipated in the literature (see references in Section~\ref{sec:qcomp} for the quantum case). This also results in a notion of \emph{boundary positivity} for dynamical maps that is stronger in general than mere \emph{positivity} and recovers in the quantum case precisely \emph{complete positivity}. Note that the requirement of complete positivity in quantum theory is conventionally obtained in a rather different way, by considering the possible dynamics induced by coupling to another system. The original spacetime perspective also clarifies the notion of \emph{state}. A state is thus just a special kind of \emph{boundary condition} (in the language of Section~\ref{sec:first}), parametrizing possible interactions between adjacent spacetime regions, in this case the past and the future. It is only thanks to the peculiar feature of a dynamical one-to-one correspondence between data on different spacelike hypersurfaces in our usual physical theories that we can use states as records of the past (or future if we so choose). The evolution of a state when a measurement is performed may thus be viewed as a Bayesian updating of this record, see Subsection~\ref{sec:sevolbayes}.

We focus now on the two ingredients of the convex operational framework that we did not obtain in Section~\ref{sec:evolution} from the spacetime version of the positive formalism. One is the notion of \emph{state of maximal uncertainty} (Subsection~\ref{sec:pou}). This can be generalized in an obvious way to the spacetime version of the positive formalism by postulating such a state in every space of boundary conditions, behaving under composition as prescribed by equation (\ref{eq:tpou}). The reason we choose not to introduce this notion in the spacetime setting was the apparent lack of need for it. In contrast, the time-evolution setting exhibits a clear utility for this notion in describing a large class of measurements. The second ingredient are the normalization conditions on probe maps that implement forward causality (Subsection~\ref{sec:causnorm}). As explained at the end of Subsection~\ref{sec:evolcompsys} these cannot be lifted to the spacetime version of the positive formalism as their consistency depends on a partial order of probes (usually) with respect to a background time.
However, at least when modeling fundamental physics, the dynamics in the form of the \emph{null probes} should contain any causality properties the theory in question might exhibit, without the need for additional normalization conditions. Indeed, this is precisely the case in quantum field theory when encoded in the \emph{amplitude formalism} (Section~\ref{sec:quantum}). There remains a question what primitive probes modeling measurement processes should be allowable in general, but it is plausible to think that this may depend on the theory under consideration. Conversely, the normalization conditions in the time-evolution version of the positive formalism as in the standard formulation of quantum theory should perhaps be thought of as capturing a key feature of the underlying microphysics without imposing the need to model the latter in detail.

Classical physics is put at the center in another reading of this work focusing on Section~\ref{sec:classical}. The implementation of our present notion of locality leads to an axiomatization of classical field theory (Subsections~\ref{sec:caxioms} and \ref{sec:cobs}) that is not in terms of sections of vector bundles (fields) and partial differential equations (PDEs), but spaces of local solutions associated to regions and hypersurfaces. This has proved useful in (and indeed was developed for) the context of \emph{quantizing} a classical theory with the target quantum theory formulated in the \emph{amplitude formalism} (Subsections~\ref{sec:tqft} and \ref{sec:qobs}) \cite{Oe:holomorphic,Oe:affine,Oe:feynobs,Oe:freefermi}. This axiomatization poses a very interesting mathematical question: Under what circumstances or with which additional data can a classical field theory in conventional form (bundles, sections and PDEs) be reconstructed from this axiomatic data? Generalizing from ordinary to statistical field theory in this formulation leads to the positive formalism (Subsections~\ref{sec:cstat} to \ref{sec:cgenprobes}). Note that the reconstruction of the ordinary theory (in terms of the Axioms of Subsection~\ref{sec:caxioms}) from the statistical theory (in terms of the Axioms of Subsection~\ref{sec:paxioms}, without non-null probes) is comparatively straightforward. Roughly speaking, the elements of a solution space are recovered from the space of statistical distributions over it as the \emph{pure} or \emph{extremal} elements, that is, the \emph{delta-distributions}.

While the traditional time-evolution framework will be sufficient for most purposes in handling statistical field theory there are potentially interesting situations which are outside its scope. Indeed, special relativistic generalizations of notions of statistical field theory are poorly explored, while general relativistic generalizations are basically non-existent. The latter involve the famous problem of how to compare different spacetimes (metrics), let alone integrate over some set of them. With the positive formalism a statistical treatment of general relativity or other theories with a dynamical metric could be attempted at least in principle. Thus we would have to consider the spaces of solutions of the Einstein equations in regions and on hypersurfaces that are differentiable manifolds, in the simplest case the 4-ball. Crucially we would limit these manifolds to be compact. Then we would consider the corresponding spaces of statistical distributions etc. Of course this still sounds like a formidable task and one could likely only make headway with severely restricted classes of solutions. Also diffeomorphisms will enter as gauge symmetries (see Subsection~\ref{sec:refqft} for some comments on gauge symmetries in the quantum case). There is a potentially much more accessible approach at a statistical treatment of general relativity or similar theories, however. This is via a version of the positive formalism with suitably modified notions of \emph{region} and \emph{boundary} proposed by Hardy, see Subsection~\ref{sec:absbdy}.

In a further reading this paper is about a \emph{generalization} of the \emph{standard formulation} of quantum theory. As such it provides a milestone in the programme of the \emph{general boundary formulation} of quantum theory. This reading is focused on Section~\ref{sec:quantum}, part of which reviews previous steps in that programme. The key innovation compared to the paper \cite{Oe:dmf} where the positive formalism was first proposed is the generalization of the notion of \emph{quantum operation}, see in particular Subsection~\ref{sec:qmesop}. With this ingredient the positive formalism incorporates all relevant aspects of quantum measurement theory, the latter being laid out and clarified through Sections~\ref{sec:first} and \ref{sec:evolution}. The general boundary formulation may thus be seen to provide a formulation of quantum theory more fundamental than the standard one.
Of particular relevance is its \emph{timelessness} manifest in both the amplitude and the positive formalism. That is, in contrast to the standard formulation no predetermined notion of time (but merely a weak notion of spacetime) is necessary to make sense of the formulation and establish its relation to the real world.
Crucially, this removes the apparent incompatibility between general relativity where the split between space and time is dynamical and quantum theory which traditionally has been identified with its standard formulation that depends on a fixed notion of time. Hopefully, this will clear the way for a fresh attack on quantum gravity with a firmer conceptual basis, compare also the respective comments in the introduction and below.

Reflecting from the present perspective on the notion of \emph{quantization} brings into focus a curious fact. In the standard formulation, a quantization prescription usually converts a classical theory expressed in terms of a phase space and observables as functions into a quantum theory in terms of a Hilbert space and observables as operators. This is illustrated by the downward arrow on the left hand side of Figure~\ref{fig:found_scheme_evol}. However, the classical and the quantum theory are conceptually and mathematically much closer to each other in the statistical setting. This is the convex operational framework depicted on the right hand side in Figure~\ref{fig:found_scheme_evol}. One might thus expect that it would be more natural to set up quantization prescriptions directly within this framework. One the level of state spaces elements of such a direct route exist in the literature. For example, in \emph{algebraic quantum field theory (AQFT)} one may start with a classical algebra of observables, deform this into a non-commutative $*$-algebra and then declare the state space to be the space of positive functionals on this algebra. This is then directly the space of mixed states.\footnote{There are important differences to our present setting. Most importantly, the algebra should be associated to a hypersurface rather than to a region as in AQFT. On the other hand, the \emph{time slice axiom} of AQFT itself points to the possibility of replacing the region with a hypersurface.} Also note that the (undeformed) algebra of observables is more or less the same as the algebra of statistical distributions. On the other hand, it seems that a direct route from classical observables to quantum operations has not been explored at all in the literature.

The interest in a direct quantization route in the convex operational framework is somewhat limited by the fact the usual quantization prescriptions followed by the transition from the Hilbert space to the density operator framework work well enough for most purposes. It should be mentioned, however, that this transition (indicated by the arrow labeled \emph{modulus-square functor} in Figure~\ref{fig:found_scheme_evol}) is functorial on state spaces and time-evolution operators, but not on observables. However, there is a standard procedure for converting observables to quantum operations ``by hand'' using the spectral decomposition (as exhibited in Subsection~\ref{sec:qmesop}). The situation with respect to quantization looks similar for the space-time frameworks depicted in Figure~\ref{fig:found_scheme_st}, except that it is much less explored in general. Again, the modulus-square functor (indicated in Figure~\ref{fig:found_scheme_st}) only transports spaces of boundary conditions (state spaces) and their dynamics, but not observables. In this case, however, there is no analog of a spectral decomposition and thus no general way to convert a local observable into a local probe. This makes the necessity of a quantization prescription working directly within the positive formalism much more urgent.

In this respect the origin of the composition rule in the amplitude formalism from the quantization via the path integral (Subsection~\ref{sec:qcomppi}) is suggestive. Is there a path integral for the positive formalism which has a similar composition rule (Subsection~\ref{sec:qstat})? Of course, using the modulus-square functor (Subsection~\ref{sec:qstat}), the probability maps of the positive formalism can be obtained as double path integrals over two copies of configuration space on the boundary. However, one might hope for a single path integral, perhaps over the first jet bundle rather than over configurations.

The positive formalism opens a new perspective in particular for quantum field theory. On the one hand, the Segal approach of encoding quantum field theory as a TQFT, i.e., in the amplitude formalism, still faces significant challenges. In particular, for theories of physical interest a sufficiently general implementation of state spaces on hypersurfaces that have boundaries has not yet been achieved. It has been suggested and in fact demonstrated for free fermions that this problem can be at least partially solved by working in the positive formalism instead \cite{Oe:locqft}. On the other hand, the positive formalism equips quantum field theory with a notion of \emph{local measurement} via \emph{probes} (Subsection~\ref{sec:qmesop}), going far beyond the limitations of the S-matrix. This should impact in particular the field of \emph{relativistic quantum information theory}. Of course, local measurements have been discussed previously in quantum field theory. However, this had to be done either by recurring to the standard formulation and thus loosing manifest locality (e.g.\ \cite{Ste:partloc}) or by explicitly modeling the measurement device as an additional quantum system (e.g.\ the Unruh-DeWitt detector \cite{Unr:bhevap,Dew:qgsynth}). Constructing probes that model local measurements in quantum field theories is a completely new challenge and it is not clear what is the best way to accomplish this. Given that we have local observables in quantum field theory, one way would be to try to apply the recipe from the standard formulation that converts observables into quantum operations. That is, as already mentioned above, we use the spectral decomposition of the observable expressed as a hermitian operator on the standard Hilbert space (compare Subsection~\ref{sec:qmesop}). However, for the standard field operators this is not quite straightforward due to their continuous spectrum and unboundedness and any regularization to make this work might mess with locality. Also, it remains to reexpress the obtained quantum operation in a manifestly local way, i.e., restrict it to a compact region. A more desirable way might be a direct quantization prescription as discussed above.

The possibility to model within the convex operational framework hybrid theories that have both classical and quantum components is well known. An example are the notions of \emph{classical-quantum} and \emph{quantum-classical channels} in quantum information theory (compare also Subsection~\ref{sec:qteleport}). The positive formalism offers the novel possibility to go beyond this non-relativistic setting and construct hybrid theories in a genuinely field theoretic setting. No recipe for doing this exists as yet, but an understanding of direct quantization within the positive formalism would most likely help, compare previous remarks.

It is worth noting that key concepts that later became part of the general boundary formulation have influenced thinking on \emph{quantum gravity} for some time. Recall from Section~\ref{sec:quantum} that \emph{topological quantum field theory (TQFT)} arose at the end of the 1980s as an abstraction of quantum field theory. This was notably the origin of the notion of \emph{locality} as used in the present paper. In the wake of these developments Witten showed that quantum gravity in $2+1$ dimensions can be formulated as a TQFT, which is moreover a solvable model \cite{Wit:gravsolv,Wit:topchange3d}. This sparked considerable interest in trying to construct also $3+1$-dimensional quantum gravity as a TQFT. At the same time the constructions of TQFTs in terms of state-sum models emerged as an approach to topological invariants in low dimensions \cite{TuVi:inv3}. It was then realized that the much older work of Ponzano and Regge \cite{PoRe:limracah} could be interpreted  as such a state-sum approach to $3$-dimensional quantum gravity. Ooguri formulated 4-dimensional $BF$-theory as a TQFT in terms of a state-sum model \cite{Oog:toplat}. Since gravity in 4 dimensions can be obtained from $BF$-theory by imposing additional constraints, this model has been taken as a starting point for developing state-sum approaches to quantum gravity. It was also realized that such state-sum models, more specifically called \emph{spin foam models} \cite{Bae:spinfoams} also arise in \emph{loop quantum gravity} \cite{Rov:qg,Thi:canqgr}. However, the physical interpretation of these models is completely different in the latter case, where they are used to calculate a Hamiltonian constraint. Recall that loop quantum gravity is an approach based on the standard formulation of quantum theory, as already mentioned in the introduction. Using spin foams in the sense of the path integral and TQFT as models for quantum gravity is thus also called the \emph{spin foam approach} to quantum gravity \cite{Per:sfqgrav}.

Recall from Subsection~\ref{sec:tqft} that TQFT in itself is merely a mathematical framework without any rules for extracting measurable quantities. Thus, none of these models based merely on TQFT has any physical content until such rules are specified. As already discussed in the introduction, the operational framework of the standard formulation (including transition probabilities and expectation values of observables) is not applicable and neither is the S-matrix of quantum field theory. Precisely this situation provided a major motivation for the development of the general boundary formulation \cite{Oe:catandclock}. Thus, in 2005 the notion of boundary measurement as reviewed in Subsection~\ref{sec:qbdymeasure} was introduced \cite{Oe:gbqft}, generalizing transition probabilities of the standard formulation to a setting without background time. Since this does not require any additional mathematical structure it provides at least one notion of measurement to this large class of models. For very brief remarks on how this might be used to describe scattering processes in quantum gravity see \cite{Oe:probgbf}. While a notion of observable analogous to that of quantum field theory can also be introduced (Subsection~\ref{sec:qobs}), its potential use for extracting measurable quantities in quantum gravity is much more limited than in quantum field theory as we have explained in Subsection~\ref{sec:qexpval}.

The results of the present work suggest a different route: Encode quantum gravity in the positive formalism. Fortunately, it is quite straightforward to transport models from the amplitude formalism, i.e., based on TQFT, to the positive formalism. This is done via the modulus-square functor as described in Subsection~\ref{sec:qstat}. Thus, a large class of models becomes available essentially immediately. However, these models have to be considered incomplete now as it remains to construct for them relevant \emph{probes} or \emph{generalized quantum operations} that encode local physical measurements. One would hope that these could be obtained by some kind of \emph{quantization prescription}, see previous comments.

\emph{Symmetries} have always played an important role in physics. In the present context it is natural to ask in particular for the potential role of spacetime symmetries that act on a \emph{spacetime system} (Subsection~\ref{sec:spacetime}). In the amplitude formalism these have played a very important role already for some time in \emph{conformal field theory}, in particular in Segal's approach \cite{Seg:cftproc,Seg:cftdef}. A basic and generic discussion of spacetime symmetries leading to representations on spaces of states and amplitudes can be found in Section~6 of \cite{Oe:gbqft}. This can be transferred almost one-to-one to the positive formalism.

In another reading of this work the focus is on viewing the development of the quantum version of the positive formalism as a partial \emph{reconstruction of quantum theory}. Thus Section~\ref{sec:first} represents a derivation of most of the structure of quantum (but as it turns out also classical) theory from first principles. The only additional purely ``quantum'' ingredient concerns the precise nature of the partially ordered vector spaces of boundary conditions. Adhering strictly to the standard formulation of quantum theory we have to take these to be the spaces of self-adjoint operators on complex Hilbert spaces. If we are more ambitious we might try something more general, such as more general anti-lattices, recall Subsection~\ref{sec:qchar}. In any case, adding the postulate of such a mathematical structure to complete the reconstruction seems rather ad hoc. An operational distinction of quantum theory would be much preferable. In contrast, almost all the structure including the axioms governing the dynamics (Subsection~\ref{sec:paxioms}) as well as the rules for predicting measurable quantities appear already in Section~\ref{sec:first}. In particular, there is no need to postulate the \emph{Born rule}, \emph{Lüders rule} etc. These all come out given that we add the mentioned ``quantum'' ingredient. Section~\ref{sec:quantum} then serves mainly to verify that we really obtain quantum theory and recover its standard formulation.

With respect to singling out quantum theory among general probabilistic theories the difference between working in the general positive formalism of Section~\ref{sec:first} or in the convex operational framework of Section~\ref{sec:evolution} appears to be of minor significance. After all, the state spaces of the latter, whose detailed structure is crucial here, are merely special instances of the spaces of boundary conditions of the former. Thus, one might hope to import approaches from the considerable literature on operational reconstructions of quantum theory, which are mostly formulated in the convex operational framework. Instead of providing a long list of references we point the interested reader to the recent book \cite{ChSp:qinfofound} and references therein.

With the positive formalism, the general boundary formulation may be claimed to provide a formulation of quantum theory that is more general and fundamental than the standard formulation. It is thus natural to ask if it implies any new insights into the nature of quantum theory and its interpretation. The implications seem to be mostly negative. That is, it appears to provide much clearer indications as to what quantum theory is not (but might have been given merely the standard formulation) rather than what it is. Let us start with the ``collapse of the wavefunction''. Firstly, the observation of single definite measurement outcome is nothing mysterious in our framework. To the contrary, this is part of the very definition of a measurement process in an operational setting. This applies equally to classical and quantum physics. Secondly, the ``instantaneous'' modification or ``collapse'' of a state in a measurement arises merely as a Bayesian updating of knowledge as explained in Subsection~\ref{sec:sevolbayes}. Again, there is no difference in principle between quantum and classical (statistical) physics here, except that in the latter case one can interpret this knowledge as knowledge about a ``true state'' (classical solution, point in phase space). Up to this point, our comments are equally applicable within the long established convex operational framework and as such nothing new. In the face of this, one might still be tempted to not exclude the possibility that when restricted to pure states the collapse might represent an actual physical process. However, in the positive formalism no such collapse can be associated with a measurement process in general. The collapse, even as a Bayesian updating, is entirely an artifact of the \emph{choice} to describe dynamics and measurement in our theories exclusively in terms of states that evolve between spacelike hypersurfaces. There is nothing that collapses in the general measurement formulas (\ref{eq:elemprob}) or (\ref{eq:elembdy}). Of course, we can always apply Bayesian reasoning and updating when considering multiple measurements taking place. However, although the measurements might be precisely localized in spacetime, the updating or ``collapse'' of our knowledge as we take into account additional measurement results cannot in general be localized in spacetime in any sensible way. To put it differently, the existence of a physical collapse would definitely falsify the positive formalism as a framework for quantum theory, as well as the amplitude formalism and probably also much of standard quantum field theory.

To further illustrate the absurdity of the physical collapse idea in the present framework consider an alternative ``space-evolution'' scenario. Instead of spacelike we choose parallel timelike hyperplanes parametrized by one spatial coordinate to describe the dynamics of a theory, but otherwise follow closely the development of the time-evolution framework in Section~\ref{sec:evolution}. This is entirely legitimate in the positive formalism (as in the amplitude formalism) and moreover, there are certainly simple quantum field theories whose dynamics do give rise to a one-to-one correspondence between data on parallel timelike hyperplanes \cite{Oe:timelike}. ``Evolution'' is now in one particular spatial direction and space and time labels on the axes in the figures of Section~\ref{sec:evolution} have to be exchanged. Interpreting the Bayesian updating of Subsection~\ref{sec:sevolbayes} as a physical collapse, this would have to take place on a plane in space and at all times. Certainly, nobody would entertain such a proposal.

We proceed to make some comments on interpretations associated with the names \emph{universal wave function}, \emph{relative state} and \emph{many worlds}, originating with the work of Everett \cite{Eve:relstate,Whe:everett}. Central to these interpretations is the claim that no special notion of measurement is necessary, but that states and their unitary evolution are sufficient to provide a complete description of quantum theory. As a first step, any measurement is described not in terms of quantum operations, but by coupling an additional system that describes the measurement device explicitly in terms of unitary dynamics. How this can be done was already shown by von~Neumann \cite{vne:mathgrundquant}. Given an initial pure state, the final state of such a ``measurement'' is a pure state that can be written as a superposition of states, each of which can be identified with the occurrence of one particular measurement outcome. Of course the final state in itself does not contain the information on how it should be decomposed into this particular superposition. Also, extracting probabilities for measurement outcomes still requires special rules distinct from those of unitary evolution. These issues aside (which have been amply discussed in the literature \cite{SBKW:manyworlds}), let us nevertheless think of the summands in this special decomposition of the final state as \emph{branches} of the wave function, describing different alternative worlds. Each measurement makes the branches branch further etc. The statement is then that the world we experience corresponds to one of those branches. Similar remarks apply here as those that we have already made when discussing the \emph{collapse}. The branching has nothing to do with the measurement process itself, but arises merely from a particular choice of describing it in terms of states that evolve in time. Consider by contrast the space-evolution scenario we have outlined above. In this scenario we could equally talk about a branching, except that our world would exhibit branching in a particular \emph{spatial} direction. We could then claim with equal justification that the world we experience corresponds to one of those spatial branches. Certainly, nobody would propose this version as the basis for a sensible approach to quantum theory. What is more, absent a specific temporal or spatial evolution picture, there is no notion of branching attributable to a measurement process at all.

As concerns \emph{hidden variables theories}, these have been shown by Bell \cite{Bel:EPR} to require \emph{non-locality} to be compatible with the violations of Bell's inequalities which have been demonstrated in numerous experiments testing quantum theory.\footnote{There remains a small and more exotic group of local hidden variable theories claimed to be not ruled out by Bell's argument, including \emph{super-determinism} \cite{Hos:testsuperdet}.} While non-locality is compatible in principle with the standard formulation of quantum theory, it is not compatible with Segal locality and thus would contradict the general boundary formulation and the positive formalism more generally. This is independent of the well known conflict of non-locality with special relativity.

On the positive side, an interpretation of quantum theory that embraces the view of state change in a measurement as a process of Bayesian updating (compare Subsection~\ref{sec:sevolbayes}) is \emph{QBism} \cite{FuMeSc:introqbism}.

One lesson to be drawn is that the special role of states, i.e., boundary conditions associated to spacelike hypersurfaces, compared to boundary conditions associated to other hypersurfaces is an artifact of the limitations of the standard formulation. Thus, any interpretation of quantum theory that relies on a special status of states is potentially in conflict with the positive formalism. This is particularly the case if an analogy is drawn between the notion of pure state in quantum theory and in classical theory, where it can be brought into correspondence with a global solution.

With respect to the unresolved issues of quantum theory we do not pretend to offer any new insight with the present work. Rather, quantum theory in the general boundary formulation just inherits these issues from quantum theory in the standard formulation. But perhaps it can contribute to clear the view as to what the issues are and what they are not.
Any operational approach to fundamental physics must face a core issue. That is, a measurement may be modeled in two fundamentally distinct ways. Either we consider it as a means to extract observable quantities from the theory. In this case we use the operational structure of the framework to encode the measurement as an ``external'' intervention. Or we consider the measurement as just any other physical process, i.e., as part of the dynamics of the (fundamental) theory considered. In this case we loose the direct, operational access to the measurement results. This dichotomy is quite unsatisfactory unless we can establish a precise correspondence between these two fundamentally distinct ways of modeling measurement. In a limited sense such a correspondence was described for quantum theory already by von~Neumann \cite{vne:mathgrundquant}. Namely, given a measurement determined by an observable, we can construct an additional system coupled to the original system that acts as a recorder of the measurement outcome in such a way that the joint system evolves unitarily. Later, the implementability of a measurement in terms of a unitarily coupled system was precisely the origin of the requirement of \emph{complete positivity} for \emph{quantum operations} \cite{Kra:statechanges}. The price to pay is that the additional system needs special boundary conditions. The necessity for these boundary conditions is clear when want to select a certain measurement outcome. However, even when no outcome is selected, i.e., we merely want to model the presence of the measurement apparatus, special boundary conditions need to be imposed. Stated differently, not only any selective, but also the non-selective quantum operation for a non-trivial measurement must be non-unitary. This precludes to a large extend the possibility to trade operational for non-operational descriptions of measurement in quantum theory.

\subsection*{Acknowledgments}

I would like to thank Lucien Hardy, Rafael Sorkin, Daniele Colosi, José Antonio Zapata, Giulio Chiribella, \v{C}aslav Brukner and Philipp Höhn for valuable discussions at various stages of this project. I am also indebted to Max Dohse for a careful reading of the manuscript leading to numerous minor corrections. This work was partially supported by UNAM-DGAPA-PAPIIT project grant IN109415 and CONACYT project grant 259258.

\appendix
\numberwithin{equation}{section}

\section{On partially ordered vector spaces}
\label{sec:povs}

We collect in this appendix some mathematical definitions and facts relevant for the present article. Mostly these concern partially ordered vector spaces and related notions as well as the special case of the space of self-adjoint operators on a Hilbert space. We refer the reader for reference to text books such as the ones of Jameson \cite{Jam:orderedlinear} and of Alfsen \cite{Alf:compactconvex}. Also the book by Schaefer \cite{Sch:tvs} contains a section on partially ordered vector spaces.

\begin{dfn}
A \emph{partially ordered set} is a set $S$ with a binary relation $\le$ satisfying (a) $a\le a$ for all $a\in S$ (\emph{reflexivity}), (b) $a\le b$ and $b\le a$ implies $a=b$ for any $a,b\in S$ (\emph{antisymmetry}), (c) $a\le b$ and $b\le c$ implies $a\le c$ for any $a,b,c\in S$ (\emph{transitivity}).
\end{dfn}

We write $b\ge a$ equivalently for $a\le b$. Also $a< b$ means $a\le b$ and $a\neq b$. Similarly $a > b$ means $a\ge b$ and $a\neq b$.

\begin{dfn}
Let $S$ be a partially ordered set. Let $a,b\in S$. An element $c\in S$ is called a \emph{minimum} of $a$ and $b$ if $c\le a$ and $c\le b$ and for any $x\in S$ with $x\le a$ and $x\le b$ we have $x\le c$. Similarly, an element $c\in S$ is called a \emph{maximum} of $a$ and $b$ if $c\ge a$ and $c\ge b$ and for any $x\in S$ with $x\ge a$ and $x\ge b$ we have $x\ge c$.
\end{dfn}

\begin{dfn}
\label{dfn:lattice}
Let $S$ be a partially ordered set. If for any pair of elements $a,b\in S$ there exists a minimum and a maximum then $S$ is called a \emph{lattice}. On the other hand, if for any  pair of elements $a,b\in S$  a minimum or a maximum exists only if $a\le b$ or $b\le a$ then $S$ is called an \emph{anti-lattice}.
\end{dfn}

\begin{dfn}
\label{dfn:povs}
Let $V$ be equipped with the structures of a real vector space and of a partially ordered set. We say that the structures are compatible iff (a) for any $a,b,c\in V$ with $a\le b$ we have $a+c\le b+c$ and (b) for any $a,b\in V$ with $a\le b$ and $\lambda>0$ we have $\lambda a\le \lambda b$. We then call $V$ a \emph{partially ordered vector space}. We say that $v\in V$ is \emph{positive} iff $v\ge 0$ and denote the set of positive elements by $V^+$.
\end{dfn}

\begin{dfn}
\label{dfn:cone}
Let $V$ be a real vector space. A subset $C\subseteq V$ is called a \emph{convex cone} iff (a) for any $a,b\in C$ we have $a+b\in C$ and (b) for any $a\in C$ and $\lambda\ge 0$ we have $\lambda a\in C$. If moreover, $C\cap -C=\{0\}$ then $C$ is called a \emph{proper cone}. On the other hand, a convex cone is called a \emph{generating cone} iff $V=C-C$.
\end{dfn}

\begin{prop}
If $V$ is a partially ordered vector space, then the set $V^+$ of positive elements forms a proper convex cone. Conversely, If $V$ is a real vector space with a proper convex cone $C\subseteq V$ then this makes $V$ into a partially ordered vector space by declaring $a\le b$ iff $b-a\in C$ for any $a,b\in V$.
\end{prop}

\begin{dfn}
Let $V$ be a partially ordered vector space. $V$ is called \emph{Archimedean ordered} iff for any $v\in V$ the existence of $w\in V^+$ with $v\le \lambda w$ for all $\lambda>0$ implies $v\le 0$.
\end{dfn}

\begin{dfn}
Let $V$ be a partially ordered vector space, $a,b\in V$ with $a\le b$. The set $[a,b]\defeq \{v\in V: a\le  v \le b\}$ is called the \emph{order interval} determined by $a$ and $b$.
\end{dfn}

\begin{dfn}
Let $V$ be a partially ordered vector space. The finest topology that makes $V$ into a locally convex topological vector space and is such that all order intervals are bounded is called the \emph{order topology}.
\end{dfn}

\begin{dfn}
\label{dfn:ou}
Let $V$ be a partially ordered vector space. An element $\ou\in V$ is called an \emph{order unit} iff for any $v\in V$ there exists $\lambda >0$ such that $v\le \lambda\ou$.
\end{dfn}

\begin{dfn}
Let $V$ be a partially ordered vector space with an order unit $\ou\in V$. We define the \emph{order-unit seminorm} for $v\in V$ by,
\begin{equation}
\| v \|\defeq\inf\{\lambda>0 : v\in [-\lambda\ou,\lambda\ou]\} .
\end{equation}
\end{dfn}

\begin{prop}
The topology generated by the order-unit seminorm is the order topology.
\end{prop}

\begin{prop}
If the order is Archimedean the order-unit seminorm is a \emph{norm}.
\end{prop}

\begin{rem}
The simplest example of a partially ordered vector space is the space $\R$ of \emph{real numbers} with the standard order. $\R$ has generating cone, is Archimedean ordered, is a lattice and taking the order unit to be $1$, the order-unit norm coincides with the \emph{absolute value}.
\end{rem}

\begin{prop}
\label{prop:cpovexa}
Let $L$ be a set and $F(L)$ denote a real vector space of real valued functions on $L$. Then, $F(L)$ is a partially ordered vector space where for $f,g\in F(L)$ we have $f\le g$ iff $f(\phi)\le g(\phi)$ for all $\phi\in L$. Moreover, $F(L)$ is Archimedean ordered. If $F(L)$ is closed under taking the positive and negative part of functions it is a \emph{lattice}. If $F(L)$ consists of bounded functions only and contains constant functions, the function $\ou(\phi)\defeq 1$ for all $\phi\in L$ is an order unit. The corresponding order-unit norm in this case is precisely the \emph{supremum norm}, for $f\in F(L)$,
\begin{equation}
\|f\|=\sup\{|f(\phi)| : \phi\in L\} .
\end{equation}
\end{prop}

\begin{prop}
\label{prop:qpovexa}
Let $\cH$ be a Hilbert space and $\rop(\cH)$ the real vector space of bounded self-adjoint operators on $\cH$. The cone $\pop(\cH)\subseteq\rop(\cH)$ of positive operators is generating and makes $\rop(\cH)$ into a partially ordered vector space. $\rop(\cH)$ is Archimedean ordered and is an anti-lattice. The identity operator $\one\in\pop(\cH)$ is an order unit. The associated order-unit norm coincides precisely with the \emph{operator norm} on $\cH$, for $A\in\rop(\cH)$,
\begin{equation}
\|A\|=\sup\{\|A(\psi)\|_{\cH}: \psi\in\cH, \|\psi\|_{\cH}=1\} .
\end{equation}
\end{prop}

\begin{dfn}
\label{dfn:posmap}
Let $V$ and $W$ be partially ordered vector spaces and $f:V\to W$ a linear map. We say that $f$ is \emph{positive linear} if and only if $f$ maps positive elements to positive elements.
\end{dfn}

\begin{prop}
Let $V$ and $W$ be partially ordered vector spaces and $f:V\to W$ a positive linear map. Then $f$ is continuous with respect to the order topologies of $V$ and $W$.
\end{prop}

\begin{dfn}
\label{dfn:spos}
Let $V,W,X$ be partially ordered vector spaces and $f:V\times W\to X$ a bilinear pairing that maps products of positive elements to positive elements. We then say that $f$ is \emph{positive bilinear}. Suppose moreover that given any $a\in V$ the inequality $f(a,b)\ge 0$ for all $b\in W^+$ implies $a\ge 0$ and that given any $b\in W$, the inequality $f(a,b)\ge 0$ for all $a\in V^+$ implies $b\ge 0$. We then say that $f$ is \emph{sharply positive}.
\end{dfn}

\begin{rem}
Note that sharp positivity implies \emph{non-degeneracy}.
\end{rem}

\begin{dfn}
\label{dfn:poips}
A partially ordered vector space $V$ equipped with a positive-definite sharply positive symmetric bilinear form $V\times V\to\R$ is called a \emph{partially ordered inner product space}.
\end{dfn}

\begin{prop}
\label{prop:qhdec}
Let $\cH_1$, $\cH_2$ be Hilbert spaces. Let $\cH_1\tens\cH_2$ denote the completed tensor product Hilbert space. Consider the map $\rop(\cH_1)\times\rop(\cH_2)\to\rop(\cH_1\tens\cH_2)$, defined in the obvious way. This map is injective and sharply positive bilinear. Moreover, restricting to Hilbert-Schmidt operators the subspace spanned by the image is dense in the Hilbert-Schmidt topology.
\end{prop}

An accurate mathematical description of physical systems with infinitely many degrees of freedom frequently requires the use of linear maps that are not everywhere defined, especially in quantum field theory, but also in classical field theory. This is usually handled using topology by e.g.\ restricting such maps to dense subspaces. The present order theoretic setting offers a simpler and more elegant approach by exploiting positivity. To this end we recall that we can do addition and positive scalar multiplication in $[0,\infty]$, the positive real numbers with positive infinity added. We define in addition to the usual operations,
\begin{align}
 & 0\cdot\infty=\infty\cdot 0= 0\\
 & \lambda\cdot\infty=\infty\cdot\lambda=\infty\quad \forall\, \lambda\in (0,\infty] \\
 & a+\infty=\infty+a =\infty\quad \forall\, a\in [0,\infty] .
\end{align}

\begin{dfn}
\label{dfn:wposmap}
Let $V$ be a partially ordered vector space with generating cone $V^+$. We say that a map $f:V^+\to[0,\infty]$ is \emph{unbounded positive linear} iff
\begin{align}
 & f(\lambda a)=\lambda f(a)\quad \forall\, a\in V^+, \forall\, \lambda\in [0,\infty) \\
 & f(a + b)=f(a)+f(b)\quad \forall\, a,b\in V^+ .
\end{align}
If the image of $f$ does not contain $\infty$, we also say that $f$ is \emph{positive linear}.
\end{dfn}

\begin{rem}
It is easy to see that the notion of positive linearity of Definition~\ref{dfn:wposmap} is equivalent to that of Definition~\ref{dfn:posmap} by canonical restriction or extension of $f$. For this the generating property of the positive cone $V^+$ is essential. We shall freely use this equivalence without necessarily making explicit the restriction or extension involved.
\end{rem}

\begin{prop}
Let $V$ be a partially ordered vector space with generating cone $V^+$ and $f:V^+\to [0,\infty]$ an unbounded positive linear map. Let $W^+\defeq f^{-1}([0,\infty))$ and set $W\defeq W^+-W^+$. Then $W$ is a partially ordered subspace of $V$ with $W^+= V^+\cap W$ its cone of positive elements which is generating. The restriction of $f$ to $W^+$ and subsequent extension to $W$ is positive linear.
\end{prop}

For the present paper the most important application of unbounded positivity arises in the notion of unbounded positive bilinear maps.

\begin{dfn}
\label{dfn:wposbmap}
Let $V$, $W$ be partially ordered vector spaces with generating cones. We say that a map $f:V^+\times W^+\to [0,\infty]$ is \emph{unbounded positive bilinear} iff
\begin{align}
 & f(\lambda a,b)=f(a,\lambda b)=\lambda f(a,b)\quad \forall\, a\in V^+, \forall\, b\in W^+, \forall\, \lambda\in [0,\infty) \\
 & f(a + b,c)=f(a,c)+f(b,c)\quad \forall\, a,b\in V^+,\forall\, c\in W^+ , \\
 & f(a,b+c)=f(a,b)+f(a,c)\quad \forall\, a\in V^+,\forall\, b,c\in W^+ ,
\end{align}
If the image of $f$ does not contain $\infty$, we also say that $f$ is \emph{positive bilinear}.
\end{dfn}

\begin{prop}
\label{prop:wpbmapdef}
Let $V$, $W$ be partially ordered vector spaces with generating cones and $f:V^+\times W^+\to [0,\infty]$ an unbounded positive bilinear map. We say that given $a\in V$ and $b\in W$, $f(a,b)$ is \emph{well-defined} if there exist $a^+,a^-\in V^+$ and $b^+,b^-\in W^+$ such that $a=a^+-a^-$ and $b=b^+-b^-$ as well as $f(a^+,b^+)<\infty$, $f(a^+,b^-)<\infty$, $f(a^-,b^+)<\infty$ and $f(a^-,b^-)<\infty$. In that case we declare the value of $f(a,b)$ to be $f(a^+,b^+)-f(a^+,b^-)-f(a^-,b^+)+f(a^-,b^-)$. If $f(a,b)$ is well-defined, its value is unique. Moreover, this value coincides with the (extension of) any positive bilinear restriction of $f$.
\end{prop}

\begin{dfn}
\label{dfn:wsp}
Let $V$, $W$ be partially ordered vector spaces with generating cones and $f:V^+\times W^+\to [0,\infty]$ an unbounded positive bilinear map. Suppose that for any $a\in V$ the inequality $f(a,b)\ge 0$ for all $b\in W^+$ where it is well-defined implies $a\ge 0$ and that for any $b\in W$, the inequality $f(a,b)\ge 0$ for all $a\in V^+$ where it is well-defined implies $b\ge 0$. We then say that $f$ is \emph{sharply positive}.
\end{dfn}

\begin{dfn}
\label{dfn:wpd}
Let $V$ be a partially ordered vector space with generating cone and $f:V^+\times V^+\to [0,\infty]$ an unbounded positive symmetric bilinear map. We say that $f$ is \emph{positive-definite} if and only if $f(a,a)> 0$ for all $a\in V\setminus\{0\}$ whenever $f(a,a)$ is well-defined.
\end{dfn}

\begin{dfn}
\label{dfn:wpoip}
A partially ordered vector space $V$ equipped with a positive-definite unbounded sharply positive symmetric bilinear form $V^+\times V^+\to [0,\infty]$ is called a \emph{partially ordered unbounded inner product space}.
\end{dfn}

\begin{prop}
\label{prop:cwpoipexa}
Let $L$ be a measure space with measure $\mu$. Let $F(L)$ denote a partially ordered vector space of real valued measurable functions on $L$ with generating cone. Then, a positive-definite unbounded positive symmetric bilinear map $F^+(L)\times F^+(L)\to [0,\infty]$ is given for all $b,c\in F(L)$ by,
\begin{equation}
 (b,c)\mapsto \int_L b(\phi) c(\phi)\,\xd\mu(\phi). 
\end{equation}
If $F(L)$ is closed under taking the positive and negative part of functions it is a lattice and sharp positivity is also satisfied. In particular, $F(L)$ becomes in this way a partially ordered unbounded inner product space.
\end{prop}

\begin{rem}
\label{rem:cwpoipexa}
A maximal subspace of $F(L)$ such that (the extension of) this pairing is positive bilinear is the subspace of \emph{square-integrable functions}. The pairing is then precisely the usual inner product on the space of square-integrable functions. Another choice yielding a positive bilinear pairing arises from pairing \emph{essentially bounded functions} with \emph{integrable functions}. The latter is also precisely the class of functions that can be paired with non-zero constant functions such as the unit function chosen as order unit in Proposition~\ref{prop:cpovexa}.
\end{rem}

\begin{prop}
\label{prop:qtrace}
Let $\cH$ be a Hilbert space. The trace defines a positive-definite unbounded sharply positive symmetric bilinear map $\pop(\cH)\times\pop(\cH)\to [0,\infty]$ via,
\begin{equation}
 (a,b)\mapsto \tr(a b) .
\end{equation}
In particular, $\rop(\cH)$ becomes in this way a partially ordered unbounded inner product space.
\end{prop}

\begin{rem}
\label{rem:qwpoipexa}
A maximal subspace of $\rop(\cH)$ such that (the extension of) this pairing is positive bilinear is the space of (self-adjoint) \emph{Hilbert-Schmidt operators}. It is then precisely the Hilbert-Schmidt inner product. On the other hand, if we only restrict one of the two spaces in the pairing, a maximal subspace of $\rop(\cH)$ such that (the extension of) this pairing is positive bilinear is the space of (self-adjoint) \emph{trace class operators}. In particular, the latter is also the maximal subspace that can be paired with the unit operator $\one$.
\end{rem}

\begin{prop}
\label{prop:cpwpos}
Let $\cH_1$, $\cH_2$ be Hilbert spaces. Let $S:\rop(\cH_1)\to\rop(\cH_2)$ be a positive linear map between the corresponding partially ordered vector spaces of self-adjoint operators. Then, $S$ is \emph{completely positive} iff there exists an unbounded positive linear map $S':\pop(\cH_1\tens\cH_2)\to [0,\infty]$ such that for all $a\in \pop(\cH_1)$ and $b\in\pop(\cH_2)$ we have
\begin{equation}
S'(a\tens b)=\tr(S(a) b) .
\end{equation}
The map $S'$ is unique if it exists.
\end{prop}

\bibliographystyle{stdnodoi} 
\bibliography{stdrefsb}
\end{document}